\documentclass[twocolumn,aps,superscriptaddress,showpacs,nofootinbib,floatfix]{revtex4}
\usepackage{epsfig,bm,feynmf}
\usepackage{graphics}
\usepackage{amsmath}
\usepackage[normalem]{ulem}  
\usepackage[dvips]{color} 

\renewcommand\sout{\bgroup \color{red} \ULdepth=-.5ex \ULset}
\begin{document}


\title{Single electrons from heavy-flavor mesons in relativistic heavy-ion collisions}


\author{Taesoo Song}\email{song@fias.uni-frankfurt.de}
\affiliation{Institute for Theoretical Physics, Johann Wolfgang Goethe Universit\"{a}t, Frankfurt am Main, Germany}
\affiliation{Frankfurt Institute for Advanced Studies, Johann Wolfgang Goethe Universit\"{a}t, Frankfurt am Main, Germany}
\affiliation{Institut f\"{u}r Theoretische Physik, Universit\"{a}t
Gie\ss en, Germany}

\author{Hamza Berrehrah}
\affiliation{Institute for Theoretical Physics, Johann Wolfgang Goethe Universit\"{a}t, Frankfurt am Main, Germany}
\affiliation{Frankfurt Institute for Advanced Studies, Johann Wolfgang Goethe Universit\"{a}t, Frankfurt am Main, Germany}

\author{Juan~M. Torres-Rincon}
\affiliation{Frankfurt Institute for Advanced Studies, Johann Wolfgang Goethe Universit\"{a}t, Frankfurt am Main, Germany}

\author{Laura Tolos}
\affiliation{Frankfurt Institute for Advanced Studies, Johann Wolfgang Goethe Universit\"{a}t, Frankfurt am Main, Germany}
\affiliation{Institut de Ciencies de l'Espai (IEEC/CSIC), Campus Universitat Autonoma de Barcelona,
Carrer de Can Magrans, s/n, E-08193 Bellaterra, Spain}

\author{Daniel Cabrera}
\affiliation{Instituto de F\'{\i}sica Corpuscular (IFIC), Centro
Mixto Universidad de Valencia - CSIC, Institutos de Investigaci\'on de
Paterna, Ap. Correos 22085, E-46071 Valencia, Spain.}

\author{Wolfgang Cassing}
\affiliation{Institut f\"{u}r Theoretische Physik, Universit\"{a}t
Gie\ss en, Germany}

\author{Elena Bratkovskaya}
\affiliation{Institute for Theoretical Physics, Johann Wolfgang Goethe Universit\"{a}t, Frankfurt am Main, Germany}
\affiliation{GSI Helmholtzzentrum f\"{u}r Schwerionenforschung GmbH, Darmstadt, Germany}

\begin{abstract}
We study the single electron spectra from $D-$ and $B-$meson
semileptonic decays in Au+Au collisions at $\sqrt{s_{\rm NN}}=$200,
62.4, and 19.2 GeV by employing the parton-hadron-string dynamics
(PHSD) transport approach that has been shown to reasonably describe
the charm dynamics at Relativistic-Heavy-Ion-Collider (RHIC) and
Large-Hadron-Collider (LHC) energies on a microscopic level. In this
approach the initial charm and bottom quarks are produced by using
the PYTHIA event generator which is tuned to reproduce the
fixed-order  next-to-leading logarithm (FONLL) calculations for
charm and bottom production. The produced charm and bottom quarks
interact with off-shell (massive) partons in the quark-gluon plasma
with scattering cross sections which are calculated in the dynamical
quasi-particle model (DQPM) that is matched to reproduce the
equation of state of the partonic system above the deconfinement
temperature $T_c$. At energy densities close to the critical energy
density ($\approx$  0.5 GeV/fm$^3$) the charm and bottom quarks are
hadronized into $D-$ and $B-$mesons through either coalescence or
fragmentation. After hadronization the $D-$ and $B-$mesons interact
with the light hadrons by employing the scattering cross sections
from an effective Lagrangian. The final $D-$ and $B-$mesons then
produce single electrons through semileptonic decay. We find that
the PHSD approach well describes the nuclear modification factor $R_{\rm AA}$ and elliptic flow
$v_2$ of single electrons in d+Au and Au+Au collisions at $\sqrt{s_{\rm
NN}}=$ 200 GeV and the elliptic flow in Au+Au reactions at $\sqrt{s_{\rm NN}}=$ 62.4 GeV
from the PHENIX collaboration, however, the large $R_{\rm AA}$ at
$\sqrt{s_{\rm NN}}=$ 62.4 GeV is not described at all. Furthermore,
we make predictions for the $R_{\rm AA}$ of $D-$mesons and of single
electrons at  the lower energy of $\sqrt{s_{\rm NN}}=$ 19.2 GeV.
Additionally, the medium modification of the azimuthal angle $\phi$
between a heavy quark and a heavy antiquark is studied. We find that
the transverse flow enhances the azimuthal angular distributions
close to $\phi=$ 0 because the heavy flavors strongly interact with
nuclear medium in relativistic heavy-ion collisions and almost flow
with the bulk matter.
\end{abstract}


\maketitle

\section{introduction}

Relativistic heavy-ion collisions are the experiments of choice to
generate hot and dense matter in the laboratory. Whereas in low
energy collisions one produces dense nuclear matter with moderate
temperature at large baryon chemical potential $\mu_B$,
ultra-relativistic collisions at  Relativistic Heavy Ion Collider
(RHIC) or Large Hadron Collider (LHC) energies produce extremely hot
matter at small baryon chemical potential. In order to explore the
phase diagram of strongly interacting matter as a function of $T$
and $\mu_B$ both type of collisions are mandatory. According to
lattice calculations of quantum chromodynamics
(lQCD)~\cite{Bernard:2004je,Aoki:2006we,Bazavov:2011nk}, the phase
transition from hadronic to partonic degrees of freedom (at
vanishing baryon chemical potential $\mu_B$=0) is a crossover. This
phase transition is expected to turn into a first order transition
at a critical point $(T_r, \mu_r)$ in the phase diagram with
increasing baryon chemical potential $\mu_B$. Since this critical
point cannot be  determined theoretically in a reliable way the beam
energy scan (BES) program performed at the  RHIC by the STAR
collaboration aims to find the critical point and the phase boundary
by gradually decreasing the collision
energy~\cite{Mohanty:2011nm,Kumar:2011us}.

Since the hot and dense matter produced in relativistic heavy-ion
collisions appears only for a couple of fm/c, it is a big challenge
for experiments to investigate its properties. The heavy flavor
mesons are considered to be promising probes in this search since
the production of heavy flavor requires a large energy-momentum
transfer. Thus it takes place early in the heavy-ion collisions, and
- due to the large energy-momentum transfer - should be described by
perturbative quantum chromodynamics (pQCD). The produced heavy
flavor then interacts with the hot dense matter
 (of partonic or hadronic nature) by exchanging energy and momentum.
As a result, the ratio of the measured number of heavy flavors in
heavy-ion collisions to the expected number in the absence of
nuclear or partonic matter, which is the definition of $R_{\rm AA}$
(cf. Section VII), is suppressed at high transverse momentum, and
the elliptic flow of heavy flavor is generated by the interactions
in noncentral heavy-ion collisions. Although it had been expected
that the $R_{\rm AA}$ of heavy flavor is less suppressed and its
elliptic flow is smaller as compared to the corresponding quantities
for light hadrons, the experimental data show that the suppression
of heavy-flavor hadrons  at high transverse momentum and its
elliptic flow $v_2$ are comparable to those of light
hadrons~\cite{ALICE:2012ab,Abelev:2013lca}. This is a puzzle for
heavy-flavor production and dynamics in relativistic heavy-ion
collisions as pointed out by many groups
~\cite{Moore:2004tg,Zhang:2005ni,Molnar:2006ci,Linnyk:2008hp,Gossiaux:2010yx,Nahrgang:2013saa,He:2011qa,He:2012df,He:2014epa,Uphoff:2011ad,Uphoff:2012gb,Cao:2011et,Cao:2015eek,Greco,Nahrgang:2016lst}.
For  recent reviews we refer the reader to Refs.
\cite{Andro,Rapp16}.

Since the heavy-flavor interactions are closely related to the
dynamics of the partonic or hadronic degrees-of-freedom due to their
mutual interactions, a proper description of the relativistic
heavy-ion collisions and their bulk dynamics is necessary. In this
study we employ the parton-hadron-string dynamics (PHSD) approach,
which differs from the conventional Boltzmann-type models in the
aspect~\cite{Cassing:2009vt} that the degrees-of-freedom for the QGP
phase are off-shell massive strongly-interacting quasi-particles
that generate their own mean-field potential. The masses of the
dynamical quarks and gluons in the QGP are distributed according to
spectral functions whose pole positions and widths, respectively,
are defined by the real and imaginary parts of their self-energies
\cite{PHSDreview}. The partonic propagators and self-energies,
furthermore, are defined in the dynamical quasiparticle model (DQPM)
in which the strong coupling and the self-energies are fitted to
lattice QCD results.

We recall that the PHSD approach has
successfully described numerous experimental data in relativistic
heavy-ion collisions from the Super Proton Synchrotron (SPS) to LHC
energies~\cite{Cassing:2009vt,PHSDrhic,Volo,PHSDreview}. More recently,
the charm production and propagation has been explicitly implemented
in the PHSD and detailed studies on the charm dynamics and
hadronization/fragmention have been performed at top RHIC and LHC
energies in comparison to the available
data~\cite{Song:2015sfa,Song:2015ykw}. In the PHSD approach the
initial charm and anticharm quarks are produced by using the PYTHIA
event generator~\cite{Sjostrand:2006za} which is tuned to the
transverse momentum and rapidity distributions of charm and
anticharm quarks from the Fixed-Order  Next-to-Leading Logarithm
(FONLL) calculations~\cite{Cacciari:2012ny}. The produced charm and
anticharm quarks interact in the QGP with off-shell partons and are
hadronized into $D-$mesons close to the critical energy density for
the crossover transition either through fragmentation or
coalescence. We stress that the coalescence is a genuine feature of
heavy-ion collisions and does not show up in p+p interactions. The
hadronized $D-$mesons then interact with light hadrons in the
hadronic phase until freeze out and subsequently undergoes semileptonic decay. We
have found that the PHSD approach, which has been applied for charm
production in Au+Au collisions at $\sqrt{s_{\rm NN}}=$200
GeV~\cite{Song:2015sfa} and in Pb+Pb collisions at $\sqrt{s_{\rm
NN}}=$2.76 TeV~\cite{Song:2015ykw}, describes the $R_{\rm AA}$ as
well as the $v_2$ of $D-$mesons  in reasonable agreement with the
experimental data from the STAR
collaboration~\cite{Adamczyk:2014uip,Tlusty:2012ix} and from the
ALICE collaboration~\cite{Adam:2015sza,Abelev:2014ipa} when
including the initial shadowing effect in the latter case.

In this work we, furthermore, extend the PHSD approach to bottom
production in relativistic heavy-ion collisions. As in case of
charm, the initial bottom pair is produced by using the PYTHIA event
generator, and the transverse momentum and rapidity distributions
are adjusted to those from the FONLL calculations. Also the
scattering cross sections of bottom quarks with off-shell partons
are calculated in the DQPM on the same basis as the $c-$quarks. The
bottom quarks are hadronized into $B-$mesons near the critical
energy density in the same way as charm quarks. Furthermore, the
scattering cross sections of $B-$mesons with light hadrons (in the
hadronic phase) are calculated from a similar effective Lagrangian
as used for $D-$mesons.

Presently, there are no exclusive experimental data for $B-$meson
production from relativistic heavy-ion collisions. The PHENIX
collaboration instead measured the single electrons which are
produced through the semileptonic decay of $D-$ and $B-$mesons in
Au+Au collisions at $\sqrt{s_{\rm NN}}=$ 200 and 62.4
GeV~\cite{Adare:2006nq,Adare:2014rly,Adamczyk:2014yew}. In this work we will study
the bottom production through the single electrons in Au+Au
collisions at $\sqrt{s_{\rm NN}}=$ 200 and 62.4 GeV by using the
extended PHSD. Additionally, we make predictions for $D-$meson and
single electron production at the much lower energy $\sqrt{s_{\rm
NN}}=$ 19.2 GeV while we compare with the experimental results
available at $\sqrt{s_{\rm NN}}=$ 200 and 62.4 GeV. Finally, we
study the medium modification of the azimuthal angle $\phi$ of a
heavy-flavor pair in relativistic heavy-ion collisions by the
interactions with the partonic or hadronic medium.

This paper is organized as follows: The production of heavy mesons
and their semileptonic decay in p+p collisions is described in
detail and compared with experimental data in Sec.~\ref{pp}. The
initial production of heavy quarks is explained in
Sec.~\ref{shadowing} including the shadowing effect in relativistic
heavy-ion collisions. We then present the heavy quark interactions
in the QGP, their hadronization and hadronic interactions,
respectively, in Sec.~\ref{QGP}, \ref{hadronization} and \ref{hg}.
Finally, we show our results in Sec.~\ref{results} in comparison
with the experimental data while a summary closes this study in
Sec.~\ref{summary}.

\section{Single electrons from heavy flavor in p+p collisions}\label{pp}

As pointed out in the introduction the charm and bottom quark pairs
are produced through initial hard nucleon-nucleon scattering in
relativistic heavy-ion collisions. We employ the PYTHIA event
generator to produce the heavy-quark pairs and modify their
transverse momentum and rapidity such that they are similar to those
from the FONLL calculations~\cite{Cacciari:2012ny}. In case of p+p collisions at
the top RHIC energy of $\sqrt{s}=200$ GeV, the transverse momentum
and the rapidity of the charm quark from the PYTHIA event generator are reduced by 10 \% and 16
\%~\cite{Song:2015sfa}, respectively, and those of the bottom quark
are unmodified. The transverse momentum of charm quarks at the
invariant energy of $\sqrt{s}=62.4$ GeV is modified to $\rm
p_T^*=p_T-0.6 \tanh(p_T/1.2~GeV)$, where $\rm p_T$ and $\rm p_T^*$,
respectively, are the original and modified transverse momenta,
while the rapidity is reduced by 15 \%~\cite{Cacciari:2012ny}. The transverse momentum and
the rapidity of a bottom quark at the same energy are, respectively,
reduced by 5 \% and enhanced by 15 \%~\cite{Cacciari:2012ny}. The `MSEL' code in PYTHIA, which enables to select a specific production channel, is
taken to be the default value 1 for charm production, and 5 for
bottom production.

\begin{figure} [h]
\centerline{
\includegraphics[width=9.5 cm]{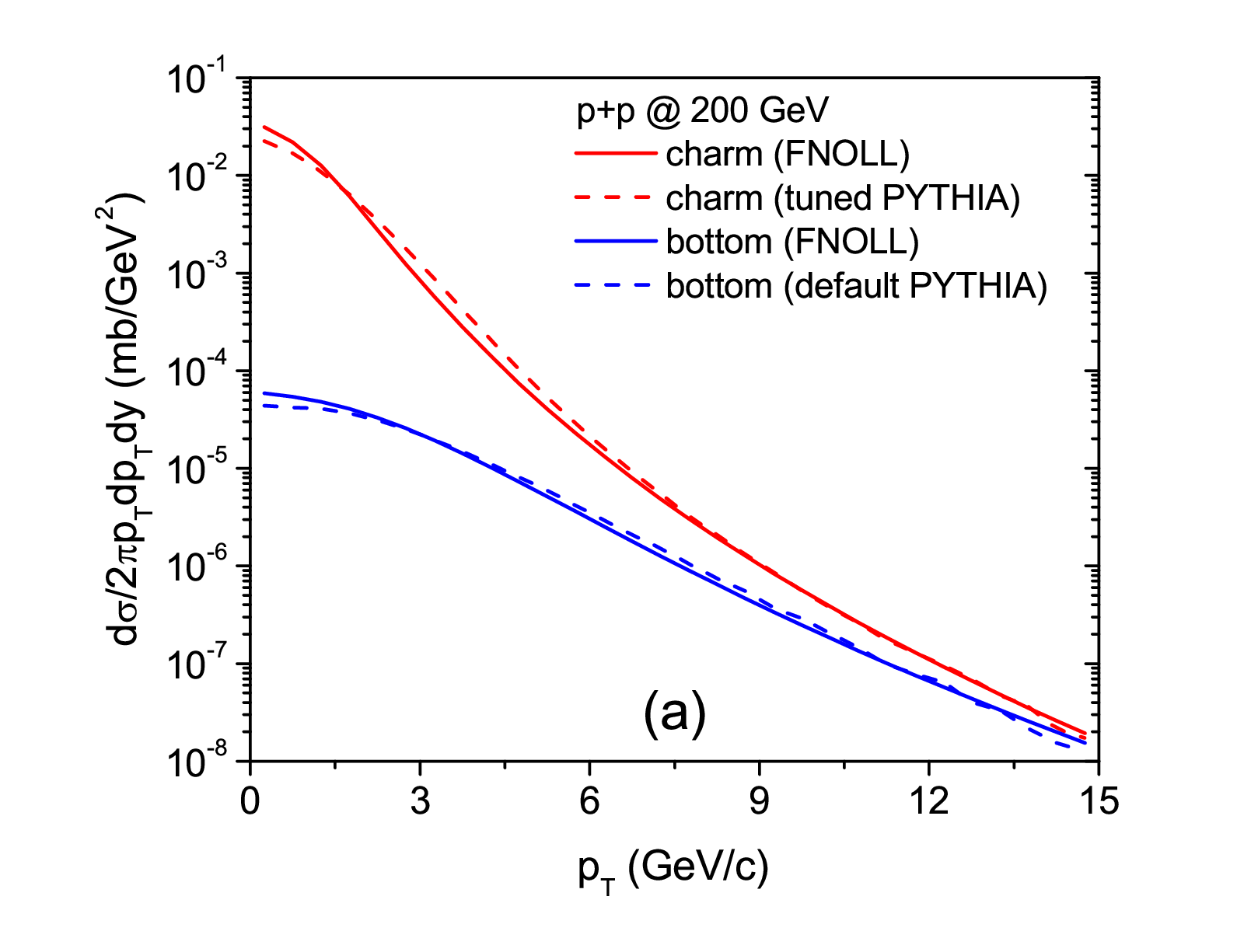}}
\centerline{
\includegraphics[width=9.5 cm]{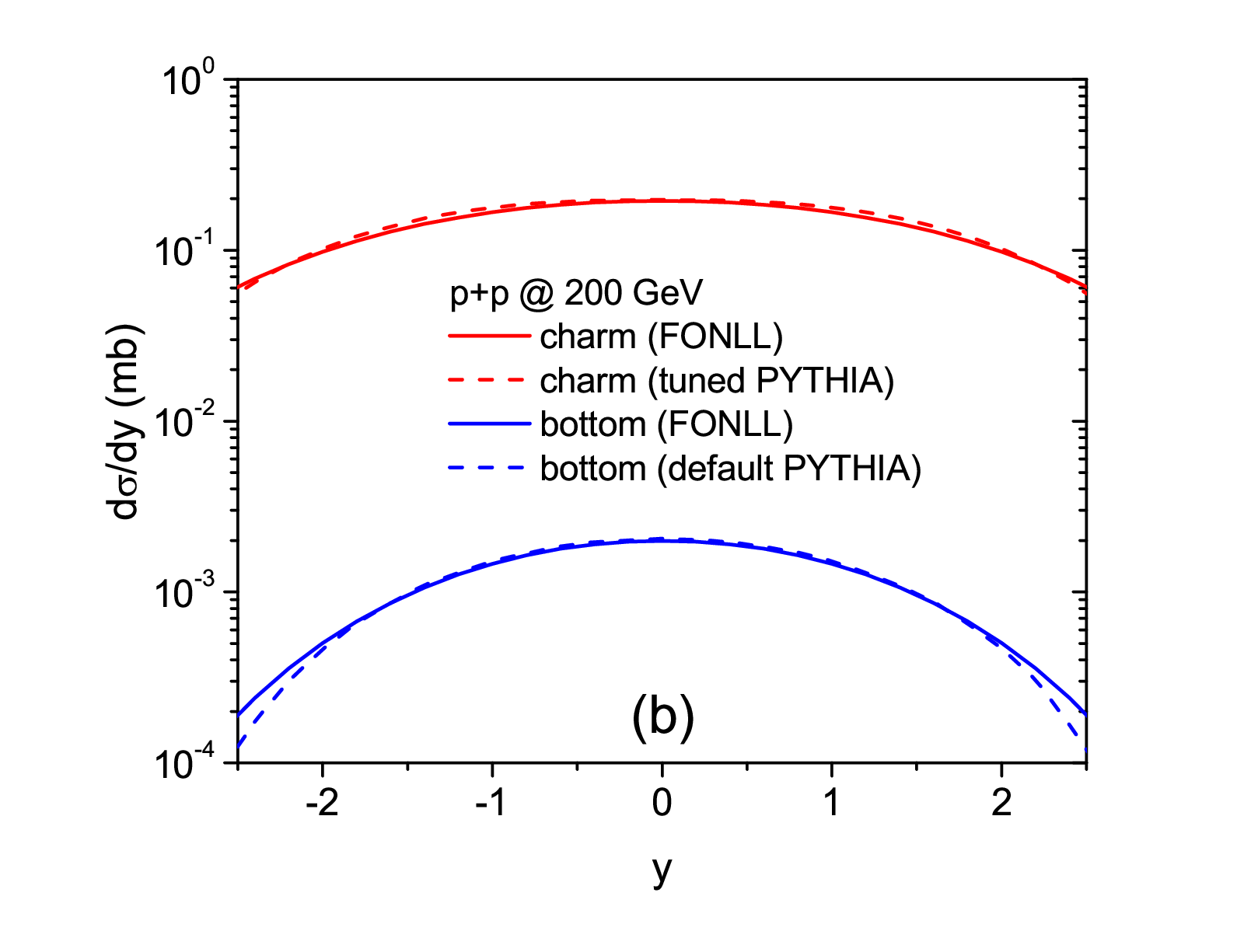}}
\caption{(Color online) The ${\rm p_T}$ spectra (a) and rapidity
distributions (b) of charm and bottom quarks in p+p collisions at
$\sqrt{s}=200$ GeV as generated by the tuned PYTHIA event generator
(dashed) in comparison to those from FONLL (solid).} \label{pp200Q}
\end{figure}

\begin{figure} [h]
\centerline{
\includegraphics[width=9.5 cm]{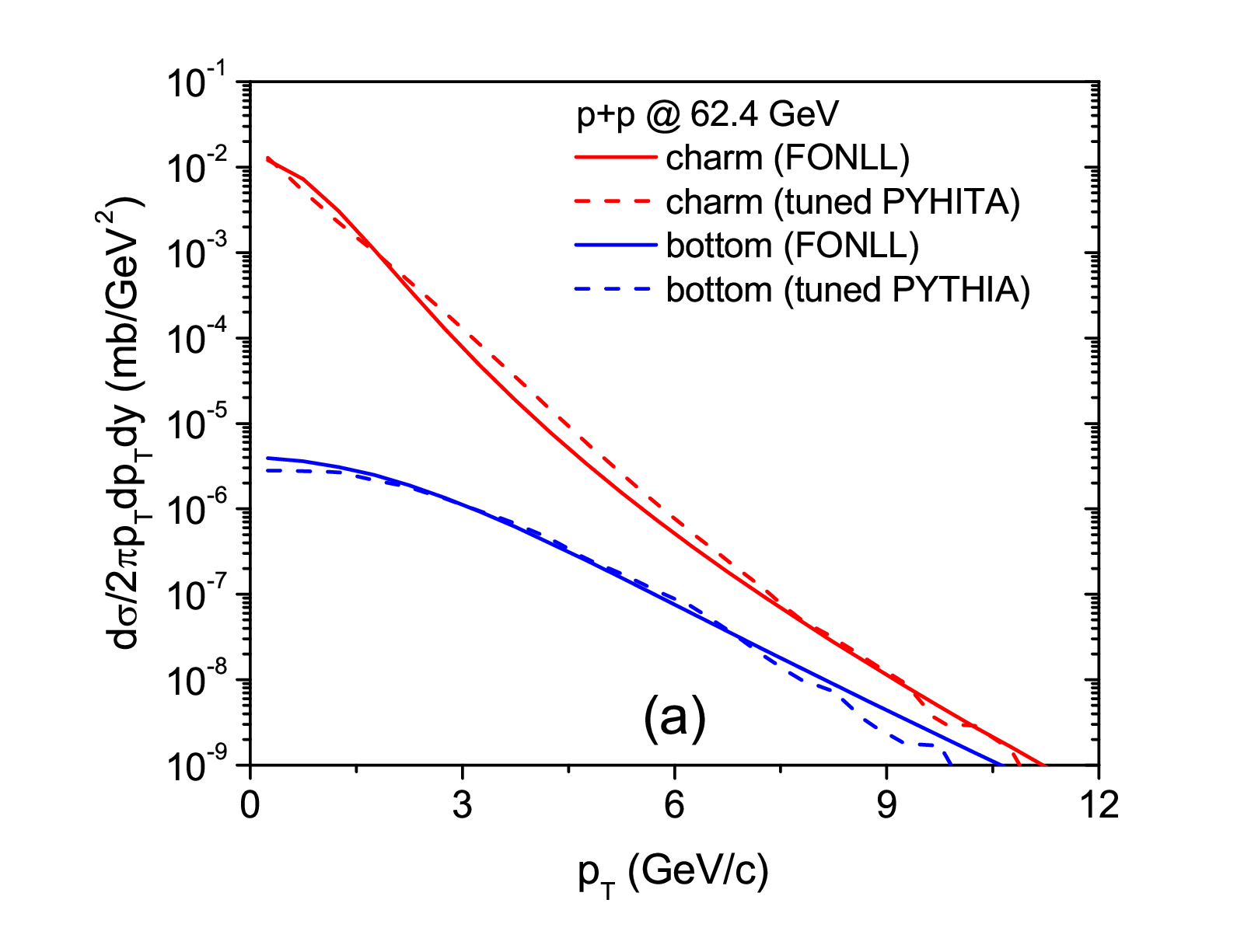}}
\centerline{
\includegraphics[width=9.5 cm]{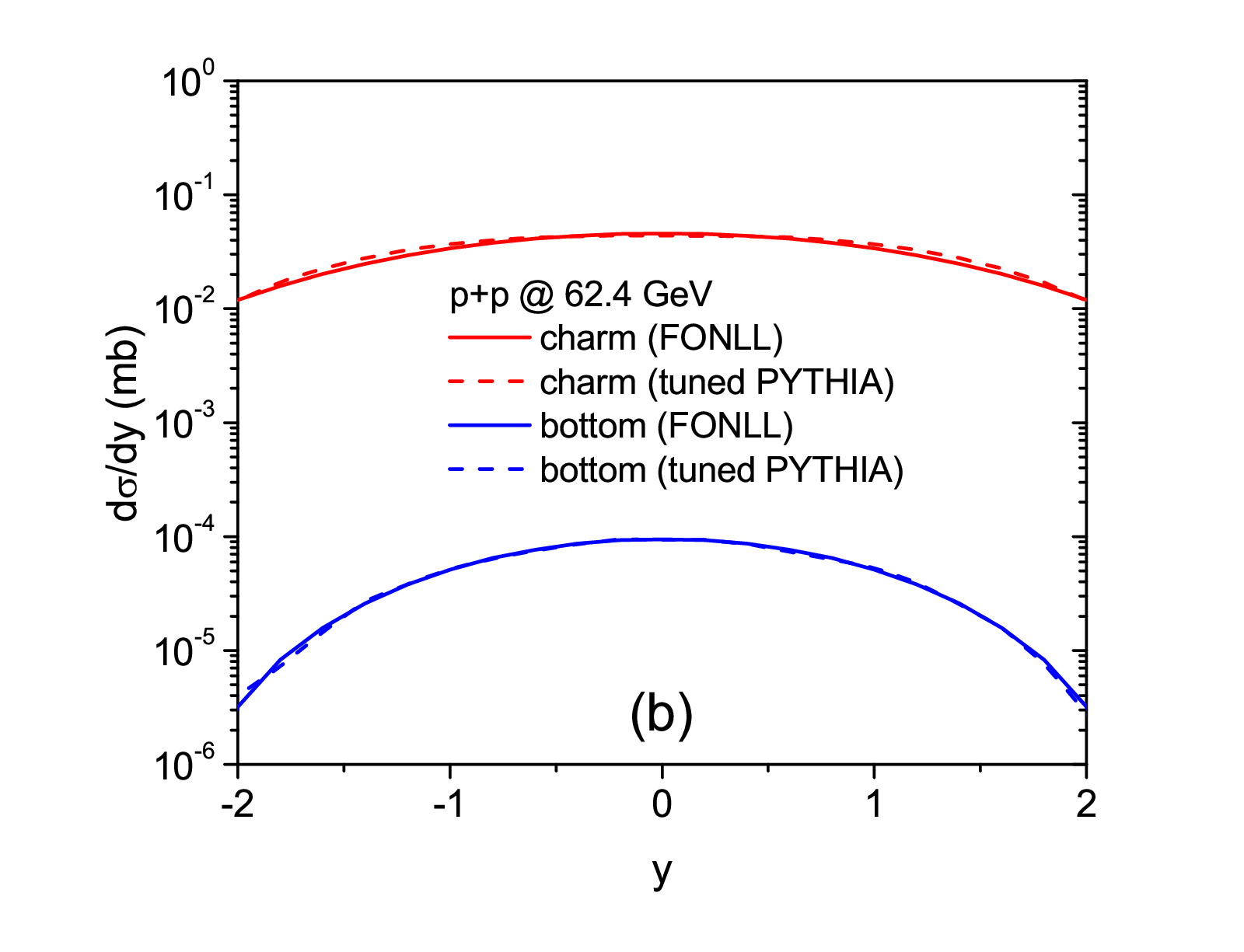}}
\caption{(Color online) The ${\rm p_T}$ spectra (a) and rapidity
distributions (b) of charm and bottom quarks in p+p collisions at
$\sqrt{s}=62.4$ GeV as generated by the tuned PYTHIA event generator
(dashed) in comparison to those from FONLL (solid).} \label{pp62Q}
\end{figure}

Figures \ref{pp200Q} and \ref{pp62Q} show the ${\rm p_T}$ spectra
and rapidity distributions of charm and bottom quarks in p+p
collisions at $\sqrt{s}=200$ GeV and 62.4 GeV,  where the dashed and
solid lines are from the tuned PYTHIA and FONLL calculations,
respectively. The FONLL calculations are rescaled such that the
total cross sections for charm production are 0.8 and 0.12 mb at
$\sqrt{s}=200$ GeV and 62.4 GeV, respectively, from the measurement of the STAR collaboration~\cite{Adamczyk:2012af} and the interpolation of the PHSD~\cite{Song:2015sfa}. The ratios of the bottom cross
section to the charm cross section are taken to be 0.75 \% and 0.145~\%
at the same energies as in the FONLL calculations~\cite{Cacciari:2012ny}. We find that
our tuned PYTHIA generator gives very similar charm and bottom
distributions as those from FONLL calculations, which fixes the
input from pQCD.

\begin{figure} [h]
\centerline{
\includegraphics[width=9.5 cm]{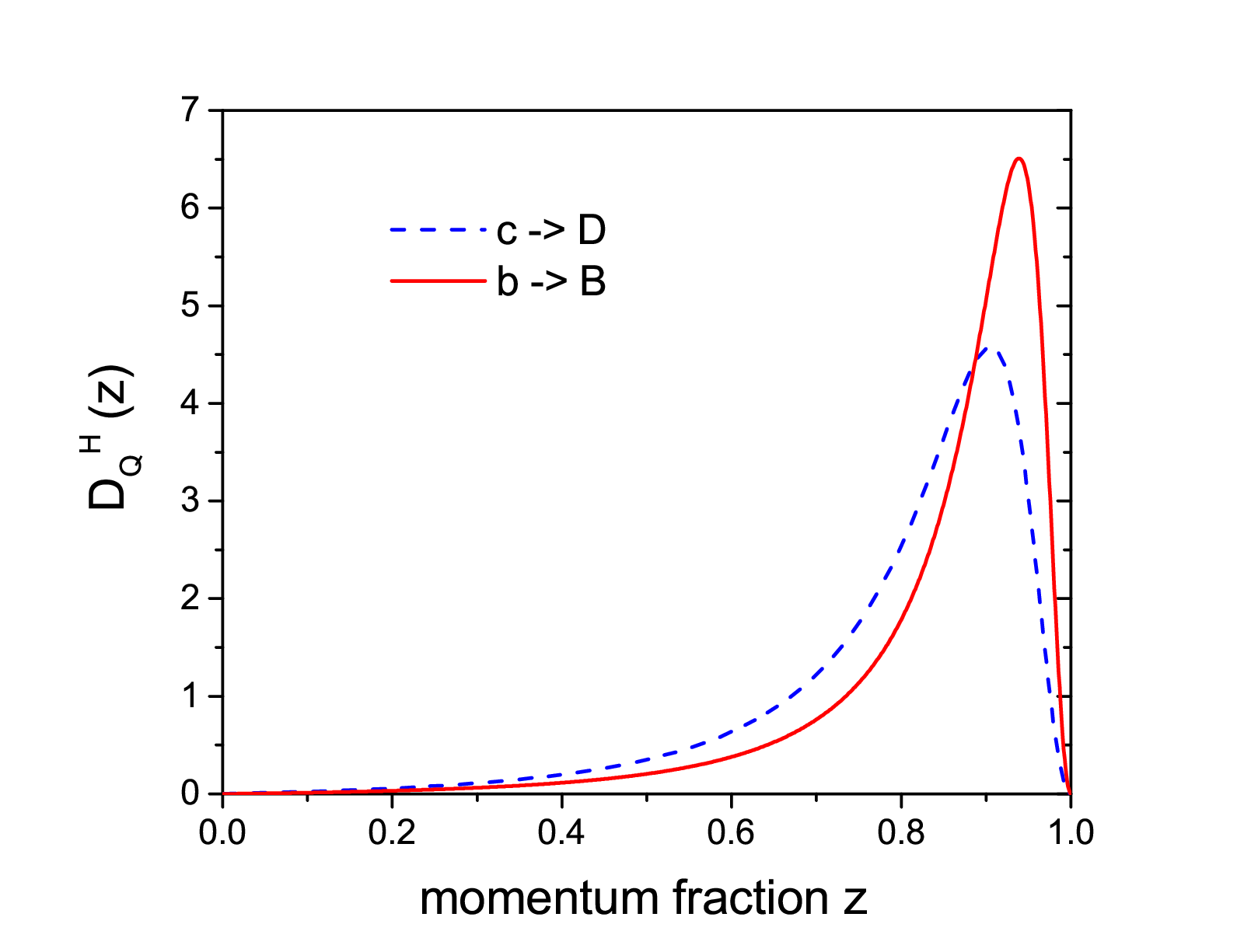}}
\caption{(Color online) Fragmentation functions of charm (dashed)
and bottom (solid) as a function of the hadron momentum fraction
$z$.} \label{peterson}
\end{figure}

Furthermore, the produced charm and bottom quarks in hard
nucleon-nucleon collisions are hadronized by emitting soft gluons,
which is denoted by `fragmentation'. We use the fragmentation
function of Peterson which reads as~\cite{Peterson:1982ak}
\begin{eqnarray}
D_Q^H(z)\sim \frac{1}{z[1-1/z-\epsilon_Q/(1-z)]^2},
\end{eqnarray}
where $z$ is the momentum fraction of the hadron $H$ fragmented from
the heavy quark $Q$ while $\epsilon_Q$ is a fitting parameter which
is taken to be $\epsilon_Q$ = 0.01 for charm~\cite{Song:2015sfa} and
0.004 for bottom~\cite{DELPHI:2011aa}. Figure~\ref{peterson} shows
the fragmentation functions of charm and bottom quarks as a function
of the hadron momentum fraction $z$. Since the charm quark is much
heavier than the soft emitted gluons, it takes a large momentum
fraction in the fragmentation into a $D-$meson. It is even more
pronounced in the case of a bottom quark.

The chemical fractions of the charm quark decay into
$D^+,~D^0,~D^{*+},~D^{*0},~D_s$, and $\Lambda_c$ are  taken to be
14.9, 15.3, 23.8, 24.3, 10.1, and 8.7
\%~\cite{Gladilin:1999pj,Chekanov:2007ch,Abelev:2012vra,Song:2015ykw},
respectively, and those of the bottom quark decay  into
$B^-,~\bar{B^0},~\bar{B^0}_s$, and $\Lambda_b$ are 39.9, 39.9, 11,
and 9.2 \%~\cite{DELPHI:2011aa}. After the momentum and the species
of the fragmented particle are decided by  Monte Carlo, the energy
of the fragmented particle is adjusted to be on-shell. Furthermore,
the $D^*$ mesons first decay into $D+\pi$ or $D+\gamma$, and then
the $D-$ and $B-$mesons produce single electrons through the
semileptonic decay~\cite{Agashe:2014kda}. For simplicity, it is
assumed that the transition amplitude for the semileptonic decay is
constant and does not depend on particle momentum. Denoting the
energy, momentum, and mass of a particle $i$ by $E_i$, $p_i$, and
$m_i$ in the decay $D\rightarrow K+e+\bar{\nu_e}$, the phase space
for the final states ($K+e+\bar{\nu_e}$) is then proportional to
\begin{eqnarray}
\frac{d^3p_K}{E_K}\frac{d^3p_e}{E_e}\frac{d^3p_{\bar{\nu}}}{E_{\bar{\nu}}}\delta^4(p_D-p_K-p_e-p_{\bar{\nu}}).
\end{eqnarray}
In the center-of-mass frame of the leptons $e$ and $\bar{\nu_e}$, it is simplified to
\begin{eqnarray}
\sim \frac{d^3p_K}{E_K}dp_e\delta(E_D-E_K-2p_e)~~~\nonumber\\
\sim \frac{d^3p_K}{E_K}~+{\rm energy~conservation},
\end{eqnarray}
because $E_e=E_{\bar{\nu}}=p_e$, assuming the leptons $e$ and
$\bar{\nu_e}$ to be massless. Finally, we perform a Lorentz boost to
the rest frame of the $D-$meson, where the solid angle of ${\bf
p}_K$ is assumed to be isotropic:
\begin{eqnarray}
\sim \frac{p_K^2}{E_K}dp_K ~+{\rm energy~conservation}.
\end{eqnarray}
The momentum $p_K$ itself is decided by a random number as follows:
\begin{eqnarray}
{\rm random~number}=\frac{\int_0^{p_K} dp (p^2/\sqrt{m_K^2+p^2})}{\int_0^{p_{\rm max}} dp (p^2/\sqrt{m_K^2+p^2})},
\end{eqnarray}
where $p_{\rm max}=(m_D^2-m_K^2)/(2m_D)$ is fixed by energy
conservation. Accordingly, once $p_K$ is fixed, the invariant mass
of the lepton pair ($e$ and $\bar{\nu_e}$), and then $p_e$ in the
center-of-mass frame of $e$ and $\bar{\nu_e}$ are fixed. The solid
angle of each particle is determined by Monte Carlo and its
energy-momentum boosted back to the p+p collision frame.

\begin{figure} [h]
\centerline{
\includegraphics[width=9.5 cm]{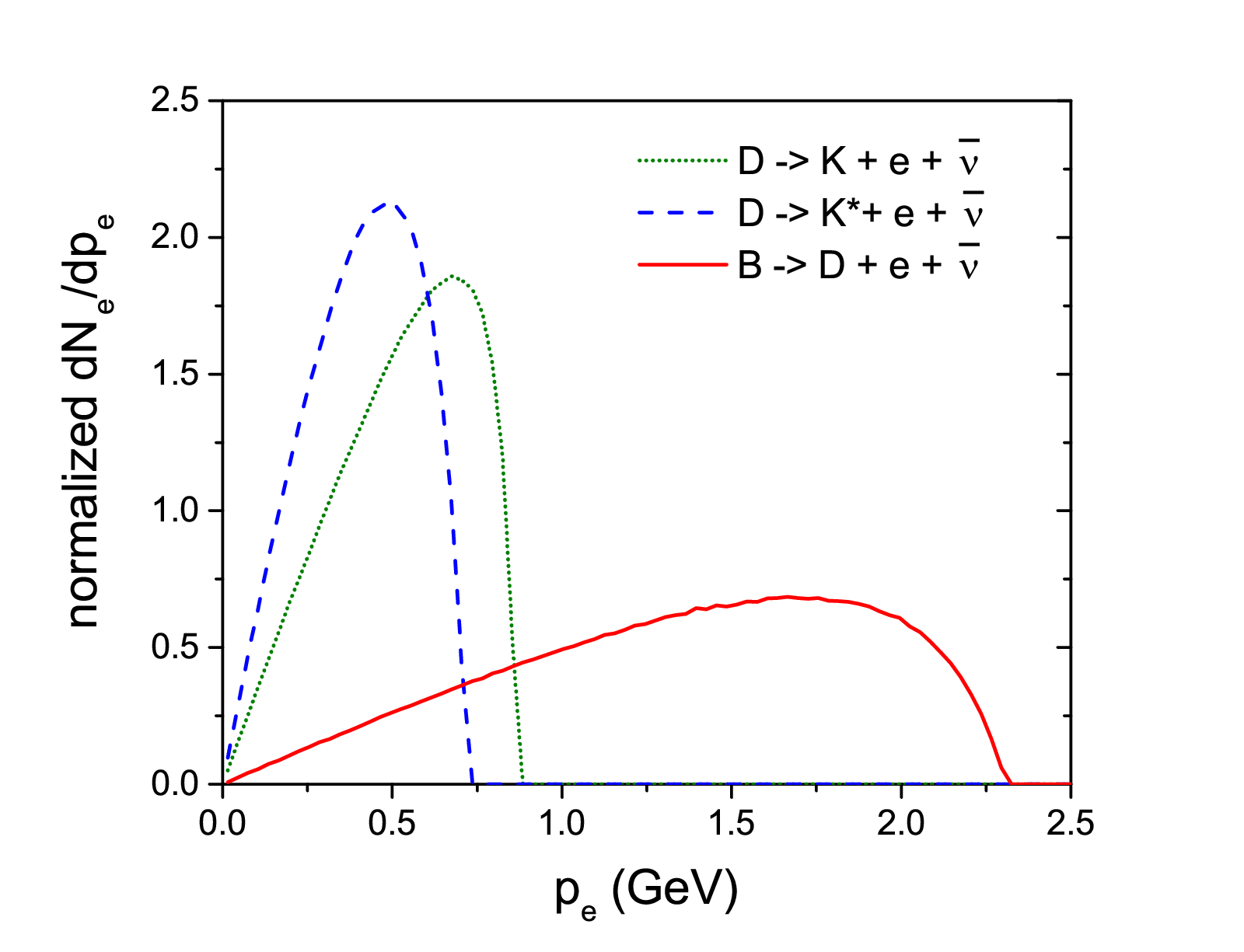}}
\caption{(Color online) The momentum distribution of single
electrons from the decays $D\rightarrow K+e+\bar{\nu_e}$ (dotted),
$D\rightarrow K^*+e+\bar{\nu_e}$ (dashed), and $B\rightarrow
D+e+\bar{\nu_e}$ (solid) in the rest frame of the heavy meson.}
\label{momentum-e}
\end{figure}

Figure~\ref{momentum-e} shows the momentum distribution of single
electrons from the semileptonic decay of heavy mesons in the rest
frame of the heavy meson. It shows that the single electron from the
decay $B\rightarrow D+e+\bar{\nu_e}$ has the largest momentum and
that from $D\rightarrow K^*+e+\bar{\nu_e}$ the lowest momentum
according to the mass difference between the mother meson and the
daughter meson. We also take into account the decay $D_s\rightarrow
\phi+e+\bar{\nu_e}$, $D_s\rightarrow \eta+e+\bar{\nu_e}$, and
$B_s\rightarrow D_s+e+\bar{\nu_e}$~\cite{Agashe:2014kda}. The
branching ratio of each decay channel is obtained from the particle
data group (PDG)~\cite{Agashe:2014kda}. This completes the
description of the semileptonic decays.

\begin{figure} [h]
\centerline{
\includegraphics[width=9.5 cm]{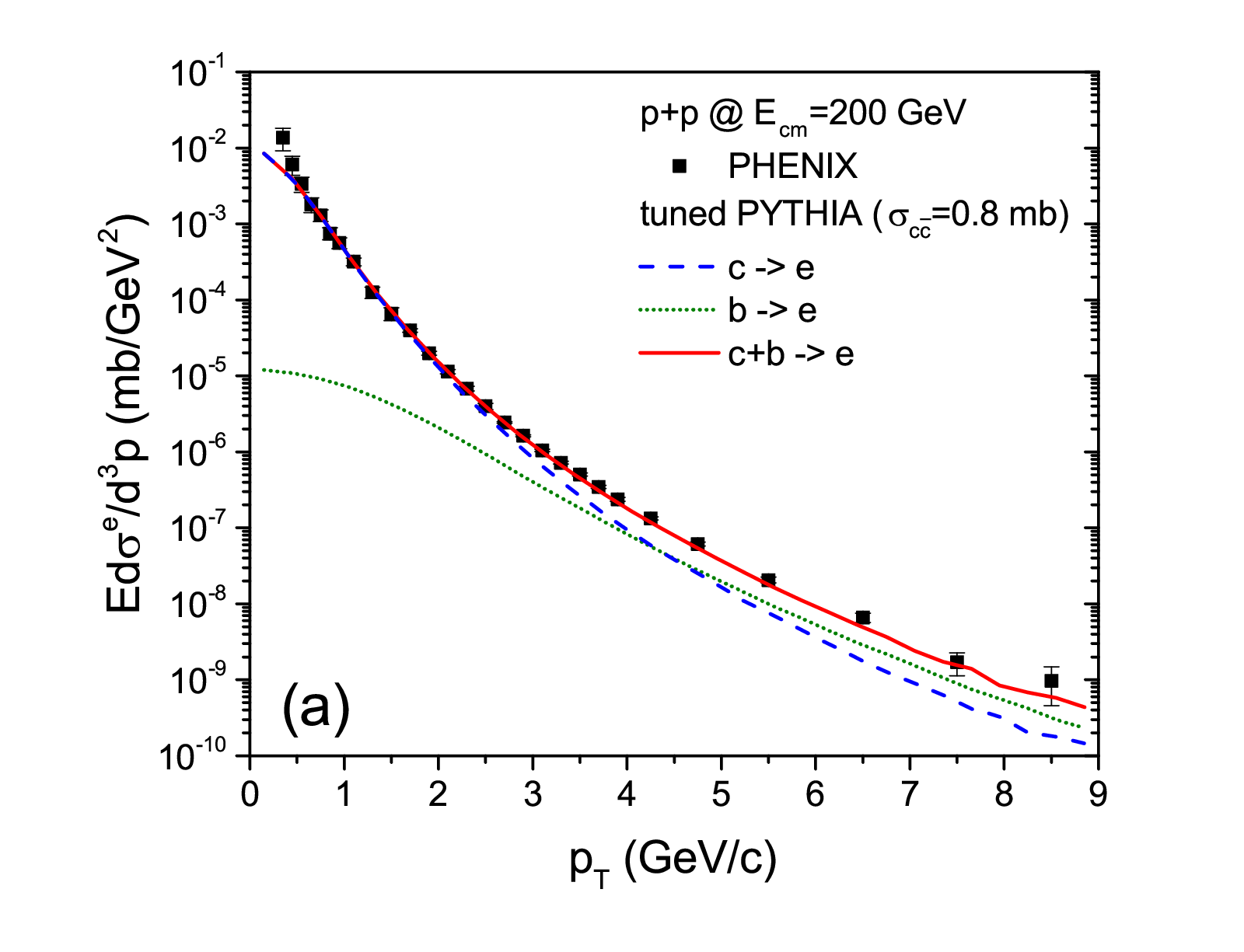}}
\centerline{
\includegraphics[width=9.5 cm]{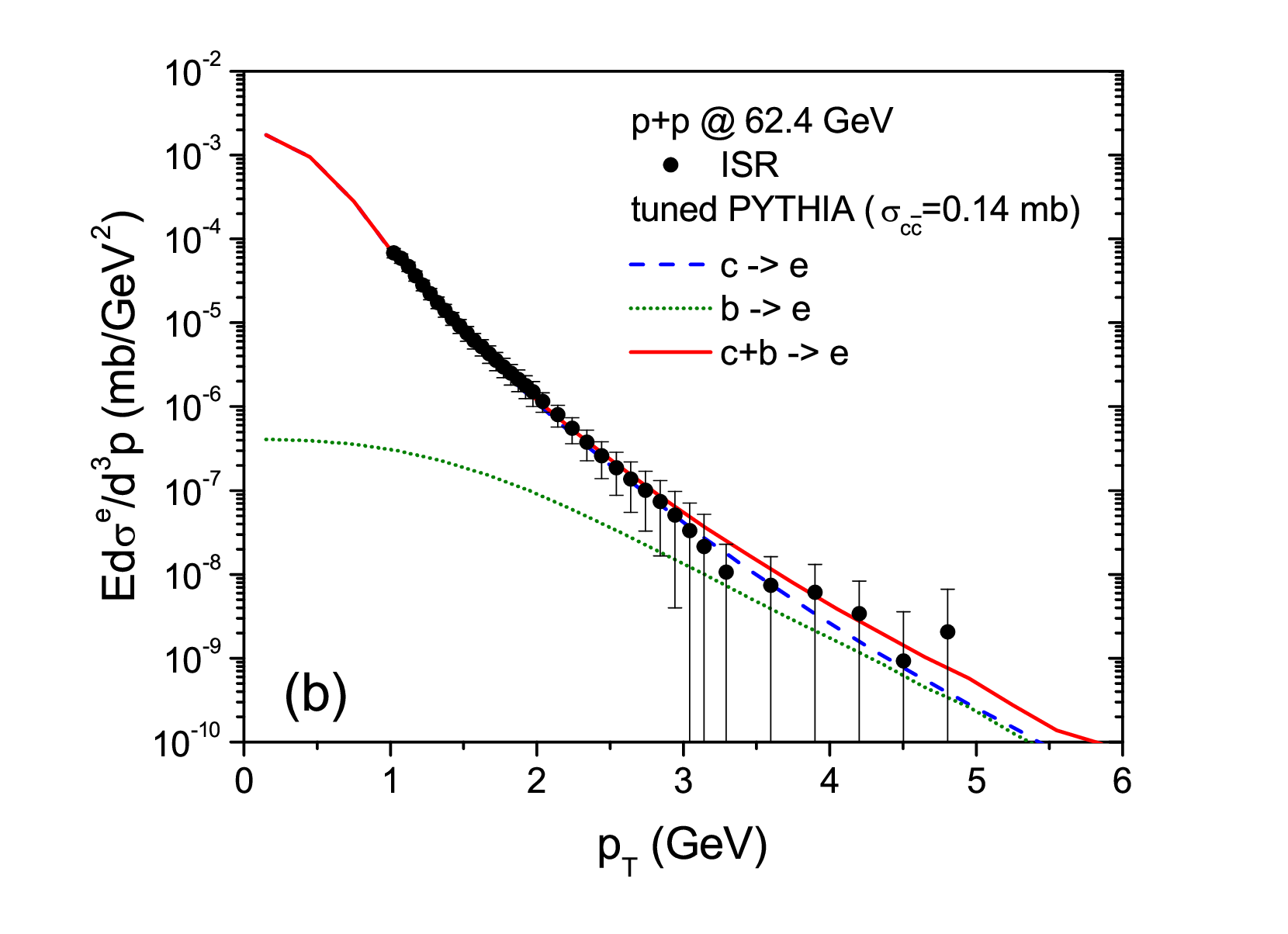}}
\caption{(Color online) The $\rm p_T$ spectrum of single electrons
from charm and bottom mesons in p+p collisions at $\sqrt{s}=200$ GeV
(a) and $\sqrt{s}$=62.4 GeV (b) in comparison to the experimental
data from Refs.~\cite{Adare:2006hc,Basile:1981dn}.} \label{ppe}
\end{figure}

In figure~\ref{ppe} we show the $\rm p_T$ spectrum of single
electrons in p+p collisions at $\sqrt{s}=200$ GeV (a) and
$\sqrt{s}=$ 62.4 GeV (b). The figure shows that our results
reproduce the experimental data at $\sqrt{s}=200$ GeV from the
PHENIX Collaboration~\cite{Adare:2006hc} as well as at
$\sqrt{s}=62.4$ GeV from the Intersecting Storage Rings
(ISR)~\cite{Basile:1981dn}. We note, furthermore, that the
contribution from $B-$meson decay is compatible to that from
$D-$meson decay around $\rm p_T \approx 4.5$ GeV at $\sqrt{s}=200$
GeV, and around $\rm p_T \approx 5.5$ GeV at $\sqrt{s}=62.4$ GeV,
respectively.

\section{cold nuclear matter effects}\label{shadowing}

The scattering cross section for heavy-quark pair production in a nucleon-nucleon collision is calculated by convoluting the partonic cross section with
the parton distribution functions of the nucleon:
\begin{eqnarray}
\sigma_{Q\bar{Q}}^{NN}(s)=\sum_{i,j}\int dx_1dx_2 f_i^N(x_1,q)f_j^N(x_2,q)\nonumber\\
\times\sigma_{Q\bar{Q}}^{ij}(x_1x_2s,q),
\label{factorize}
\end{eqnarray}
where $f_i^N(x,q)$ is the distribution function of the parton $i$ with the
energy-momentum fraction $x$ in the nucleon at scale $q$. The
momentum fractions
$x_1$ and $x_2$ are calculated from the transverse mass ($M_T$) and the
rapidity ($y$) of the final-state particles by
\begin{eqnarray}
x_1=\frac{M_T}{E_{\rm cm}}e^{y},~~~x_2=\frac{M_T}{E_{\rm cm}}e^{-y},
\label{x1x2}
\end{eqnarray}
where $E_{\rm cm}$ is the nucleon-nucleon collision energy in the center-of-mass frame.

As it is well known the parton distribution function (PDF) is modified in a nucleus to
\begin{eqnarray}
f_i^{N^*}(x,q)=R_i^A(x,q)f_i^N(x,q),
\label{shadow}
\end{eqnarray}
where $N^*$ indicates a nucleon in nucleus $A$, and $R_i^A(x,q)$ is
the ratio of the PDF of $N^*$ to that of a free nucleon. The ratio
$R_i^A(x,q)$ for a heavy nucleus $A$ is lower than 1 at small
momentum fraction $x$, and becomes larger than 1 with increasing
$x$. The former phenomenon is called `shadowing' and the latter
`antishadowing'. When increasing $x$ further, the ratio reaches a
maximum and then decreases again, which is called the European Muon
Collaboration (EMC) effect~\cite{Bodek:1983qn}. Finally, the ratio
increases again close to $x=1$ due to Fermi motion. The EPS09
package~\cite{Eskola:2009uj} parameterizes this behavior of
$R_i^A(x,q)$ and fits the heights and positions of the local extrema
of the ratio to the experimental data from deep inelastic $l+$A
scattering, Drell-Yan dilepton production in p+A collisions, and
inclusive pion production in d+Au and p+p collisions at RHIC. We
also employ the EPS09 package~\cite{Eskola:2009uj} in our PHSD
calculations \cite{Song:2015ykw}.

Furthermore, the (anti-)shadowing effect is supposed to depend on
the impact parameter in heavy-ion collisions such that it is strong
in central collisions and weak in peripheral collisions. Therefore,
we modify the ratio to~\cite{Song:2015ykw}
\begin{eqnarray}
R_i^A(r_\bot,x,q)=\frac{4}{3}\sqrt{1-\frac{r_\bot^2}{R_A^2}}~R_i^A(x,q),
\label{impact}
\end{eqnarray}
where $R_A$ and $r_\bot$ are, respectively, the radius of the
nucleus $A$ and the transverse distance of the heavy-quark pair
production from the nucleus center, while $R_i^A(x,q)$ is given by
EPS09.

\begin{figure} [h]
\centerline{
\includegraphics[width=9.5 cm]{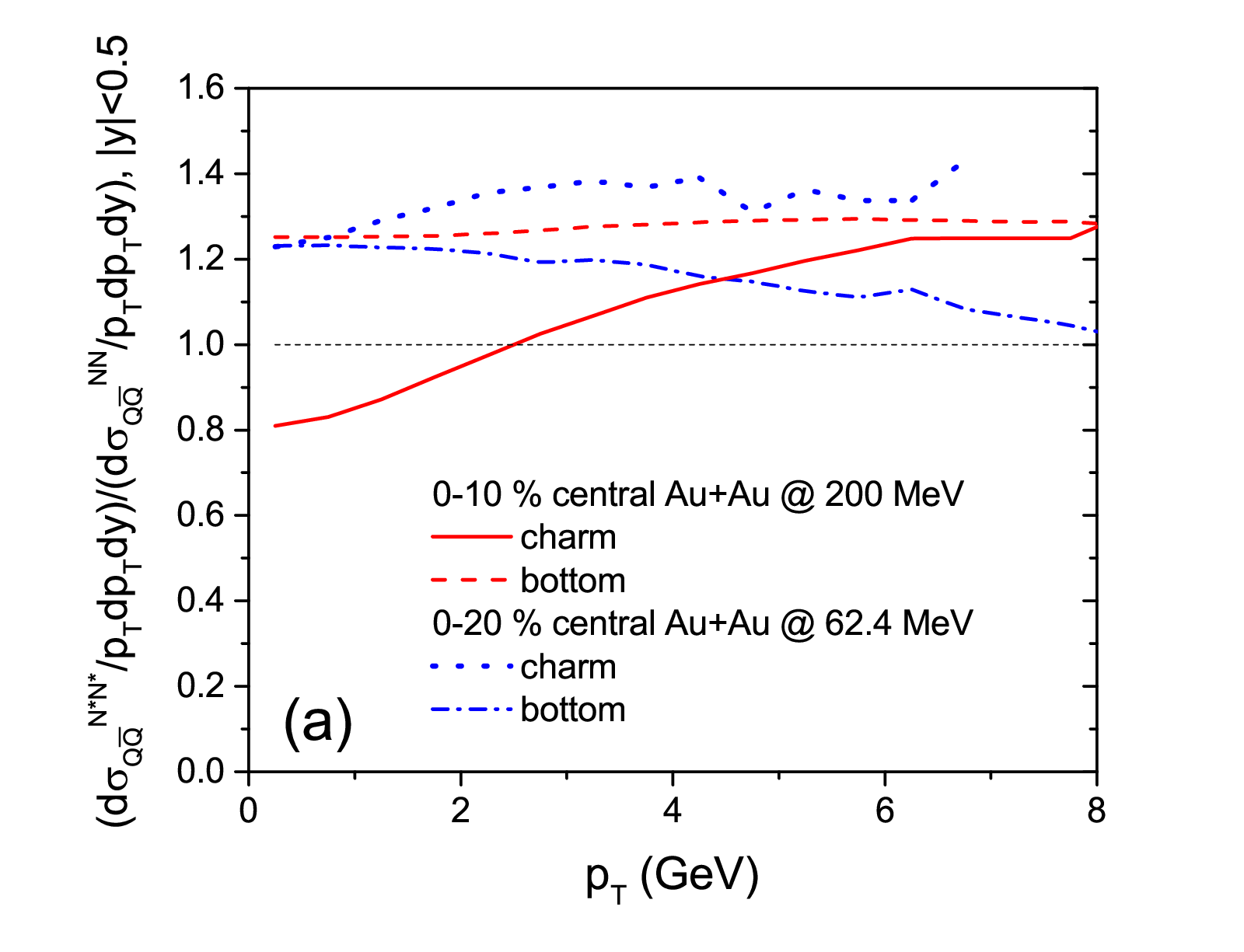}}
\centerline{
\includegraphics[width=9.5 cm]{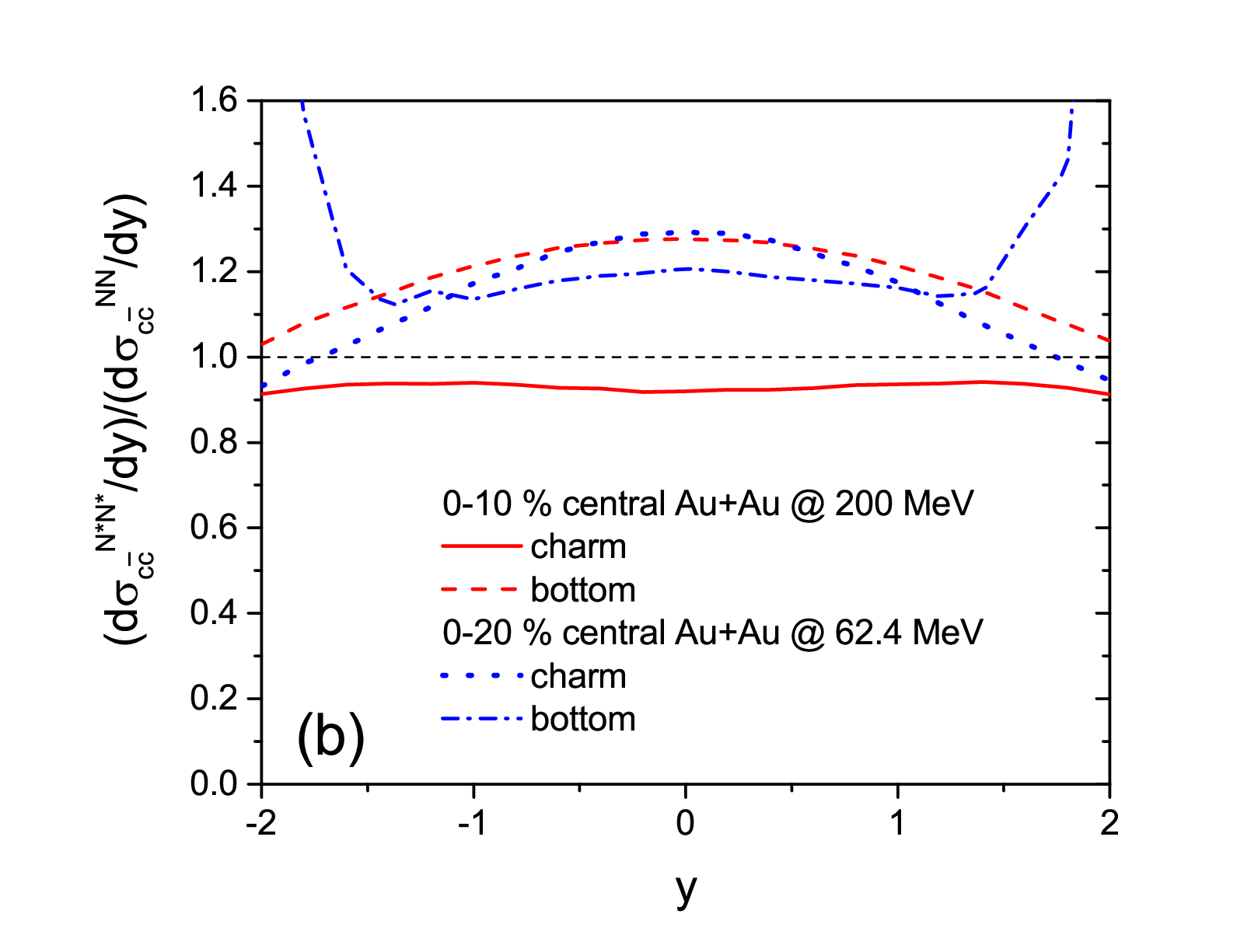}}
\caption{(Color online) The (anti)shadowing effect from EPS09 on
charm and bottom production in 0-10 \% central Au+Au collisions at
$\sqrt{s}=200$ GeV and in 0-20 \% central Au+Au collisions at
$\sqrt{s}=62.4$ GeV as a function of $\rm p_T$ (a) and of y
(b).} \label{shadowingf}
\end{figure}

Substituting $f_i^{N}$ in Eq.~(\ref{factorize}) by Eq.~(\ref{shadow}), the cross section
for heavy quark production is modified to
\begin{eqnarray}
\sigma_{Q\bar{Q}}^{N^*N^*}(s)=\sum_{i,j}\int dx_1dx_2 R_i^A(x_1,q)R_j^A(x_2,q) \nonumber\\
\times f_i^N(x_1,q)f_j^N(x_2,q)\sigma_{Q\bar{Q}}^{ij}(x_1x_2s,q).
\label{factorize2}
\end{eqnarray}
Figure \ref{shadowingf} shows the ratio of the cross sections with
(anti)shadowing to without (anti)shadowing. The scale $q$ is taken
to be the average of the transverse mass of the heavy quark and that
of the heavy antiquark. The total cross sections for charm and
bottom production, respectively, decreases by 8 \% and increases by 21 \% in 0-10
\% central Au+Au collisions at $\sqrt{s}=200$ GeV, and increases by 18 \% and 21 \% in 0-20 \% central Au+Au collisions at $\sqrt{s}=64.2$ GeV.
The ratio of the charm cross sections increases with increasing
transverse momentum in 0-10 \% central Au+Au collisions at
$\sqrt{s}=200$ GeV, because $R_i^A(x,q)$ increases with increasing
$x$ in the (anti)shadowing region. In the case of bottom production,
the corresponding $x$ is larger and located close to the maximum of
the antishadowing region. Therefore, the dependence of the
(anti)shadowing effect on $\rm p_T$ or on $x$ is monotonous. Since
the momentum fraction $x$ corresponding to bottom production at
$\sqrt{s}=$ 200 GeV is similar to that corresponding to charm
production at $\sqrt{s}=$ 62.4 GeV, the (anti)shadowing effect on
both are similar in $\rm p_T$ as well as in rapidity $y$. Finally,
the ratio of bottom production at $\sqrt{s}=$ 62.4 GeV decreases
with increasing $\rm p_T$, because the corresponding momentum
fraction $x$ moves towards the EMC effect region, where $R_i^A(x,q)$
decreases with increasing $x$.

Additionally, we study the Cronin effect on the production of initial heavy quarks in relativistic heavy-ion collisions.
A heavy-flavor pair is produced by a hard scattering of partons, mainly, gluons.
On the other hand, the gluons may interact with other nucleons before participating in heavy-flavor production.
This scattering enhances the transverse momentum of gluons and consequently also the transverse momentum  of the heavy-quark pair.
As a result, the distribution in transverse momentum of heavy flavor is smeared in the collisions of nuclei A and B as~\cite{Zhao:2007hh,He:2014epa}
\begin{eqnarray}
f_{AB}^{Q\bar{Q}}(\vec{p}_T)=\frac{1}{\pi\langle p_T^{~\prime 2}\rangle} \int d^2 p_T^{~\prime} \exp\bigg[\frac{- p_T^{~\prime 2}}{\langle p_T^{~\prime 2}\rangle}\bigg]f_{NN}^{Q\bar{Q}}(\vec{p}_T-\vec{p}_T^{~\prime})\nonumber\\
\end{eqnarray}
with the broadening width in transverse momentum~\cite{Wang:1998hs}
\begin{eqnarray}
\langle p_T^{~\prime 2}\rangle=\frac{0.225\ln^2(\mu/{\rm GeV})}{1+\ln(\mu/{\rm GeV})}(N_A^{\rm coll}+N_B^{\rm coll})~(\rm GeV^2),
\end{eqnarray}
where the scale $\mu$ is taken to be twice the heavy-quark mass and $N_A^{\rm coll}$ and $N_B^{\rm coll}$ are the numbers of nucleon-nucleon binary collisions of two nucleons in nucleus A and B before they produce a heavy-quark pair.

We assume that the transverse momentum change $\vec{p}_T^{~\prime}$ is shared equally by the heavy quark and the heavy antiquark (from the same pair).
This is different by a factor $1/\sqrt{2}$ from other studies where the heavy quark and heavy antiquark are completely independent of each other, even though they stem from the same pair~\cite{He:2014epa,Alberico:2011zy}. This difference reduces the importance of the Cronin effect on open heavy flavor in heavy-ion collisions.

\begin{figure} [h]
\centerline{
\includegraphics[width=9.5 cm]{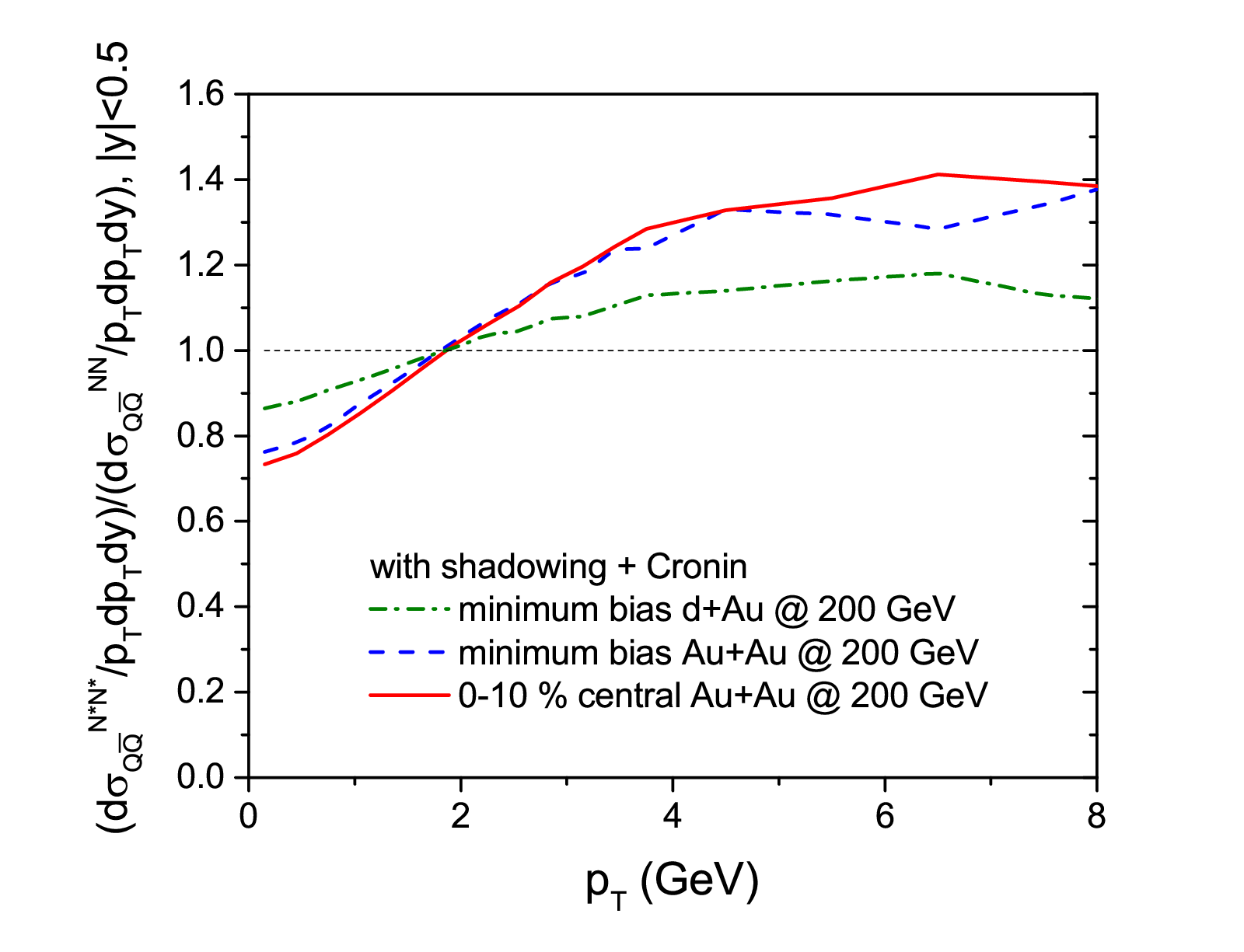}}
\caption{(Color online) The modifications of the charm transverse momentum due to the shadowing and Cronin effects in minimum bias d+Au and Au+Au collisions and 0-10 \% central Au+Au collisions at $\sqrt{s}=200$ GeV.}\label{croninf}
\end{figure}

Figure~\ref{croninf} shows the modifications of the charm transverse momentum due to the shadowing and Cronin effects in heavy-ion collisions at $\sqrt{s}=200$ GeV.
Since the path length of nucleons, which participate heavy flavor production, is twice as large in Au+Au collisions as compared to d+Au collisions and both, shadowing and Cronin effects, are proportional to the path length, the cold nuclear matter effects are roughly twice in minimum bias Au+Au collisions compared to those in minimum bias d+Au collisions, as shown in  figure~\ref{croninf}.
We can also see that the cold nuclear matter effects are stronger in central Au+Au collisions than in minimum bias reactions.
Comparing with figure~\ref{shadowingf} (a), we note that the Cronin effect enhances charm distribution additionally by about 10 \% at intermediate transverse momentum.

\begin{figure} [h]
\centerline{
\includegraphics[width=9.5 cm]{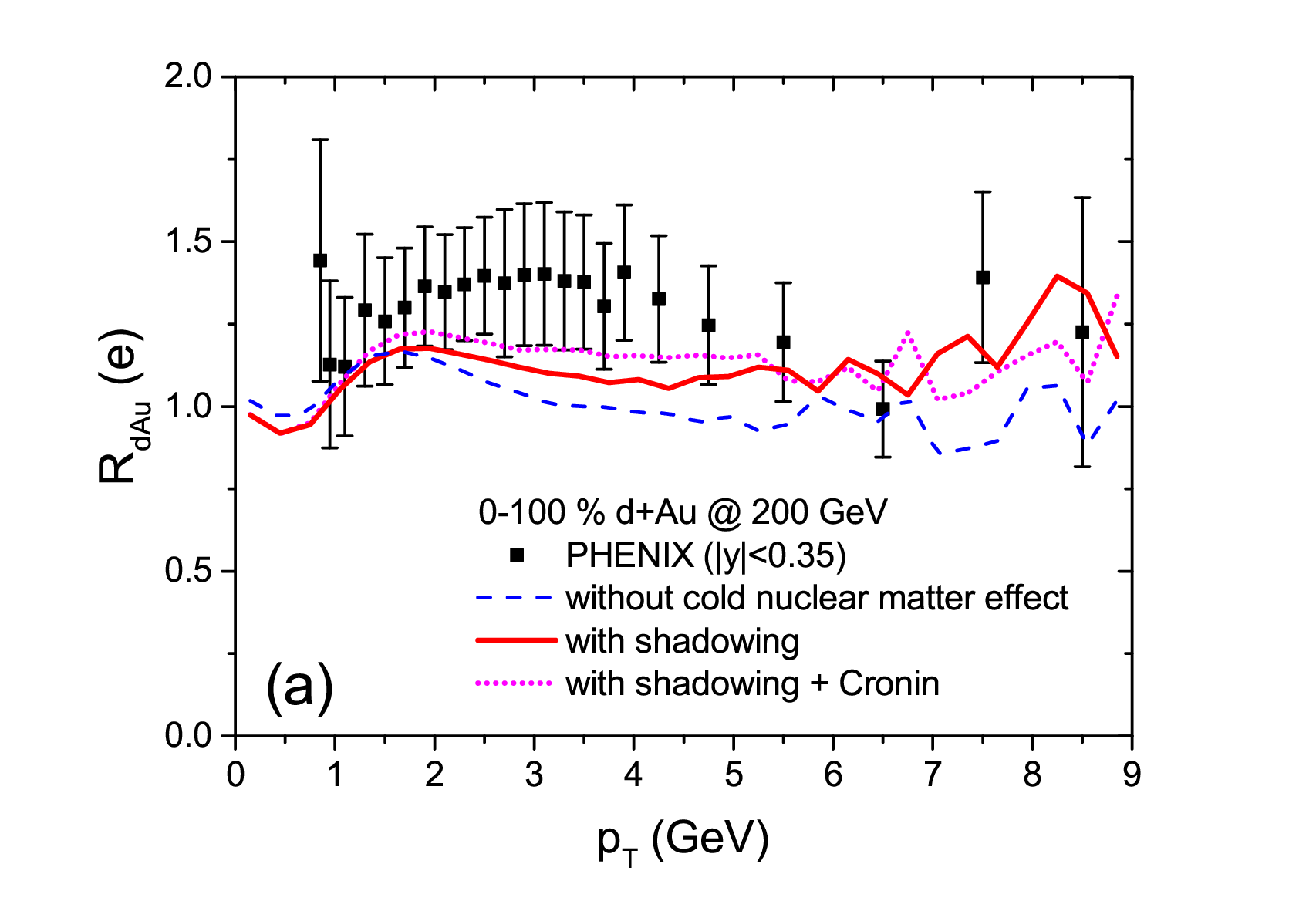}}
\centerline{
\includegraphics[width=9.5 cm]{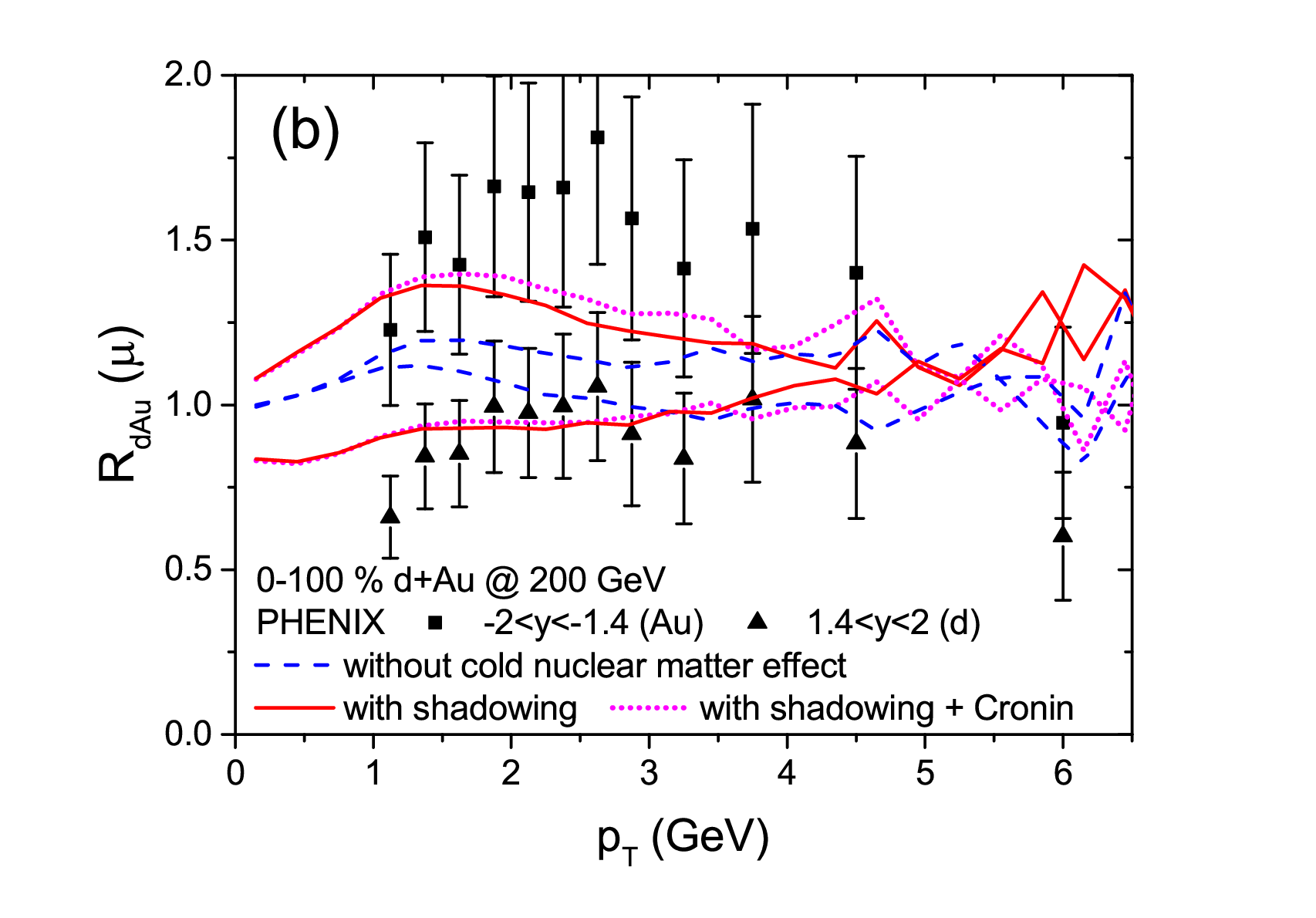}}
\caption{(Color online) $R_{\rm dAu}$ of single electrons at mid-rapidity and that of single muons at forward/backward rapidities in minimum-bias d+Au collisions at $\sqrt{s_{\rm NN}}=200$ GeV.
Dashed, solid, and dotted lines correspond to the $R_{\rm dAu}$ without cold nuclear matter effect, with the shadowing effect, and with both shadowing and Cronin effects, respectively.
Experimental data are given by the PHENIX collaboration~\cite{Adare:2012yxa,Adare:2013lkk}.}\label{dAuf}
\end{figure}

The light p+A or d+A collisions are well suited to investigate the cold nuclear matter effects, because  thermal effects in such collisions will be of minor importance.
We compare our cold nuclear matter effects with the experimental data for d+Au collisions from the PHENIX Collaboration~\cite{Adare:2012yxa,Adare:2013lkk}.

Figure~\ref{dAuf} a) shows the $R_{\rm dAu}$ of single electrons at midrapidity ($|y|<0.35$) and figure~\ref{dAuf} b) that of single muons at forward ($1.4<y<2$, d-direction) and backward ($-2<y<-1.4$, Au-direction). The kinematics for a single muon decay from D or B mesons is exactly the same as that for a single electron except that the maximum momentum of the daughter meson is substituted by
\begin{eqnarray}
p_{\rm max}=\frac{\sqrt{\{m_D^2-(m_K+m_\mu)^2\}\{m_D^2-(m_K-m_\mu)^2\}}}{2m_D}\nonumber
\end{eqnarray}
or
\begin{eqnarray}
p_{\rm max}=\frac{\sqrt{\{m_B^2-(m_D+m_\mu)^2\}\{m_B^2-(m_D-m_\mu)^2\}}}{2m_B}, \nonumber
\end{eqnarray}
respectively.

Figure~\ref{dAuf} shows that the interactions of heavy flavor with cold nuclear matter generate a maximum in $R_{\rm dAu}$ around $p_{\rm T}=$1.5 GeV.
More interaction and larger coalescence probability of heavy flavor in Au-direction split the $R_{\rm dAu}$ in forward and backward rapidities, but the separation is not enough to explain the experimental data.
The shadowing effects help the separation, because the heavy flavors at forward rapidities are contributed by gluons with small x in the Au nucleus and those in backward rapidities by gluons with large x.
Since $R_i^A(x,q)$ in Eq.~(\ref{shadow}) increases with larger x, the production of heavy flavor is suppressed at forward rapidity and enhanced at backward rapidity.
Figure~\ref{dAuf} also shows that the Cronin effect helps our results to come closer to the experimental data at mid-rapidity as well as at forward/backward rapidities.

\section{Heavy-quark interactions in the QGP}\label{QGP}

In PHSD the baryon-baryon and baryon-meson collisions at high-energy
produce strings. If the local energy density is above the critical
energy density ($\sim$ 0.5 GeV/fm$^3$), the strings melt into quarks
and antiquarks with masses  determined by the temperature-dependent
spectral functions from the DQPM~\cite{Cassing:2008nn}. Massive
gluons are formed through flavor-neutral quark and antiquark fusion
in line with the DQPM. In contrast to normal elastic scattering,
off-shell partons may change their mass after the elastic scattering
according to the local temperature $T$ in the cell (or local
space-time volume) where the scattering happens. This automatically
updates the parton masses as the  hot and dense  matter expands,
i.e. the local temperature decreases with time. The same holds true
for the reaction chain from gluon decay to quark+antiquark ($g
\rightarrow q + {\bar q}$) and the inverse reaction ($q + {\bar q}
\rightarrow g$) following detailed balance.

Due to the finite spectral width of the partonic degrees-of-freedom,
the parton spectral function has time-like as well as space-like
parts. The time-like partons propagate in space-time within the
light-cone while the space-like components are attributed to a
scalar potential energy density~\cite{Cassing:2009vt}. The gradient
of the potential energy density with respect to the scalar density
generates a repulsive force in relativistic heavy-ion collisions and
plays an essential role in reproducing experimental flow data and
transverse momentum spectra (see Ref. \cite{PHSDreview} for a
review).

However, the spectral function of a heavy quark or heavy antiquark
cannot be fitted from lattice QCD data on thermodynamical properties
because the contribution from a heavy quark or heavy antiquark to
the lattice entropy is small. Our recent study shows that the
scattering cross sections of heavy quark moderately depend on the
spectral function of heavy quark, and the repulsive force for charm
quarks -- as originating from the scalar potential energy density --
is disfavored by experimental data \cite{Song:2015ykw}. This is
expected since the width of the spectral function for a charm quark
is very small compared to the pole mass such that space-like
contributions to the (potential) energy density are practically
vanishing. Therefore, we assume in this study that the heavy quark
has a constant (on-shell) mass: the charm quark mass is taken to be
1.5 GeV and the bottom quark mass as 4.8 GeV, but the light
quarks/antiquarks as well as gluons are treated fully off-shell.

The heavy quarks and antiquarks produced in early hard collisions -
as described above - interact with the dressed lighter off-shell
partons in the QGP. The cross sections for the heavy-quark
scattering with massive off-shell partons have been calculated by
considering explicitly the mass spectra of the final state particles
in Refs.~\cite{Berrehrah:2013mua,Berrehrah:2014kba}. The elastic
scattering of heavy quarks in the QGP is treated by including the
non-perturbative effects of the strongly interacting quark-gluon
plasma (sQGP) constituents, i.e. the temperature-dependent coupling
$g(T/T_c)$ which rises close to $T_c$, the multiple scattering etc.
The multiple strong interactions of quarks and gluons in the sQGP
are encoded in their effective propagators with broad spectral
functions (imaginary parts). As pointed out above, the effective
propagators, which can be interpreted as resummed propagators in a
hot and dense QCD environment, have been extracted from lattice data
in the scope of the DQPM~\cite{Cassing:2008nn}. We recall that the
divergence encountered in the $t$-channel scattering is cured
self-consistently, since the infrared regulator is given by the
finite DQPM gluon mass and width. For further details we refer the
reader to Refs.~\cite{Berrehrah:2013mua,Berrehrah:2014kba}.

\begin{figure} [h]
\centerline{
\includegraphics[width=9.5 cm]{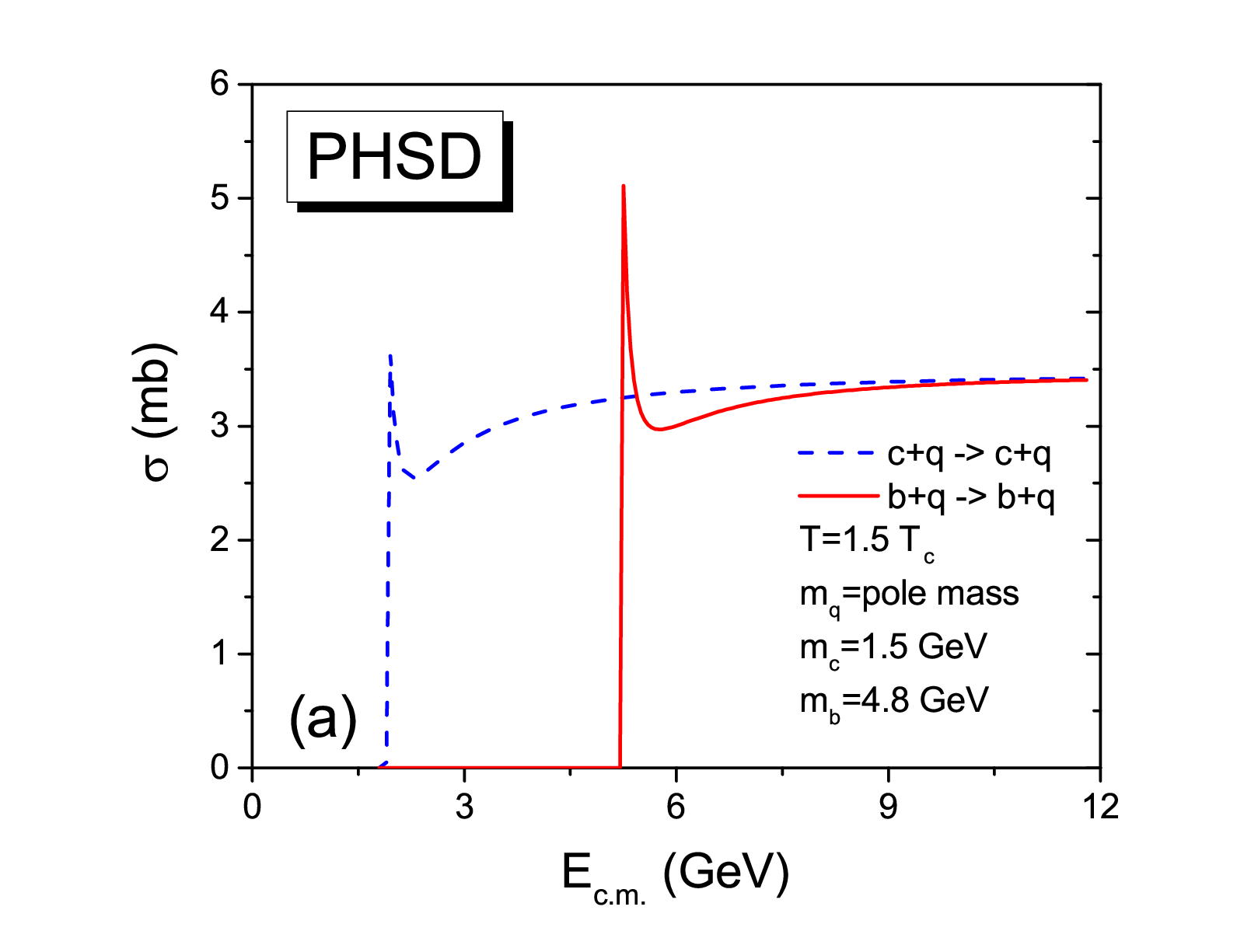}}
\centerline{
\includegraphics[width=9.5 cm]{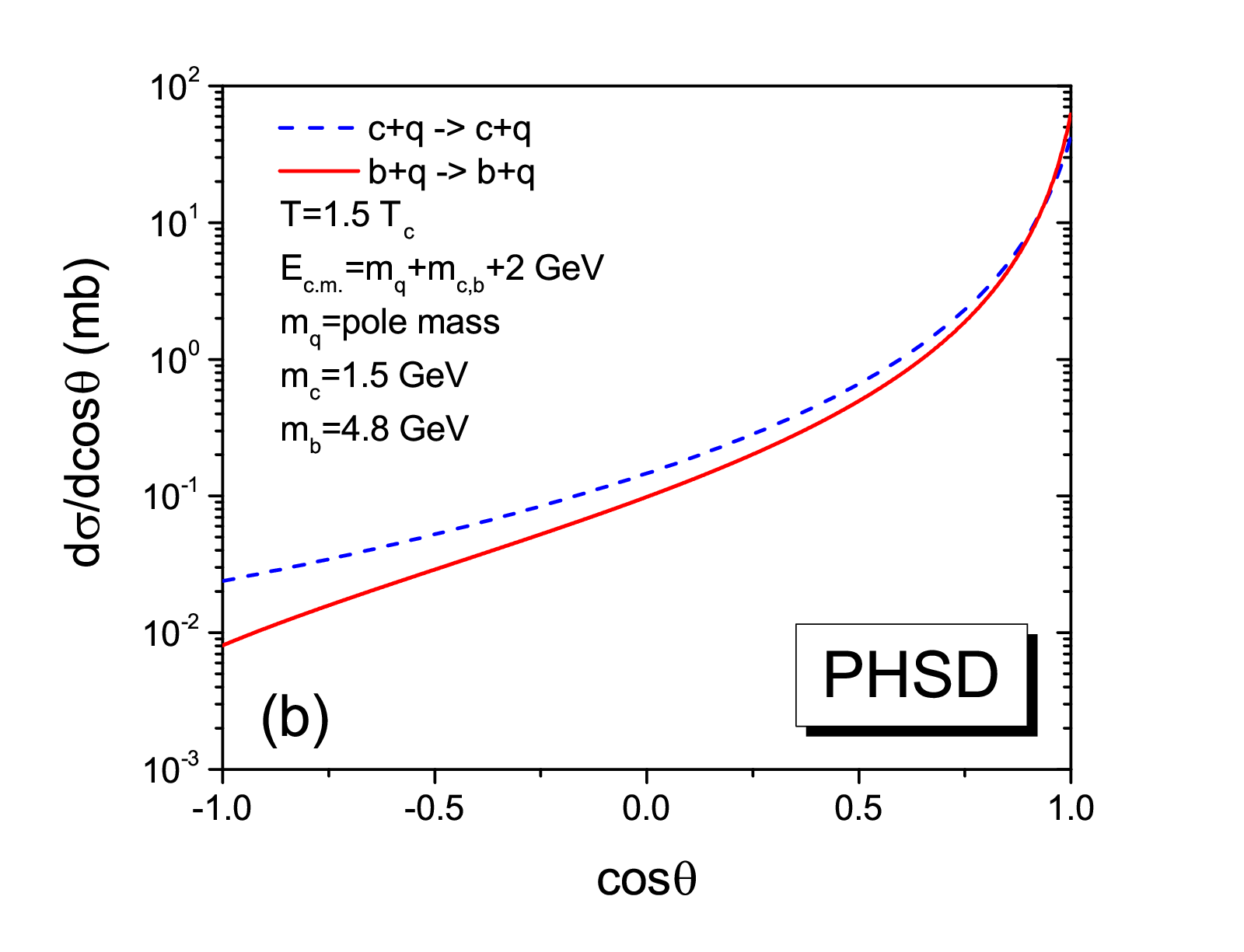}}
\caption{(Color online) The total (a) and differential (b)
scattering cross sections for the reactions $c+q \rightarrow c+q$
and $b+q \rightarrow b+q$ employed in the PHSD have been multiplied by a factor of two~\cite{Berrehrah:2013mua,Berrehrah:2014kba}.
} \label{sigma}
\end{figure}

Figure~\ref{sigma} compares the total (a) and differential (b)
scattering cross sections of charm and bottom quarks with a light
quark at $T=1.5~T_c$. It shows that the total cross section of a
charm quark is similar to that of the bottom quark apart from
different threshold energies. However, the differential scattering
cross section of a bottom quark is more peaked in forward direction,
compared to that of a charm quark. This is expected because it is
harder to change the direction of motion of a bottom quark in
elastic scattering due to the larger mass.

We note that the scattering cross sections of heavy quark used in the PHSD calculations are twice larger than from the Born diagrams of the DQPM in order to achieve consistency with the lQCD data from Refs.~\cite{Song:2015ykw,Berrehrah:2016vzw}, however, they differ substantially from the pQCD scenario~\cite{Berrehrah:2016led}.

\section{Heavy-quark hadronization}\label{hadronization}


The heavy-quark hadronization in heavy-ion collisions is
realized via `dynamical coalescence' and fragmentation. Here
`dynamical coalescence' means that the probability to find a
coalescence partner is defined by Monte Carlo in the vicinity of the
critical energy density $0.4\le \epsilon \le 0.75$ GeV/fm$^3$ as
explained below. We note that such a dynamical realization of
heavy-quark coalescence is in line with the dynamical hadronization
of light quarks in the PHSD and differs from the `spontaneous'
coalescence used in our early work \cite{Song:2015sfa} when
heavy-quarks are forced to hadronize at a critical energy density
$\epsilon_c=0.5$~GeV/fm$^3$ via coalescence or fragmentation  by
Monte Carlo. Indeed, the dynamical realization gives some window in
energy density to find the proper light partner and leads to an
enhancement of  the heavy-quark fraction that hadronizes via
coalescence.

In PHSD all antiquarks neighboring in phase space are
candidates for the coalescence partner of a heavy quark. From the
distances in coordinate and momentum spaces between the heavy quark
and light antiquark (or vice versa), the coalescence probability is
given by~\cite{Song:2016lfv}
\begin{eqnarray}
f(\boldsymbol\rho,{\bf k}_\rho)=\frac{8g_H}{6^2}
\exp\left[-\frac{\boldsymbol\rho^2}{\delta^2}-{\bf k}_\rho^2\delta^2\right],
\label{meson}
\end{eqnarray}
where $g_H$ is the degeneracy of the heavy meson, and
\begin{eqnarray}
\boldsymbol\rho=\frac{1}{\sqrt{2}}({\bf r}_1-{\bf r}_2),\quad{\bf k}_\rho
=\sqrt{2}~\frac{m_2{\bf k}_1-m_1{\bf k}_2}{m_1+m_2},
\label{coalescence}
\end{eqnarray}
with $m_i$, ${\bf r}_i$ and ${\bf k}_i$ denoting the mass, position and momentum
of the quark or antiquark $i$ in the center-of-mass frame, respectively. The
width parameter $\delta$ is related to the root-mean-square radius of the
produced heavy meson through
\begin{eqnarray}
\langle r^2 \rangle=\frac{3}{2}\frac{m_1^2+m_2^2}{(m_1+m_2)^2}\delta^2,
\end{eqnarray}
where $m_1$ and $m_2$ are respectively the masses of quark and antiquark.
Since this prescription gives a  larger coalescence probability at
low transverse momentum, the radius is taken to be 0.9 fm for a
charm quark as well as for a bottom quark~\cite{Song:2015sfa}. We
also include the coalescence of charm quarks into highly excited
states, $D_0^*(2400)^0$, $D_1(2420)^0$, and $D_2^*(2460)^{0,\pm}$
and the coalescence of bottom quarks into $B_1(5721)^{+,0}$,
$B_2^*(5747)^{+,0}$, and $B(5970)^{+,0}$, which are respectively
assumed to immediately decay to $D$ (or $D^*$) and $\pi$ and to $B$
(or $B^*$) and $\pi$ after hadronization as described in
Ref.~\cite{Song:2015sfa}.

Summing up the coalescence probabilities from all candidates,
whether the heavy quark or heavy antiquark hadronizes by coalescence
or not, and which quark or antiquark among the candidates will be
the coalescence partner, is decided by Monte Carlo. If a random
number is above the sum of the coalescence probabilities, it is
tried again in the next time step till the local energy density is
lower than 0.4 $\rm GeV/fm^3$. The heavy quark or heavy antiquark,
which does not succeed to hadronize by coalescence, then hadronizes
through fragmentation as in p+p collisions.

\begin{figure} [h]
\centerline{
\includegraphics[width=9.5 cm]{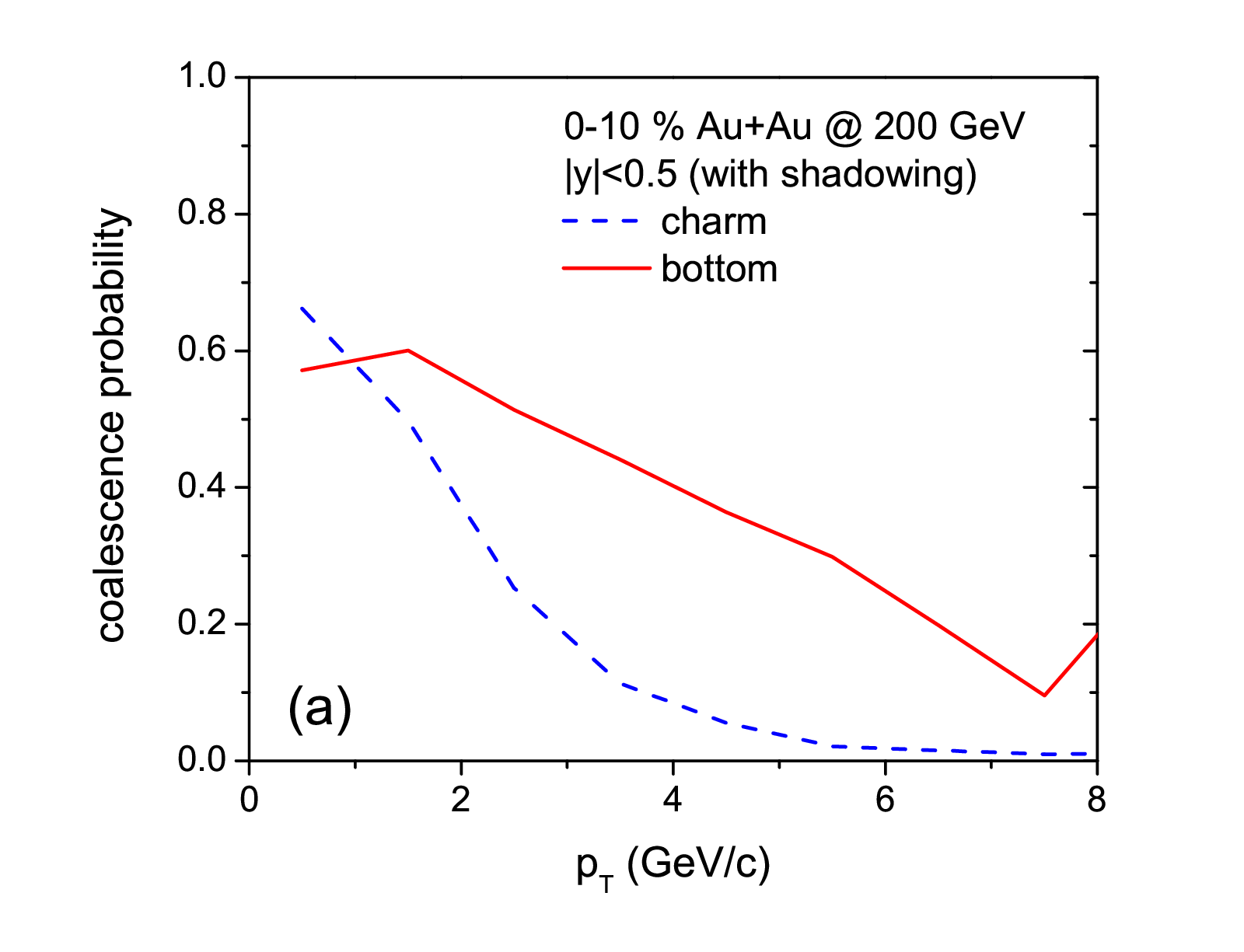}}
\centerline{
\includegraphics[width=9.5 cm]{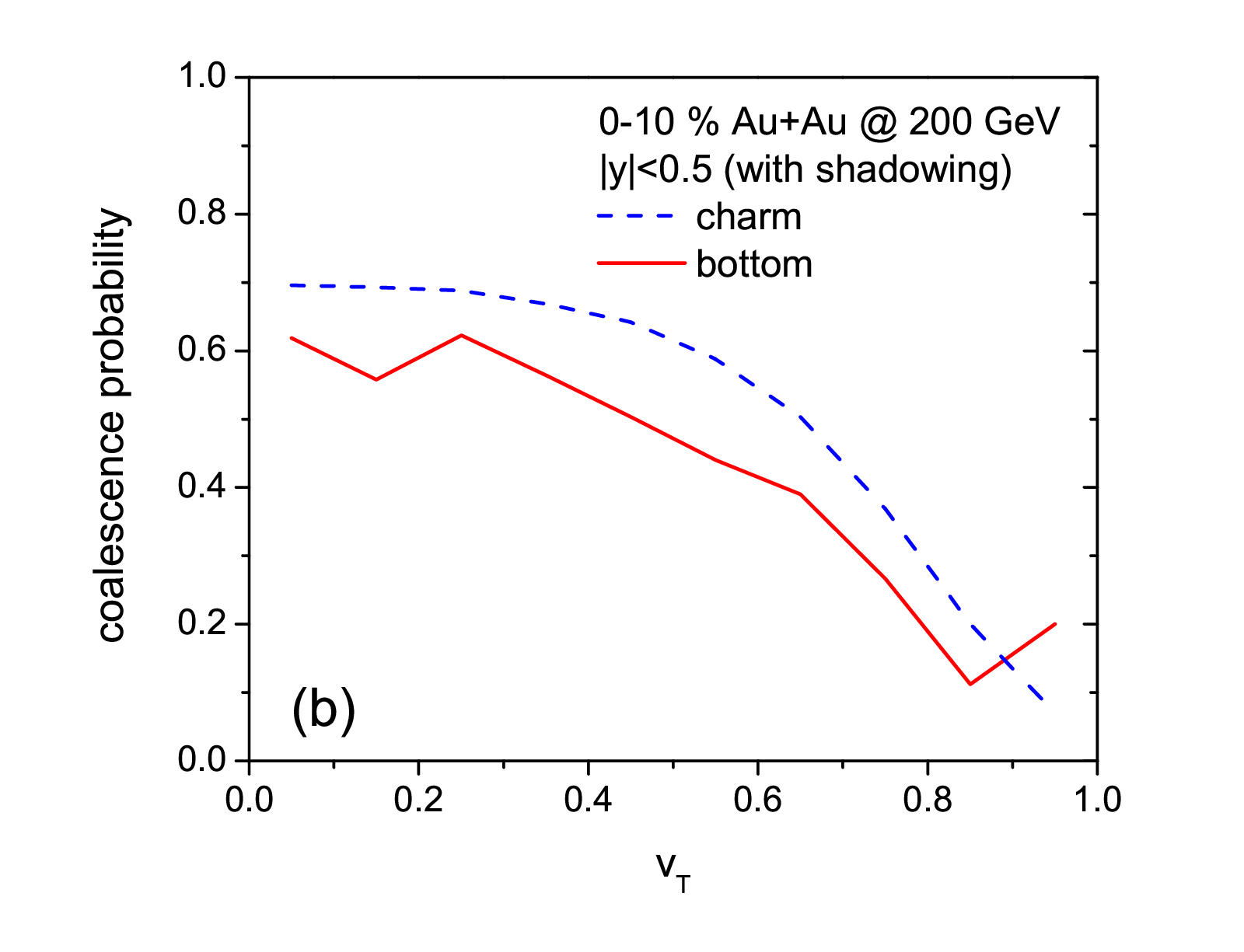}}
\caption{(Color online) Coalescence probability of charm and bottom
quarks at midrapidity ($|y|<0.5$) as functions of transverse
momentum (a) and of transverse velocity (b) in 0-10 \% central Au+Au
collisions at $\sqrt{s_{\rm NN}}=$200 GeV taking into account the
shadowing effect.} \label{coalpro}
\end{figure}

Figure~\ref{coalpro} shows the coalescence probabilities of charm
and bottom quarks at midrapidity ($|y|<0.5$) as functions of
transverse momentum (a) and of transverse velocity (b) in 0-10 \%
central Au+Au collisions at $\sqrt{s_{\rm NN}}=$200 GeV. Since a
heavy quark with a large transverse momentum has a smaller chance to
find a coalescence partner close by in phase space, the coalescence
probability decreases with increasing transverse momentum. It
appears from the upper figure (a)  that the coalescence probability
of a bottom quark is larger than that of a charm quark. It emerges,
however, because the bottom quark is much heavier than the charm
quark. The lower figure (b) clearly shows that the coalescence
probability is similar for a bottom or charm quark, when it is
expressed as a function of the transverse velocity $v_T$.

\section{Interactions of charm and bottom mesons with the hadronic medium}\label{hg}

After the hadronization of heavy quarks and their subsequent decay
into $D, D^*, B$ and $B^*$ mesons, the final stage of the evolution
concerns the interaction of these states with the hadrons conforming
the expanding bulk medium. A realistic description of the
hadron-hadron scattering ---potentially affected by resonant
interactions--- includes collisions with the states
$\pi,K,\bar{K},\eta,N,\bar{N},\Delta,\bar{\Delta}$. A description of
their interactions has been developed in
Refs.~\cite{GarciaRecio:2008dp,Abreu:2011ic,Romanets:2012hm,Abreu:2012et,GarciaRecio:2012db,Garcia-Recio:2013gaa,Tolos:2013kva,Torres-Rincon:2014ffa,Tolos:2013gta}
using effective field theory. Moreover, after the application of an
effective theory, one should implement to the scattering amplitudes
a unitarization method to better control the behavior of the cross
sections at moderates energies.

The details of the interaction for the four heavy states follows
quite in parallel by virtue of the ``heavy-quark spin-flavor
symmetry''. It accounts for the fact that if the heavy masses are
much larger than any other typical scale in the system, like
$\Lambda_{QCD}$, temperature and the light hadron masses, then the
physics of the heavy subsystem is decoupled from the light sector,
and the former is not dependent on the mass nor on the spin of the
heavy particle. This symmetry is exact in the ideal limit $m_Q
\rightarrow \infty$, with $m_Q$ being the mass of the heavy quark
confined in the heavy hadron. In the opposite limit $m_Q \rightarrow
0$, one can exploit the chiral symmetry of the QCD Lagrangian to
develop an effective realization for the light particles. This
applies to the pseudoscalar meson octet ($\pi,K,\bar{K},\eta$).
Although both symmetries are broken in nature (as in our approach,
when implementing physical masses), the construction of the
effective field theories incorporates the breaking of these
symmetries in a controlled way. In particular, it provides a
systematic expansion in powers of $1/m_H$ (inverse heavy-meson mass)
and powers of $p, m_l$ (typical momentum and mass of the light
meson). Following these ideas, we use two effective Lagrangians for
the interaction of a heavy meson with light mesons and with baryons,
respectively.

In the scattering with light mesons, the scalar ($D$,$B$) and vector
($D^*, B^*$) mesons are much heavier than the pseudoscalar meson
octet ($\pi,K,\bar{K},\eta$). The latter have, in addition, masses
smaller than the chiral scale $\Lambda_{\chi} \simeq 4 \pi f_\pi$,
where $f_\pi$ is the pion decay constant. In this case one can
exploit standard chiral perturbation theory for the dynamics of the
(pseudo) Goldstone bosons, and add the heavy-quark mass expansion up
to the desired order to account for the interactions with heavy
mesons. In our case the effective Lagrangian is kept to
next-to-leading order in the chiral expansion, but to leading order
in the heavy-quark expansion~\cite{Abreu:2011ic,Abreu:2012et}. From
this effective Lagrangian one can compute the tree-level amplitude
(or potential), which describes the scattering of a heavy meson off
a light meson as worked out in
Refs.~\cite{Tolos:2013kva,Torres-Rincon:2014ffa}. At leading order
in the heavy-quark expansion one gets a common result for all heavy
mesons due to the exact heavy-flavor symmetry and heavy-quark spin
symmetry (HQSS). The potential reads explicitly
\begin{eqnarray} \label{eqq}
 V^{meson}_{ij} &=& \frac{C_{0,ij}}{4f_\pi^2} (s-u) + \frac{2C_{1,ij} h_1}{3f_\pi^2} \ \\
 &+& \frac{2C_{2,ij}}{f_\pi^2} h_3 (p_2 \cdot p_4)  \nonumber \\
 &+& \frac{2C_{3,ij}}{f_\pi^2} h_5 [(p_1 \cdot p_2)(p_3 \cdot p_4) + (p_1 \cdot p_4)(p_2 \cdot p_3)]  \ ,  \nonumber
\end{eqnarray}
where $C_{n,ij}$ are numerical coefficients (fixed by chiral
symmetry) which depend on the incoming $i$ and outgoing $j$ channels
---and also on the quantum numbers $IJSC/B$ (isospin, total angular
momentum, strangeness and charm/bottom). In Eq.~(\ref{eqq}) $f_\pi$ is
the pion decay constant in the chiral limit, and $h_i$ are the
low-energy constants at NLO in the chiral expansion (see
Ref.~\cite{Tolos:2013kva,Torres-Rincon:2014ffa} for details).
Finally, $s, u$  denote the Mandelstam variables and $p_a$ the
four-momentum of the $a-$particle in the scattering ($1,2
\rightarrow 3,4$).

For the heavy meson--baryon interaction we use an effective
Lagrangian based on a low-energy realization of a $t-$channel vector
meson exchange between mesons and baryons. In the low-energy limit
the interaction provides a generalized Weinberg-Tomozawa contact
interaction as worked out in Refs.
~\cite{GarciaRecio:2008dp,Romanets:2012hm,GarciaRecio:2012db,Garcia-Recio:2013gaa}.
The effective Lagrangian obeys SU(6) spin-flavor symmetry in the
light sector, plus HQSS in the heavy sector (which is preserved
either the heavy quark is contained in the meson or in the baryon).
The tree-level amplitude reads
 \begin{eqnarray} \label{ell}
  V^{baryon}_{ij} &=& \frac{D_{ij} }{4f_if_j} (2 \sqrt{s} - M_i -M_j) \nonumber \\
  & \times & \sqrt{\frac{M_i+E_i}{2M_i}} \sqrt{\frac{M_j+E_j}{2M_j}} \ ,
 \end{eqnarray}
where $D_{ij}$ are numerical coefficients which depend on the
initial and final channels ($i,j$), as well as all the quantum
numbers $IJSC/B$. In Eq.~(\ref{ell}) $f_i$ is the meson decay constant
in the $i$ channel, and $M_i, E_i$ are the baryon mass and energy in
the C.M. frame. From the form of this potential it is evident that
HQSS is again maintained. We note again that in both $V^{meson}$ and
$V^{baryon}$, HQSS is eventually broken when using physical values
for the heavy masses.

The tree-level amplitudes for meson-meson and meson-baryon
scattering have strong limitations in the energy range in which they
should be applied. It is limited for those processes in which the
typical momentum transfer is low, and below any possible resonance.
To increase the applicability of the scattering amplitudes and
restore exact unitarity for the scattering-matrix elements, we apply
a unitarization method, which consists in solving a coupled-channel
Bethe-Salpeter equation for the unitarized scattering amplitude
$T_{ij}$ using the potential as a kernel,
\begin{equation}
 T_{ij} = V_{ij} + V_{ik} G_k T_{kj} \ ,
\end{equation}
where $G_k$ is the diagonal meson-meson (or meson-baryon) propagator
which is regularized by dimensional regularization in the
meson-meson (or meson-baryon) channel. We adopt the ``on-shell''
approximation to the kernel of the Bethe-Salpeter equation to reduce
it into a set of algebraic equations. We refer the reader to
Refs.~\cite{GarciaRecio:2008dp,Romanets:2012hm,GarciaRecio:2012db,Garcia-Recio:2013gaa,Tolos:2013kva,Torres-Rincon:2014ffa}
for technical details.

The unitarization procedure allows for the possibility of generating
resonant states as poles of the scattering amplitude $T_{ij}$. Even
when these resonances are not explicit degrees-of-freedom, and we do
not propagate them in our PHSD simulations, they are automatically
incorporated into the two-body interaction. This is an important
fact, because such (intermediate) resonant states will strongly
affect the scattering cross section of heavy mesons due to the
presence of resonances, subthreshold states (bound states), and
other effects like the opening of a new channel when a resonance is
forming (Flatt\'e effect).

To mention some particular examples, in the interaction of $D, D^*,
B, B^*$ mesons with light mesons we generate broad resonances in the
$S,I,J^\pi=0,1/2,0^+ (1^+)$ channels. In the charm sector we
identify them with the experimentally-observed states $D_0^*(2400)$
and $D_1(2430)$ that can decay in an $s$-wave into $D\pi$ and
$D^*\pi$, respectively. In the bottom sector we obtain analogous
states $B_0(5530)$ and $B_1(5579)$, not yet identified by
experiment. We also find a series of bound states in the channel
$S,I,J^\pi=1, 0, 0^+ (1^+)$ which are identified with the
$D_{s0}^*(2317)$ and the $D_{s1} (2460)$ states. Their bottom
relatives $B_{s0}^* (5748)$ and $B_{s1}^*(5799)$ are again
predictions.

In the meson-baryon channel, we find the experimental $\Lambda_c
(2595)$ and $\Lambda_c(2625)$ charm resonances in the
$S,I,J^\pi=0,0,1/2^- (3/2^-)$ sector.  In our model, the $\Lambda_c
(2595)$ couples dominantly to $DN$ and $D^*N$, while the
$\Lambda_c(2625)$ to $D^*N$. Their bottom homologues are associated
with the experimental $\Lambda_b(5912)$ and $\Lambda_b(5920)$
baryons seen by the LHCb collaboration~\cite{GarciaRecio:2012db}. We
finally mention the subthreshold states in the $S,I,J^\pi=0,1,3/2^-$
channel, the $\Sigma_c(2550)$ and $\Sigma_b(5904)$ (that strongly
couple to the $D \Delta$ and $\bar{B} \Delta$ channels,
respectively). These states are the counterparts of the experimental
$\Sigma^*(1670)$ in the strange sector, but have not been yet
observed, so they can be taken as predictions for future
measurements. Many other resonant states (especially in the
meson-baryon sector) are found in the remaining scattering channels.

The resulting cross sections for the binary scattering of
$D,D^*,B,B^*$ (with any possible charged states) with
$\pi,K,\bar{K},\eta,N,\bar{N},\Delta,\bar{\Delta}$ are implemented
in the PHSD code considering both elastic and inelastic channels.
Around 200 different channels are taken into account. Although the
unitarization method helps to extend the validity of the tree-level
amplitudes into the resonant region, one cannot trust the final
cross sections for higher energies. Beyond the resonant region we
adopt constant cross sections inspired by the results of the Regge
analysis in the energy domain of several GeV~\cite{Agashe:2014kda},
where one expects an almost flat energy dependence of the cross
sections.

\begin{figure} [h]
\centerline{
\includegraphics[width=9.5 cm]{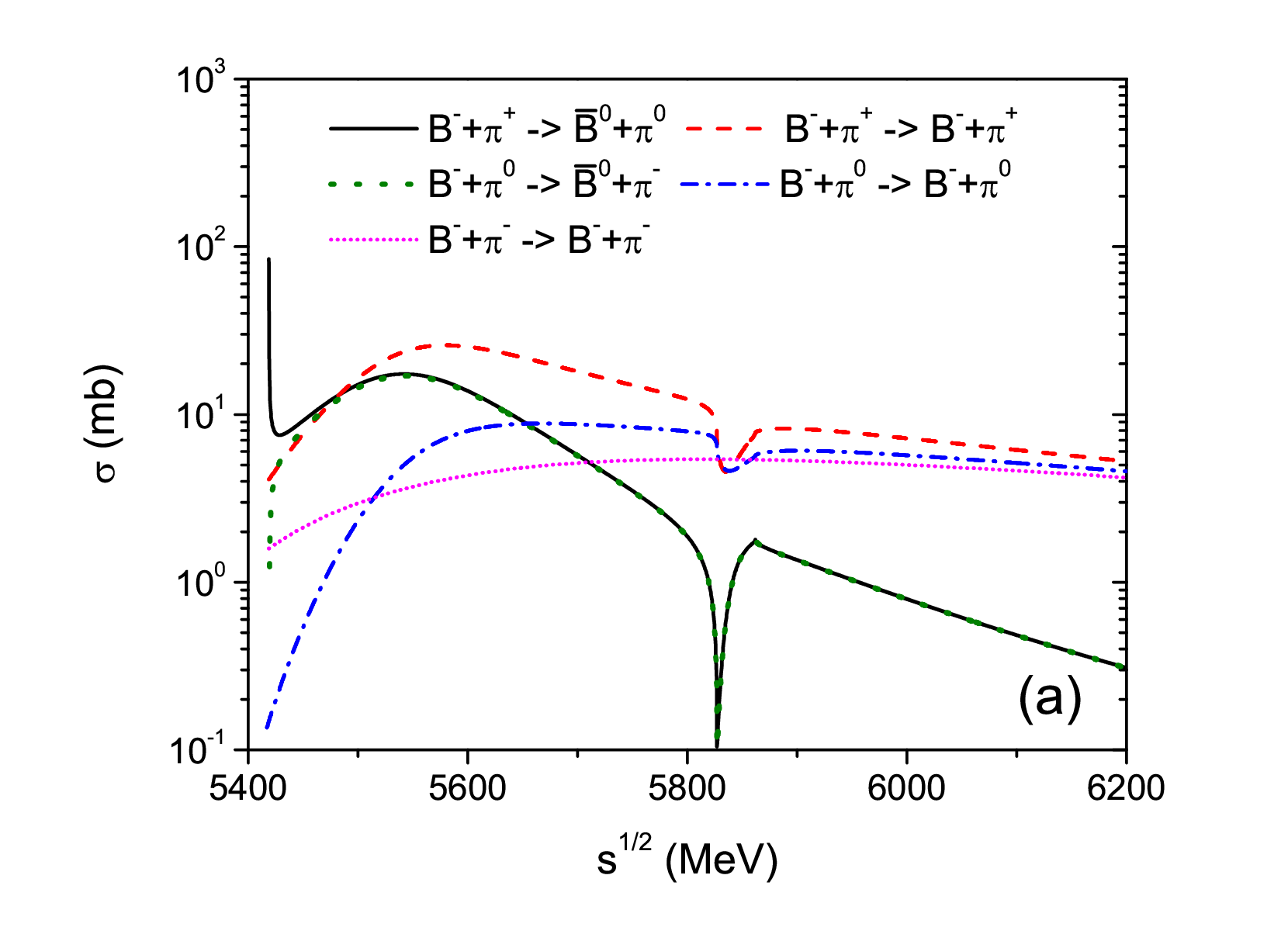}}
\centerline{
\includegraphics[width=9.5 cm]{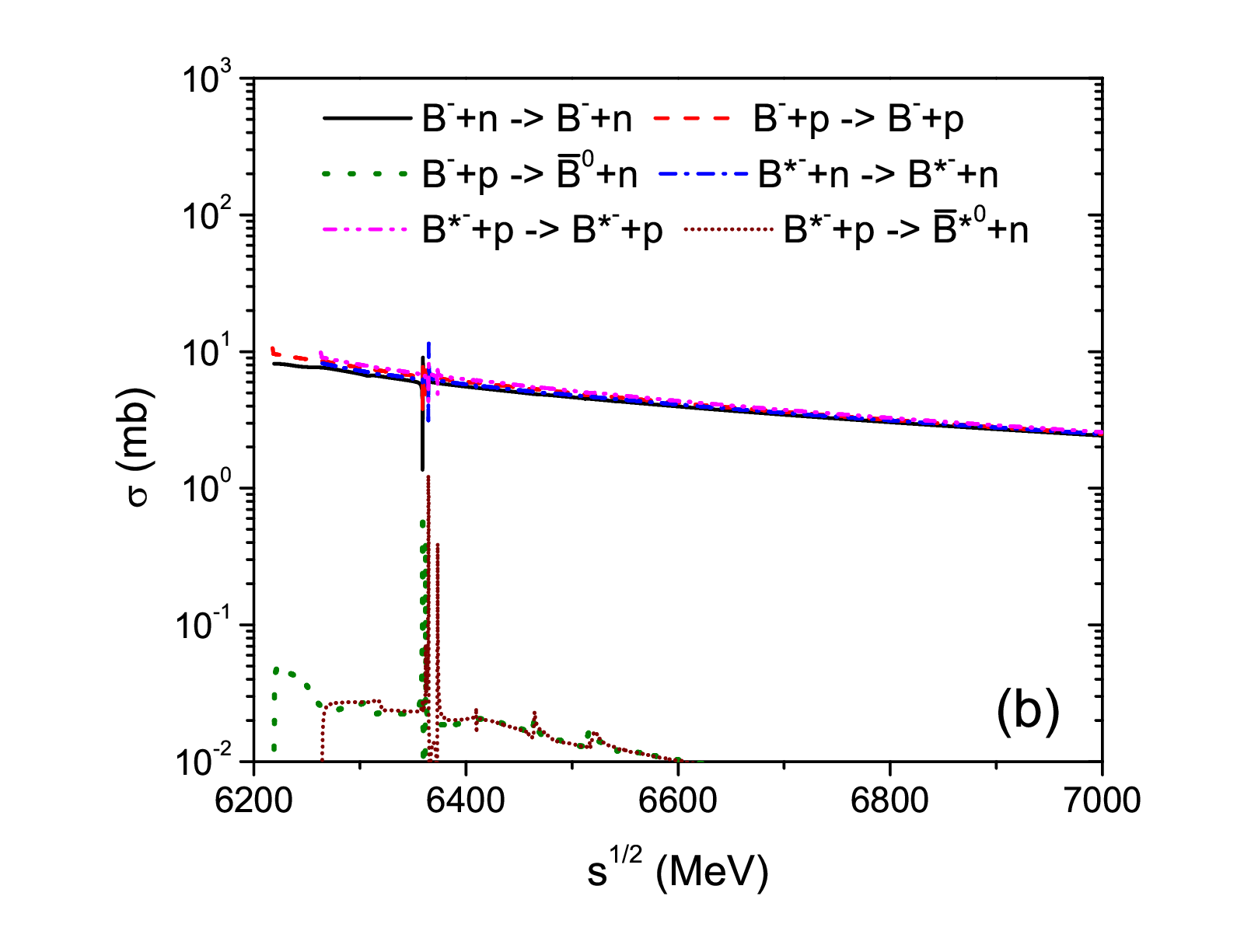}}
\caption{(Color online) Several examples of $B$ meson scattering cross sections with pion (a) and nucleon (b).} \label{BH}
\end{figure}

In figure~\ref{BH} we present several examples of $B^{(*)-}$ meson scattering
cross sections with pions and nucleons. The cross sections show a
non-smooth behaviour with energy, due to the presence of several mesonic
and baryonic beauty states generated dynamically, as described above. As an example, in the scattering with pions we observed
the very broad resonant peak of $B_0(5530)$. The clear dip of some of the
cross sections around 5830 MeV is due to the opening of the coupled-channel
$B-\eta$ at $s^{1/2}=m_B+m_\eta$ (Flatt\'e effect). In the $B-n$ and $B-p$
cross sections, we observe the presence of baryonic states around 6360 MeV
in both $I=0$ and $I=1$ channels. The position and the width of these
states as well as the coupling of these states to the main channels have
been carefully analyzed in Ref.~\cite{Tolos:2013kva,Torres-Rincon:2014ffa} in connection with several transport
coefficients in heavy-ion collisions.

\section{Results for heavy-ion reactions}\label{results}

So far we have described the interactions of the heavy flavor
produced in relativistic heavy-ion collisions with partonic and
hadronic degrees-of-freedom. Since the  matter produced in heavy-ion
collisions is extremely dense, the interactions with the bulk matter
suppresses heavy flavors at high-$\rm p_T$. On the other hand, the
partonic or nuclear matter is accelerated outward (exploding), and a
strong flow is generated via the interactions of the bulk particles
and the repulsive scalar interaction for partons. Since the heavy
flavor strongly interacts with the expanding matter, it is also
accelerated outwards. Such effects of the medium on the heavy-flavor
dynamics are expressed in terms of the nuclear modification factor
defined as
\begin{eqnarray}
R_{\rm AA}({\rm p_T})\equiv\frac{dN_{\rm AA}/d{\rm p_T}}{N_{\rm binary}^{\rm AA}\times dN_{\rm pp}/d{\rm p_T}},
\label{raa}
\end{eqnarray}
where $N_{\rm AA}$ and $N_{\rm pp}$ are, respectively, the number of
particles produced in heavy-ion collisions and that in p+p
collisions, and $N_{\rm binary}^{\rm AA}$ is the number of binary
nucleon-nucleon collisions in the heavy-ion collision for the
centrality class considered. Note that if the heavy flavor does not
interact with the medium in heavy-ion collisions, the numerator of
Eq.~(\ref{raa}) will be similar to the denominator. For the same
reason, an $R_{\rm AA}$ smaller (larger) than one in a specific
$\rm p_T$ region implies that the nuclear matter suppresses
(enhances) the production of heavy flavors in that transverse
momentum region.

In noncentral heavy-ion collisions the produced  matter expands
anisotropically due to the different pressure gradients between in
plane and out-of plane. If the heavy flavor interacts strongly with
the nuclear matter, then it also follows this anisotropic motion to
some extend. The anisotropic flow is expressed in terms of the
elliptic flow $v_2$ which reads
\begin{eqnarray}
v_2({\rm p_T})\equiv\frac{\int d\phi \cos2\phi (dN_{\rm AA}/d{\rm p_T}d\phi)}{2\pi dN_{\rm AA}/d{\rm p_T}},
\end{eqnarray}
where $\phi$ is the azimuthal angle of a particle in momentum space.

In the following Subsections, we will present our results on the
production of heavy flavors and single electrons in Au+Au collisions
at $\sqrt{s_{\rm NN}}=$200 GeV, 62.4 GeV, make predictions for even
lower energies, and discuss the azimuthal angular correlations
between a heavy-flavor meson and its anti-flavor meson.

\subsection{Au+Au at $\sqrt{s_{\rm NN}}=$200 GeV}

\begin{figure} [h]
\centerline{
\includegraphics[width=9.5 cm]{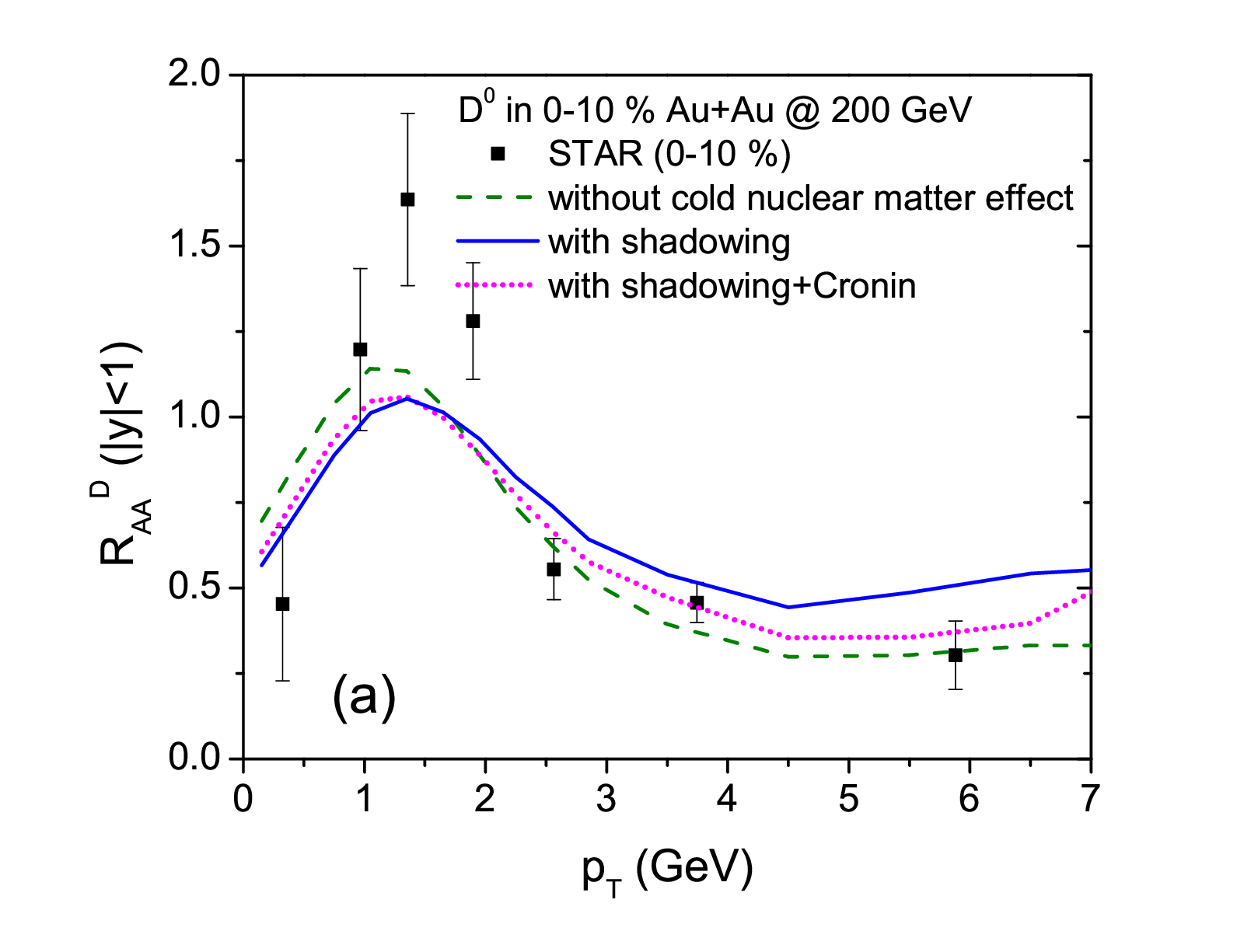}}
\centerline{
\includegraphics[width=9.5 cm]{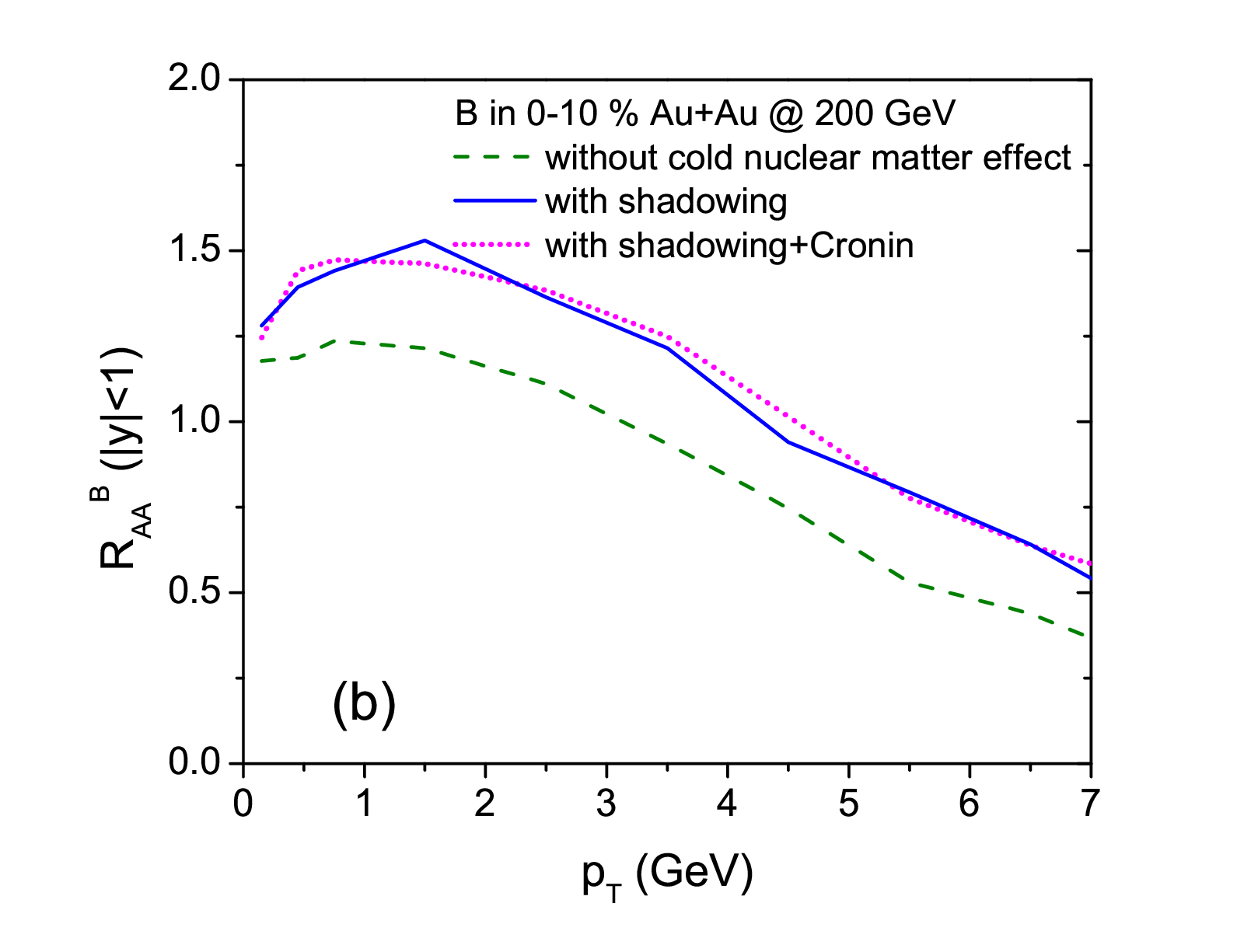}}
\caption{(Color online) $R_{\rm AA}$ of $D^0-$mesons (a) and of
$B-$mesons (b) with (solid) and without (dashed) shadowing effect in
0-10 \% central Au+Au collisions at $\sqrt{s_{\rm NN}}=$200 GeV in
comparison to the experimental data from the STAR
collaboration~\cite{Adamczyk:2014uip}. The dotted lines are $R_{\rm AA}$ including both shadowing and Cronin effects.} \label{raa200DB}
\end{figure}

The upper (a) and lower (b) panels of figure~\ref{raa200DB} are, respectively,
the $R_{\rm AA}$ of $D-$mesons and of $B-$mesons in 0-10 \% central
Au+Au collisions at $\sqrt{s_{\rm NN}}=$200 GeV from the PHSD
calculations. The shadowing effect is excluded in the dashed lines
and is included in the solid lines. Furthermore, in figure
\ref{raa200DB} (a) the $R_{\rm AA}$ of $D-$mesons are compared with
the experimental data from the STAR
collaboration~\cite{Adamczyk:2014uip}. We note that the $R_{\rm AA}$
of $D-$mesons without shadowing effect is slightly different from
our previous results in Ref.~\cite{Song:2015sfa}, because the
elastic backward scattering has been improved and the coalescence of
charm quark takes place continuously as in the later
work~\cite{Song:2015ykw}. As shown in figure~\ref{shadowingf}, the
shadowing effect decreases the charm production by $\sim$8 \% and
increases the bottom production by $\sim$20 \%. Apparently the
$R_{\rm AA}$ of $B-$mesons is much larger than that of $D-$mesons at
the same transverse momentum. However, this is attributed to the
larger bottom mass than charm mass as demonstrated in
figure~\ref{coalpro} before.

\begin{figure} [h]
\centerline{
\includegraphics[width=9.5 cm]{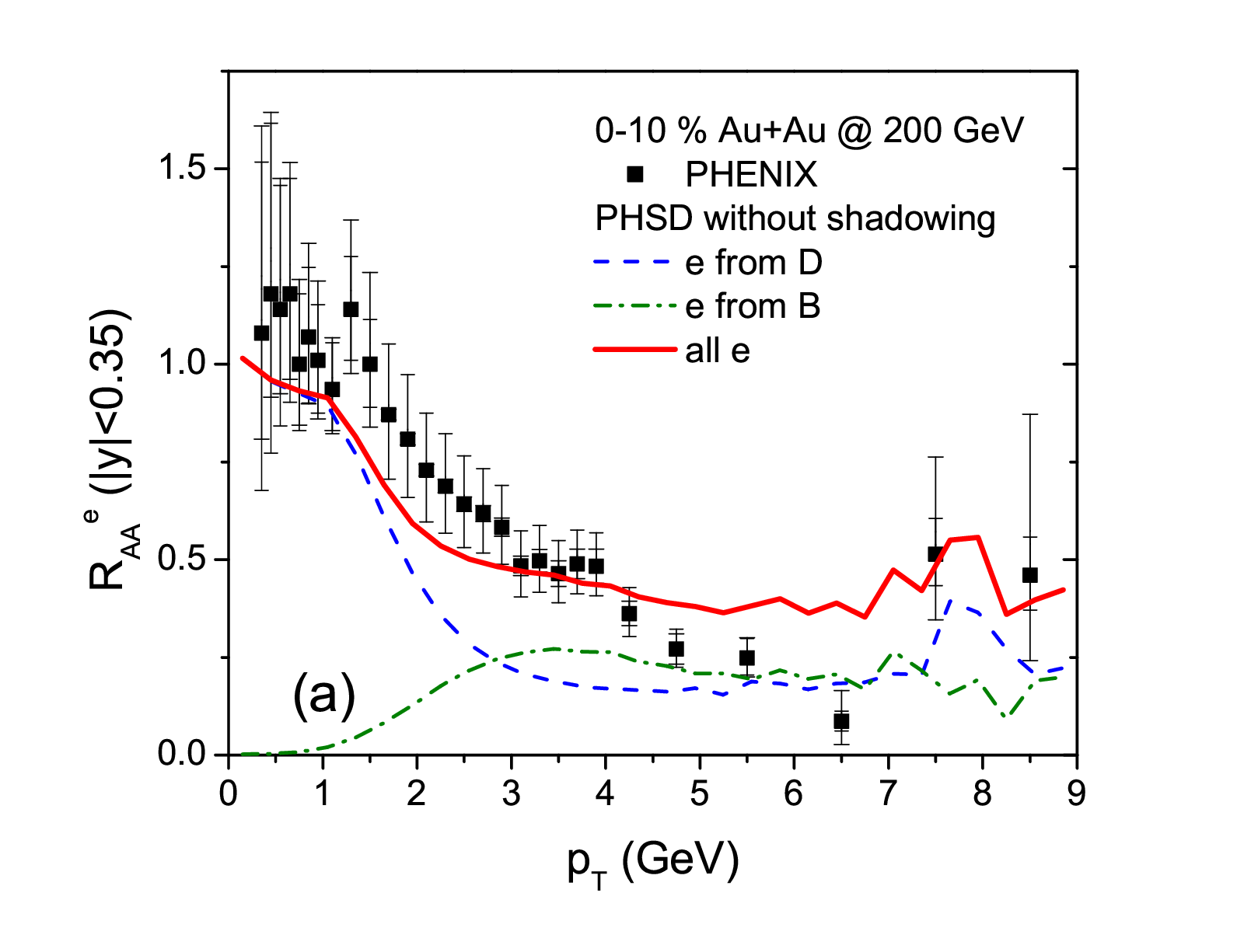}}
\centerline{
\includegraphics[width=9.5 cm]{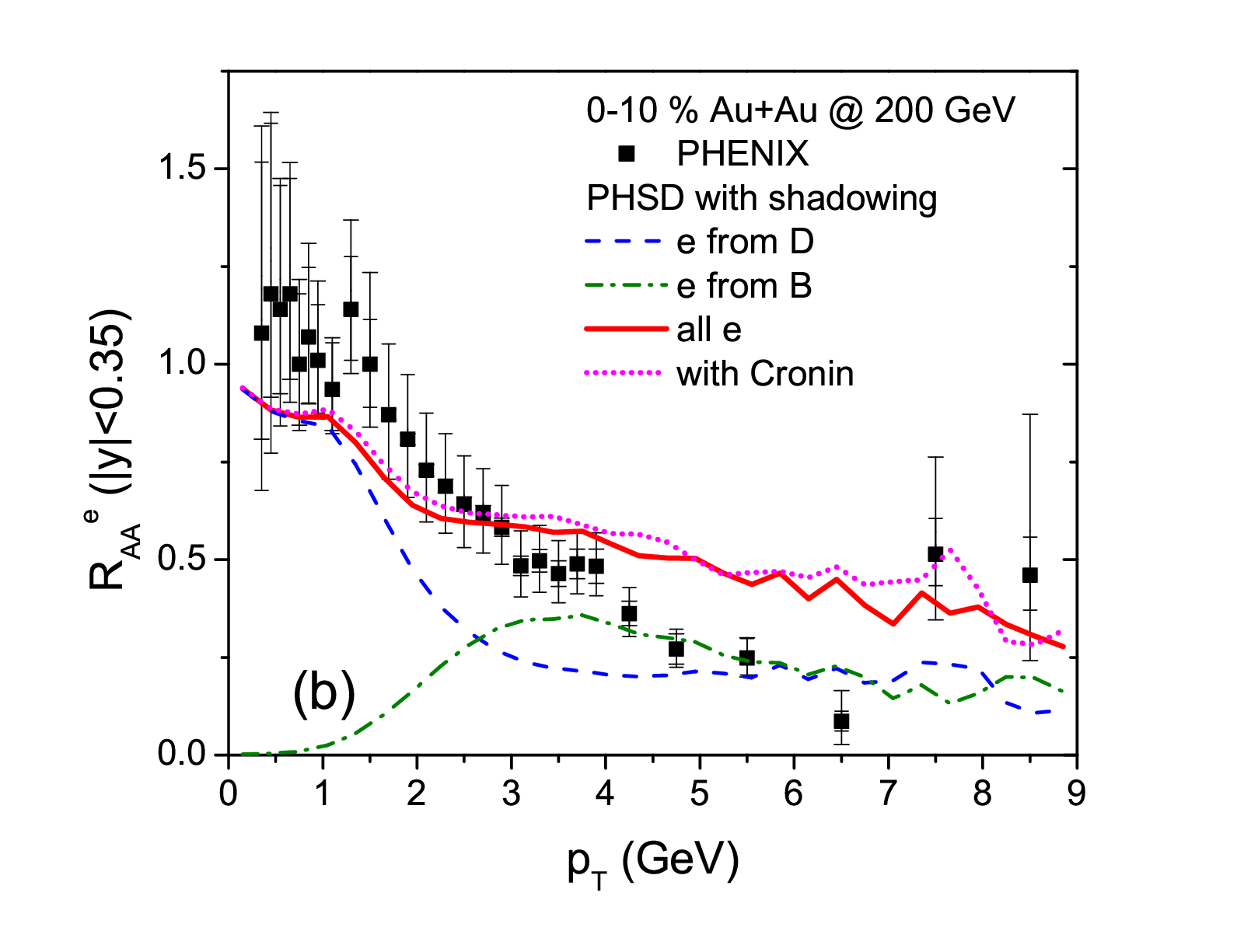}}
\caption{(Color online) $R_{\rm AA}$ of single electrons from the
semi-leptonic decay of $D-$mesons (dashed) and of $B-$mesons
(dot-dashed) and the sum of them (solid) with (b) and without (a)
shadowing effect in 0-10 \% central Au+Au collisions at
$\sqrt{s_{\rm NN}}=$200 GeV in comparison to the experimental data
from the PHENIX collaboration~\cite{Adare:2014rly}. The dotted line in (b) is the $R_{\rm AA}$ including both shadowing and Cronin effects.} \label{raa200e}
\end{figure}

Figure~\ref{raa200e} shows the $R_{\rm AA}$ of single electrons from
$D-$meson and $B-$meson semileptonic decays, which correspond to the
dashed and dot-dashed lines, respectively, while the solid lines are
sum of them in 0-10 \% central Au+Au collisions at $\sqrt{s_{\rm
NN}}=$200 GeV. The upper figure (a) is the $R_{\rm AA}$ without
shadowing effect, and the lower one (b) includes the shadowing
effect, which enhances the bottom production and suppresses the
charm production at low transverse momentum in line with the
discussion above. We find that the single electrons from $B$ decay
have a larger contribution than that from $D$ decay above ${\rm p_T}
\approx 2.7-2.8$ GeV. In p+p collisions, the contribution from $B$
decay starts to be larger than that from $D$ decay at about ${\rm
p_T} \approx 4$ GeV as shown in figure~\ref{ppe}. The reason for the
dominance of $B$ decay at lower transverse momentum in Au+Au
collisions is that the $R_{\rm AA}$ of $B-$mesons is larger than
that of $D-$mesons at high transverse momentum as shown in
figure~\ref{raa200DB}.
The dotted lines in figure~\ref{raa200DB} and~\ref{raa200e} are, respectively, the $R_{\rm AA}$ of heavy mesons and single electrons
including both shadowing and Cronin effects. Although the Cronin effect enhances the $R_{\rm AA}$, it is not significant.

\begin{figure} [h]
\centerline{
\includegraphics[width=10 cm]{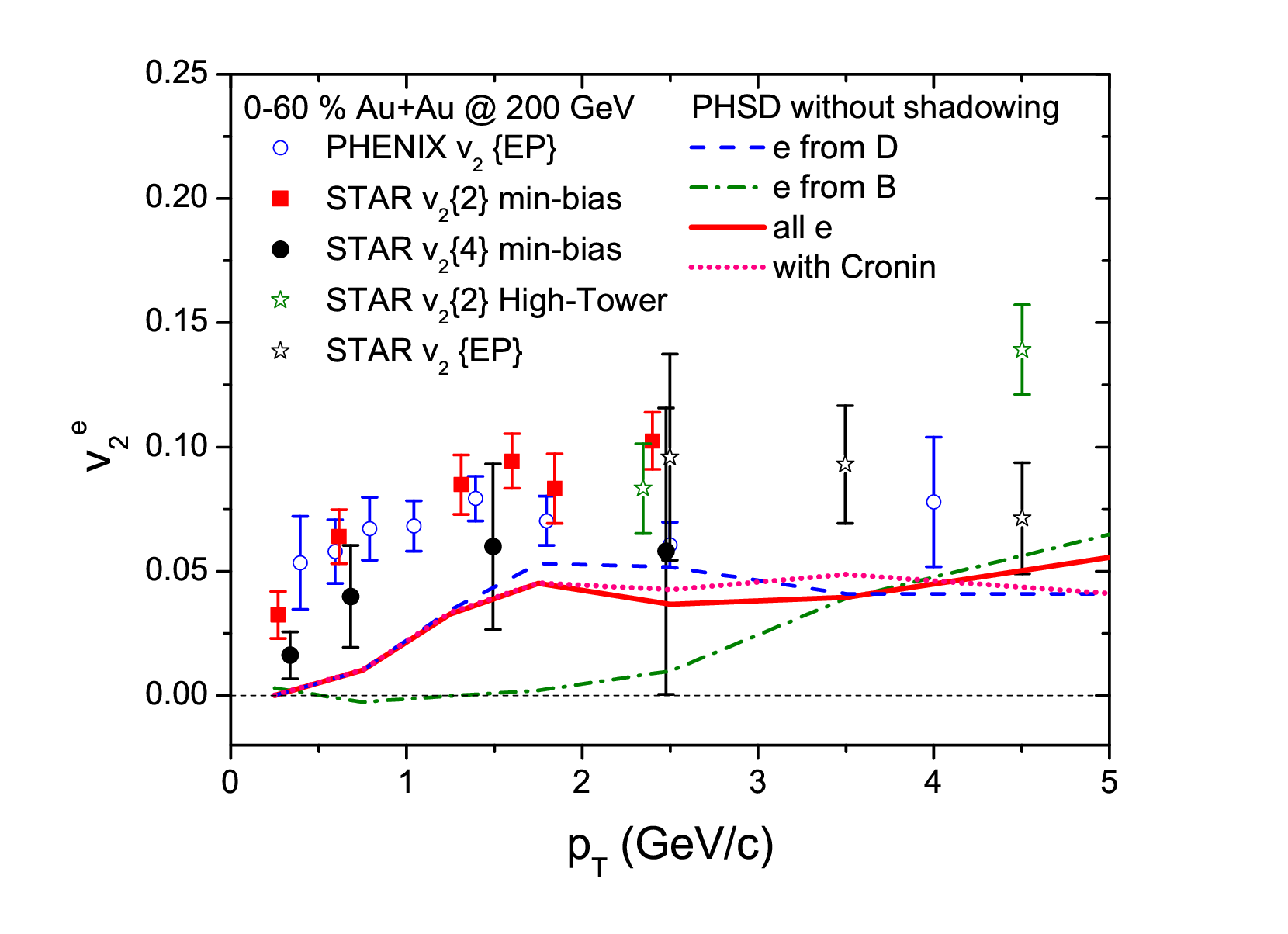}}
\caption{(Color online) The elliptic flow $v_2$ of single electrons
from the semi-leptonic decay of $D-$mesons (dashed) and of
$B-$mesons (dot-dashed) and of both of them (solid) with shadowing effect in 0-60 \% central Au+Au collisions at
$\sqrt{s_{\rm NN}}=$200 GeV in comparison to the experimental data
from the PHENIX and STAR collaborations~\cite{Adare:2006nq,Adamczyk:2014yew}. The dotted line is the $v_2$ of single electrons including both shadowing and Cronin effects.} \label{v2e200}
\end{figure}

We present in figure~\ref{v2e200} the elliptic flow $v_2$ of single
electrons with shadowing effect in 0-60 \% central
Au+Au collisions at $\sqrt{s_{\rm NN}}=$200 GeV. The dashed and
dot-dashed lines are, respectively, the $v_2$ of single electrons
from $D-$meson and $B-$meson decays. Since the $B-$meson is much
more massive, the elliptic flow from $B-$meson decay starts to grow
from much higher transverse momentum. The red lines are the elliptic
flow $v_2$ of all single electrons. Figures~\ref{raa200e} and
\ref{v2e200} show that PHSD can approximately reproduce the
experimental data on the $R_{\rm AA}$ but slightly underestimates the $v_2$ of single electrons
from $D-$mesons and $B-$mesons at $\sqrt{s_{\rm NN}}=$200 GeV. They
also show that the shadowing effect is not so critical in
reproducing experimental data, which is different from the LHC
energies~\cite{Song:2015ykw}.

\subsection{Au+Au at $\sqrt{s_{\rm NN}}=$62.4 GeV}

The beam energy scan (BES) program at RHIC has been carried out by
colliding Au nuclei at various energies down to $\sqrt{s_{\rm
NN}}=$7.7 GeV. The aim of the program is to find information on the
phase boundary and hopefully the critical point in the QCD phase
diagram as pointed out in Refs.~\cite{Mohanty:2011nm,Kumar:2011us}.
It is expected that if the trajectories of the produced nuclear
matter in the QCD phase diagram pass close to the critical point,
some drastic changes of observables could be measured in
experiments. Since the PHENIX and STAR collaborations recently measured the
single electrons from heavy flavor decay at $\sqrt{s_{\rm NN}}=$62.4
GeV~ \cite{Adare:2014rly,Adamczyk:2014yew}, which is much lower than the maximum
energy at RHIC, we first address this system in the present
subsection.

\begin{figure} [h]
\centerline{
\includegraphics[width=9.5 cm]{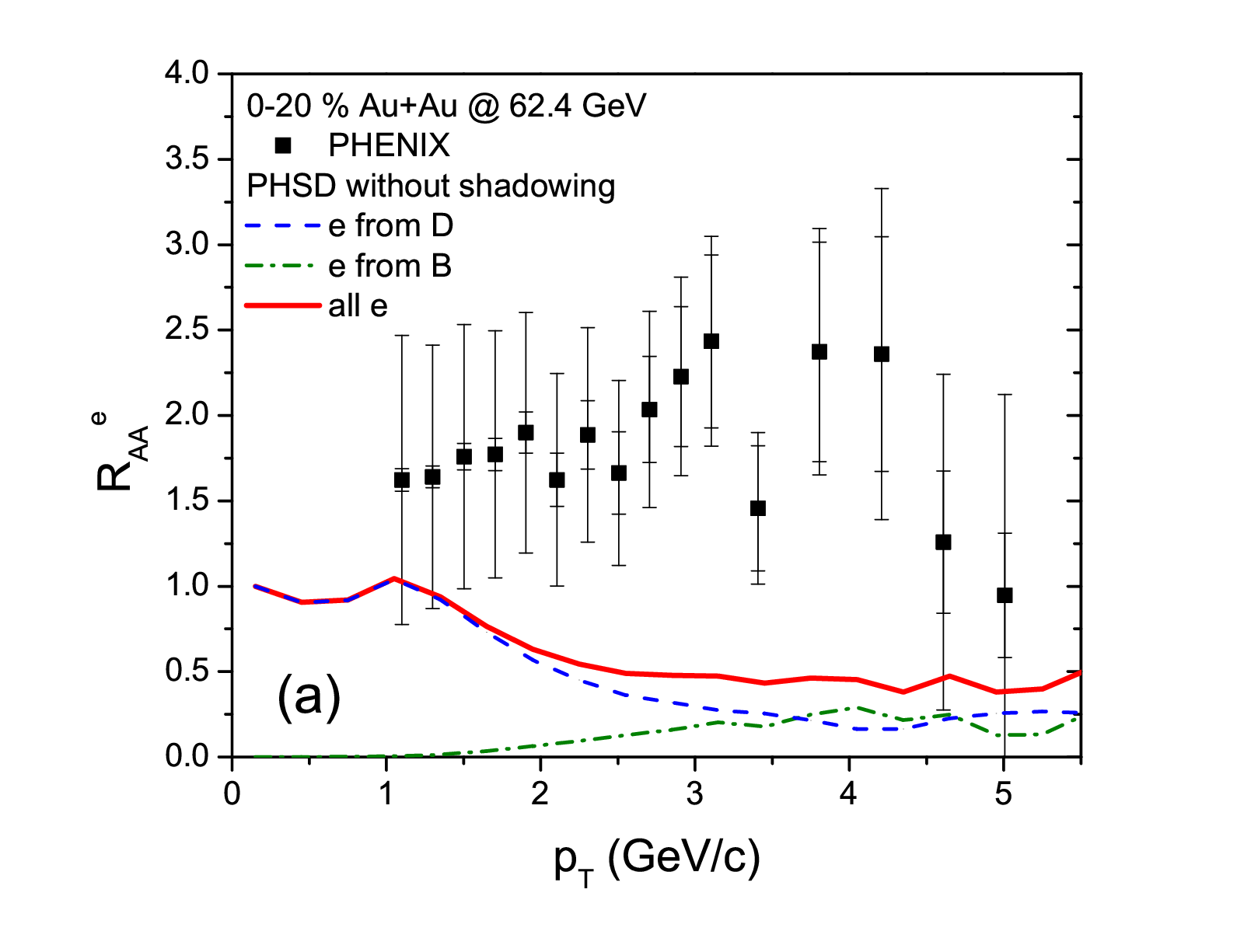}}
\centerline{
\includegraphics[width=9.5 cm]{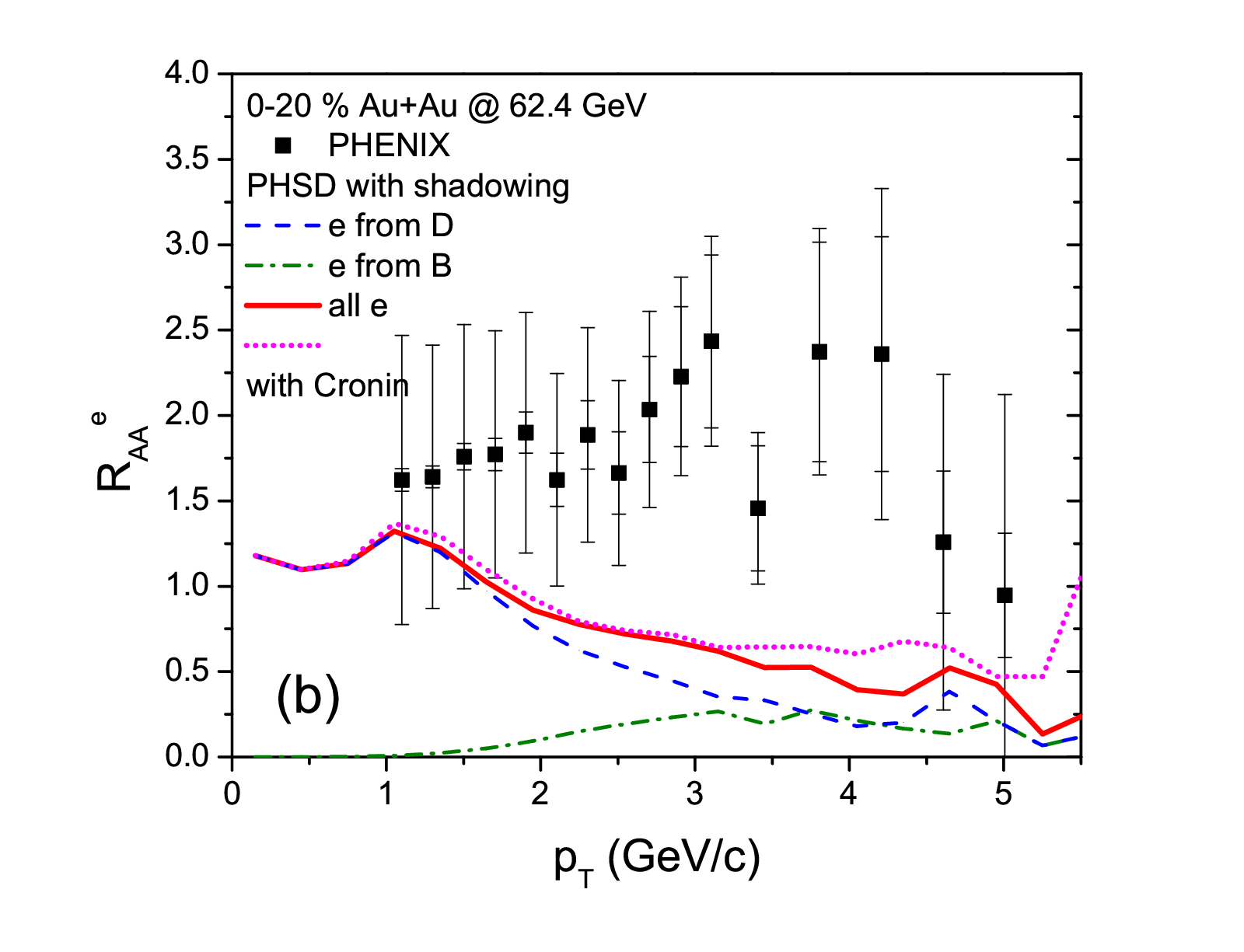}}
\caption{(Color online) $R_{\rm AA}$ of single electrons from the
semi-leptonic decay of $D-$mesons (dashed) and of $B-$mesons
(dot-dashed) and the sum of them (solid) with (b) and without (a)
shadowing effect in 0-20 \% central Au+Au collisions at
$\sqrt{s_{\rm NN}}=$62.4 GeV in comparison to the experimental data
from the PHENIX collaboration~\cite{Adare:2014rly}. The dotted line in (b) is the $R_{\rm AA}$ including both shadowing and Cronin effects.} \label{raa62e}
\end{figure}

\begin{figure} [h]
\centerline{
\includegraphics[width=10 cm]{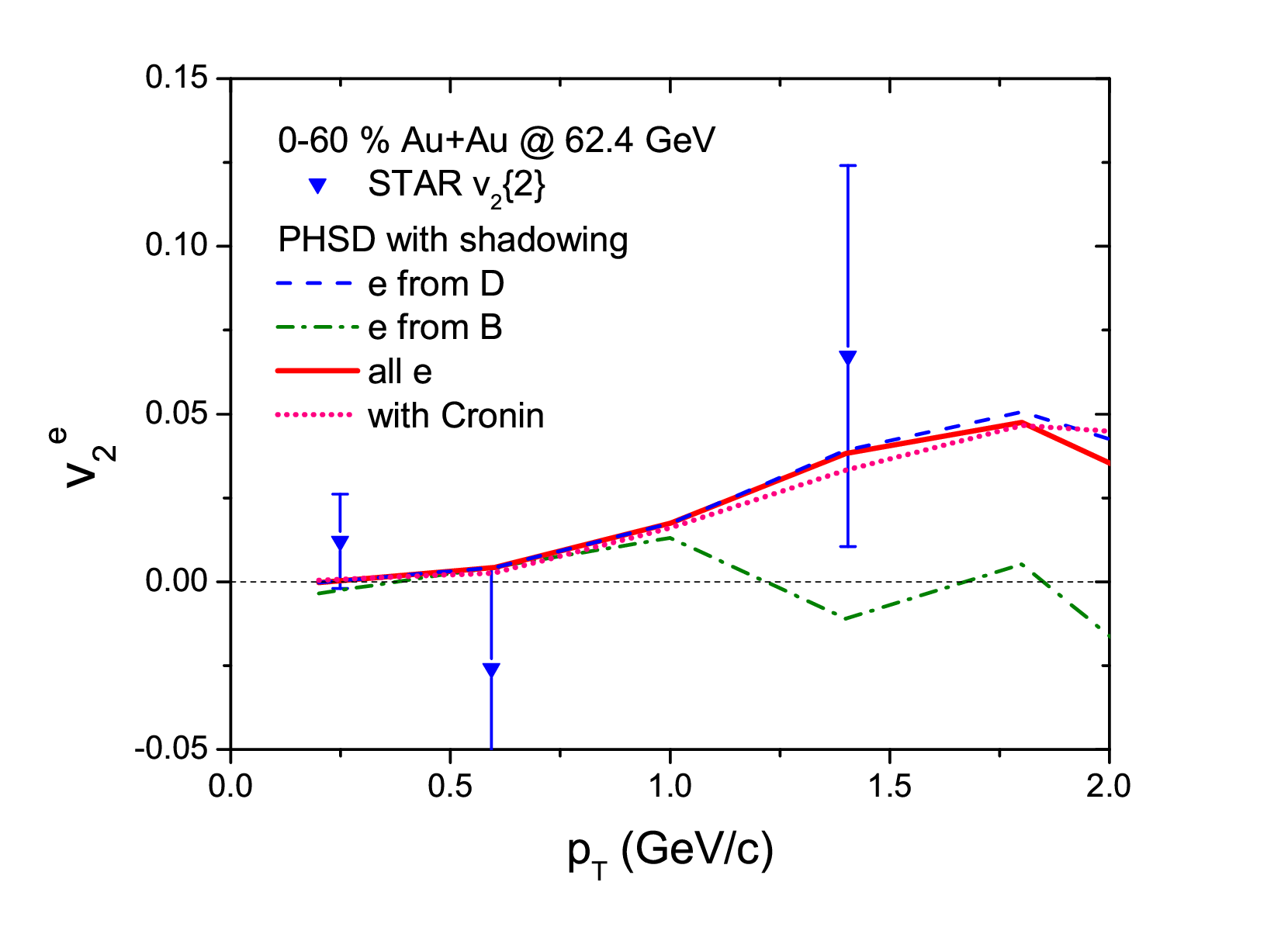}}
\caption{(Color online) The elliptic flow $v_2$ of single electrons
from the semi-leptonic decay of $D-$mesons (dashed) and of
$B-$mesons (dot-dashed) and of both of them (solid) with shadowing effect in 0-60 \% central Au+Au collisions at
$\sqrt{s_{\rm NN}}=$62.4 GeV in comparison to the experimental data
from the STAR collaboration~\cite{Adamczyk:2014yew}. The dotted line is the $v_2$ of single electrons including both shadowing and Cronin effects.} \label{v2e62}
\end{figure}

Figures~\ref{raa62e} and \ref{v2e62} show, respectively, the
$R_{\rm AA}$ and elliptic flow $v_2$ of single electrons from the
semileptonic decay of heavy flavors in 0-20 \% and 0-60 \% central
Au+Au collisions at $\sqrt{s_{\rm NN}}=$62.4 GeV. The upper figure
(a) is the results without shadowing effect and the lower one (b)
with the shadowing effect. As at $\sqrt{s_{\rm NN}}=$200 GeV,
the contribution from $D-$meson decay is important at low transverse
momentum and superseeded by the contribution from $B$ decay above 3
GeV. The contribution from $B$ decay becomes dominant at higher
transverse momentum than at $\sqrt{s_{\rm NN}}=$200 GeV, because the
ratio of the scattering cross section for bottom production to that
for charm production is much lower at $\sqrt{s_{\rm NN}}=$62.4 GeV.
The latter ratio is 0.75 \% at $\sqrt{s_{\rm NN}}=$200 GeV and 0.145
\% at $\sqrt{s_{\rm NN}}=$62.4 GeV according to the FONLL
calculations~\cite{Cacciari:2012ny}.

Our PHSD results in figure~\ref{raa62e} underestimate $R_{\rm AA}$
besides touching the lower error bars of the experimental data at
low and high $\rm p_T$. Although the shadowing and Cronin effects enhance $R_{\rm AA}$ at low $\rm p_T$, there is still a large discrepancy
between the experimental data and our results in the range of $\rm
p_T$ between 2.5 and 4 GeV, which clearly lacks an explanation.
We mention that a similar pattern of results has been shown
in Ref.~\cite{He:2014epa}.

In spite of the difficulty in reproducing $R_{\rm AA}$, the elliptic
flow $v_2$ of single electrons at $\sqrt{s_{\rm NN}}=$62.4 GeV is
well described by the PHSD approach irrespective whether the shadowing effect is included or not, as
shown in figure~\ref{v2e62}. The $v_2$ of single electrons from
$B-$meson decay is small at low transverse momentum for the same
reason as at $\sqrt{s_{\rm NN}}=$200 GeV.

\subsection{Predictions at lower energies}

Presently there are only few available experimental data on open heavy
flavors below $\sqrt{s_{\rm NN}}=$62.4 GeV from the BES program.
Accordingly, we will make a prediction on the production of
$D-$mesons and of single electrons at $\sqrt{s_{\rm NN}}=$19.2 GeV
and compare again with the results at $\sqrt{s_{\rm NN}}=$200 and
62.4 GeV in order to obtain some excitation
function.

\begin{figure} [h]
\centerline{
\includegraphics[width=9.5 cm]{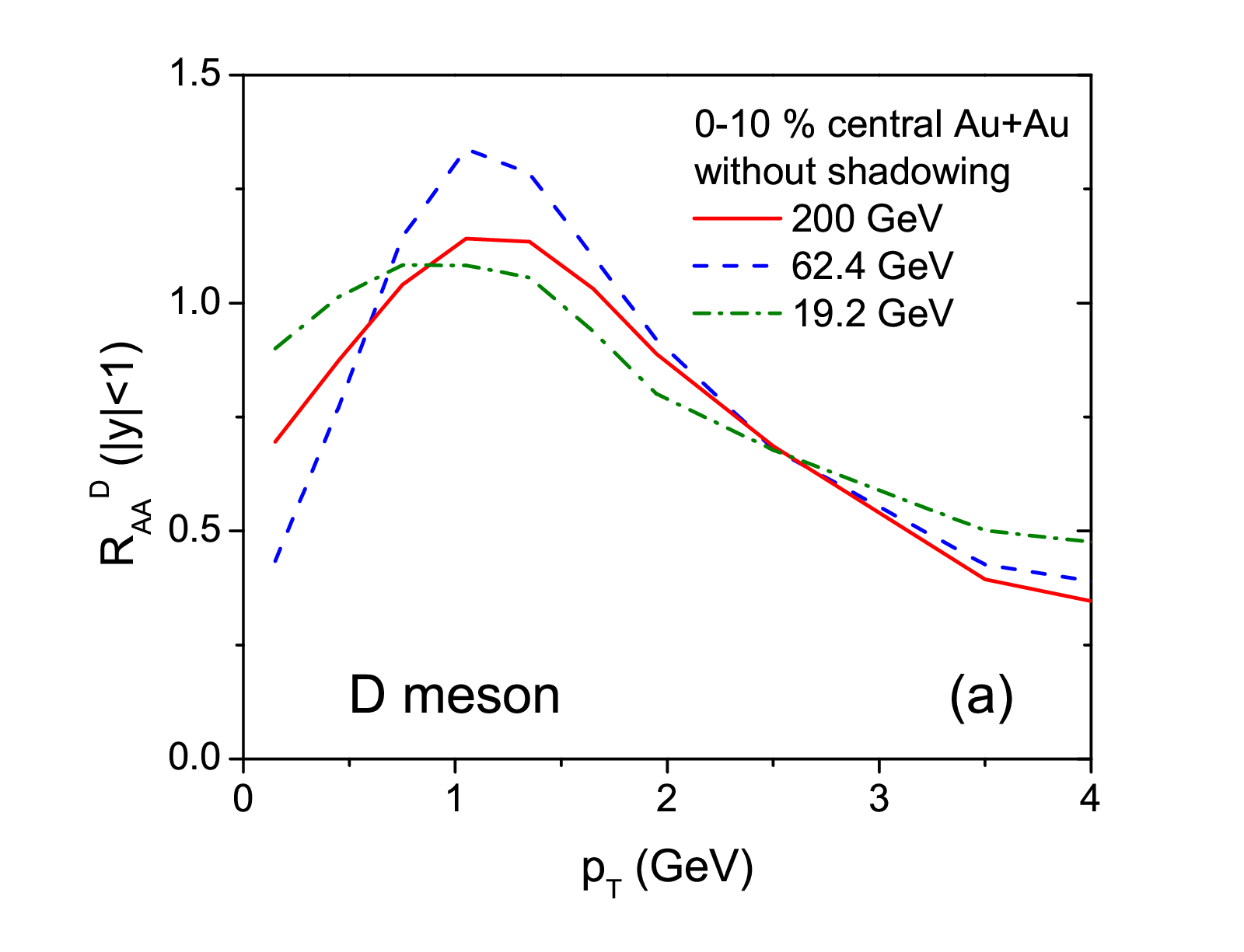}}
\centerline{
\includegraphics[width=9.5 cm]{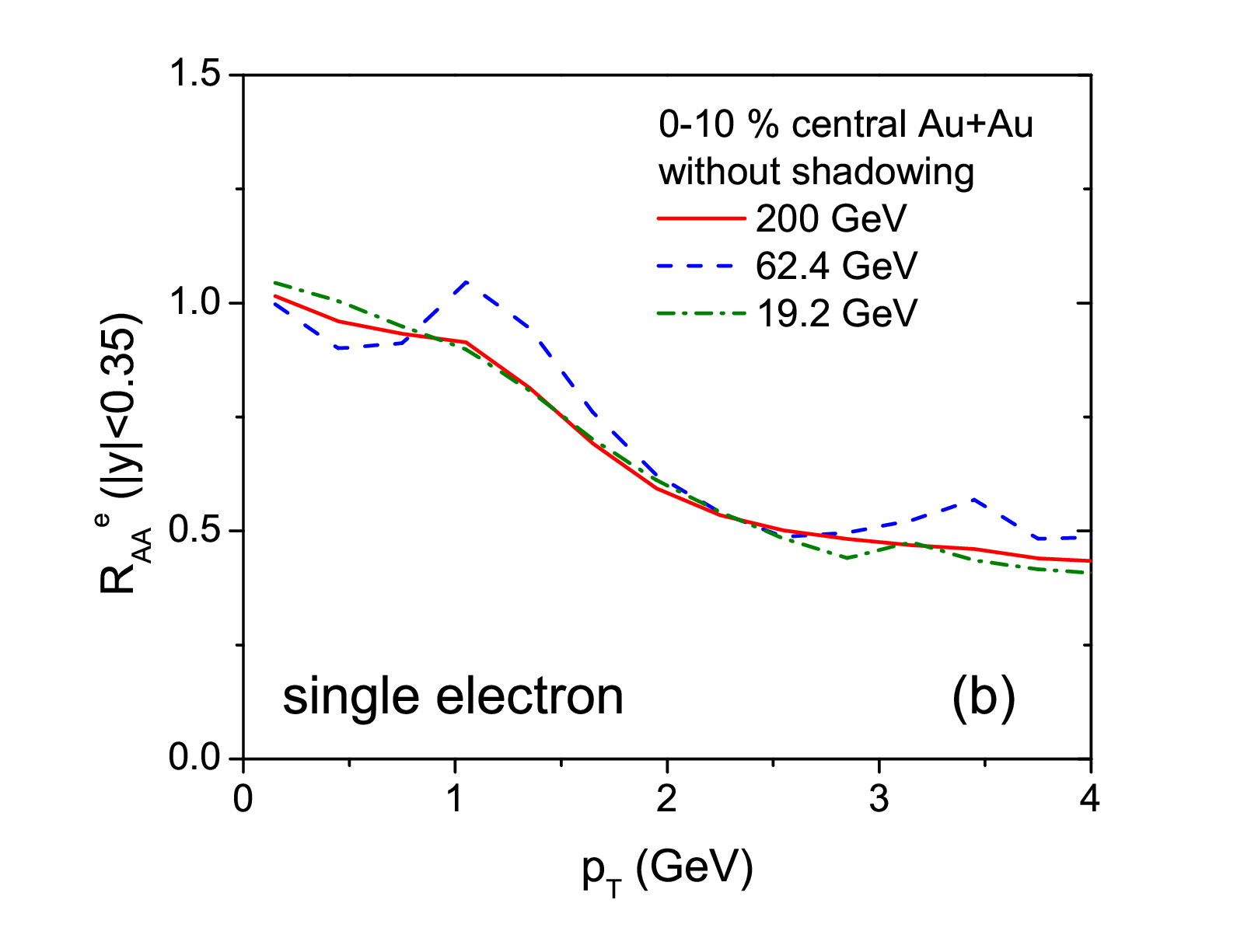}}
\caption{(Color online) $R_{\rm AA}$ of $D-$mesons (b) and of single
electrons (a) without shadowing effect in 0-10 \% central Au+Au
collisions at $\sqrt{s_{\rm NN}}=$200 GeV (solid), 62.4 GeV (dashed)
and 19.2 GeV (dot-dashed) from the PHSD approach.} \label{raa3E}
\end{figure}

The upper (a) and lower (b) panels of figure~\ref{raa3E},
respectively, show the $R_{\rm AA}$ of $D-$mesons and that of single
electrons in 0-10 \% central Au+Au collisions at $\sqrt{s_{\rm
NN}}=$200, 62.4, and 19.2 GeV. For simplicity, the shadowing effect
is not taken into account in these PHSD calculations. Comparing the
$R_{\rm AA}$ of $D-$mesons at $\sqrt{s_{\rm NN}}=$200 and 62.4 GeV,
the peak of the $R_{\rm AA}$ at 200 GeV is slightly shifted to
higher $\rm p_T$ than at 62.4 GeV. On the other hand, the $R_{\rm
AA}$ of $D-$mesons is more highly peaked at $\sqrt{s_{\rm NN}}=$62.4
GeV and the same higher peak is seen in the $R_{\rm AA}$ of single
electrons in the lower panel (b) of figure~\ref{raa3E}.
We note that the higher peak of the $R_{\rm AA}$ at $\sqrt{s_{\rm NN}}=$62.4
GeV is attributed to the initial spectrum of heavy quarks shown in figure \ref{pp200Q} and \ref{pp62Q}.
For example, if the same charm quarks produced in p+p collisions at $\sqrt{s_{\rm NN}}=$200 GeV are used as the initial charm quarks in Au+Au collisions at $\sqrt{s_{\rm NN}}=$200 and 62.4 GeV, the enhancement of the $R_{\rm AA}$ peak at $\sqrt{s_{\rm NN}}=$62.4 GeV is not observed any more.
The peak is rather shifted to slightly lower transverse momentum, compared to the $R_{\rm AA}$ at $\sqrt{s_{\rm NN}}=$200 GeV.
As for the
$R_{\rm AA}$ at $\sqrt{s_{\rm NN}}=$19.2 GeV, it is peaked at lower
$\rm p_T$ for $D-$mesons as well as for single electrons because of
the smaller transverse flow.
Comparing two panels of figure~\ref{raa3E} shows that the clear structure of $R_{\rm AA}$ of $D$ mesons at low transverse momentum smears in the $R_{\rm AA}$ of single electrons.

\begin{figure} [h]
\centerline{
\includegraphics[width=9.5 cm]{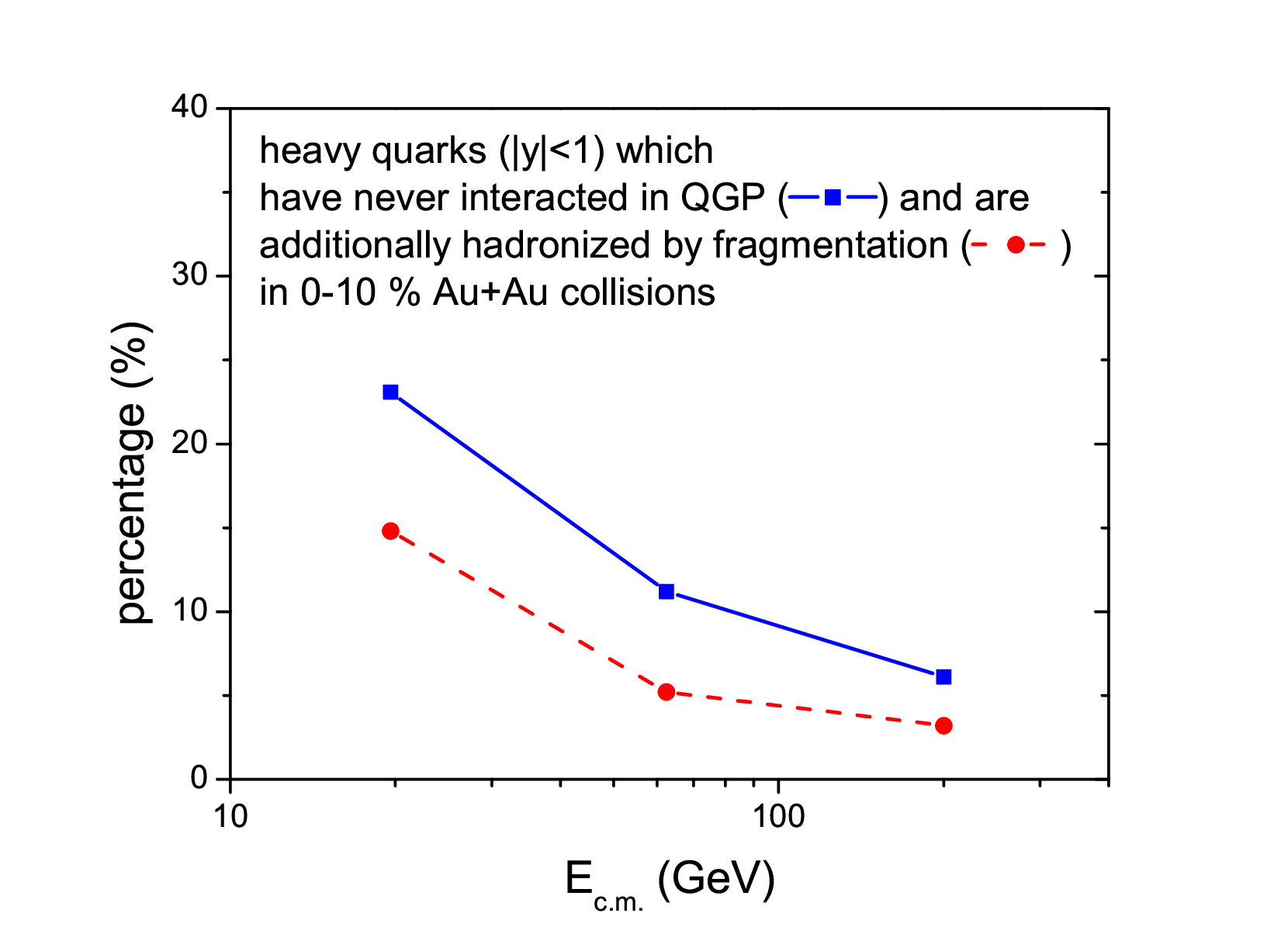}}
\caption{(Color online) The percentages of heavy quarks at mid-rapidity ($|y|<1$) with no
interactions in the QGP (solid) and additional hadronization by
fragmentation (dashed) in 0-10 \% central Au+Au collisions as a
function of collision energy.} \label{interaction}
\end{figure}

If a heavy quark is produced in the corona region, it will escape
from the interaction zone produced in heavy-ion collisions. We show
in figure~\ref{interaction} (by the solid line) the percentages of
heavy quarks at mid-rapidity ($|y|<1$) with no interactions with other partons in 0-10 \%
central Au+Au collisions as a function of collision energy. The
percentages are 23.1, 11.2, and 6.1 \% at $\sqrt{s_{\rm NN}}=$19.2, 62.4, and 200 GeV, respectively, demonstrating the decreasing role
of the corona with bombarding energy that goes along with an
increasing partonic fraction of the `fireball'. Since parton
coalescence is some kind of medium effect --  not existing in p+p
collisions --  we should interpret it as being generated by
interactions with the medium. The dashed line in the figure is the
percentage of heavy flavors at mid-rapidity ($|y|<1$) with no interactions in the QGP and
which are additionally hadronized by fragmentation. The percentages
drop down from 14.8, 5.2, and 3.2 \% , respectively. This again
demonstrates the effects from a larger (and longer) QGP phase of the
interaction zone with increasing bombarding energy.



\begin{figure*}[t]
\centerline{
\includegraphics[width=9.5 cm]{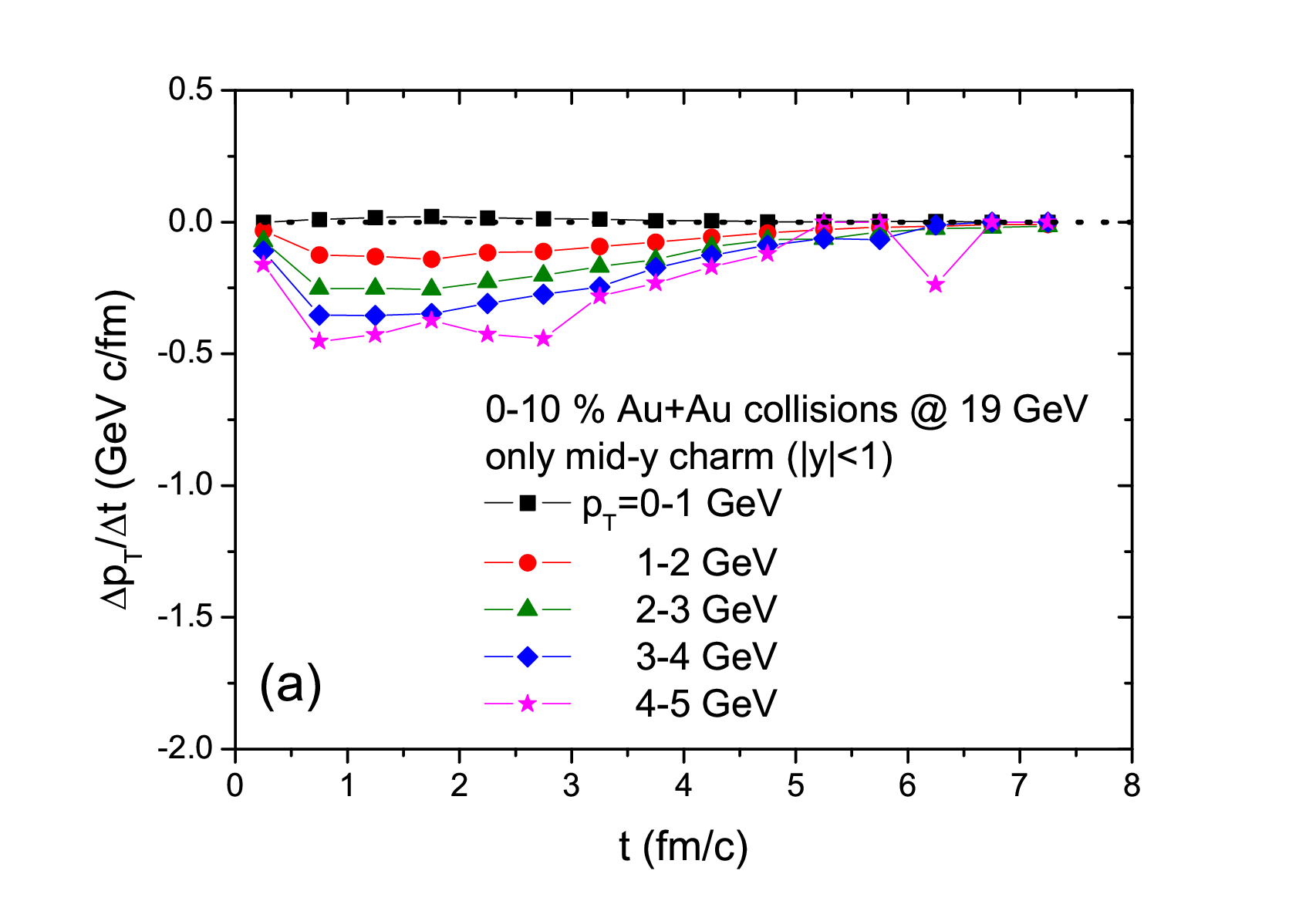}
\includegraphics[width=9.5 cm]{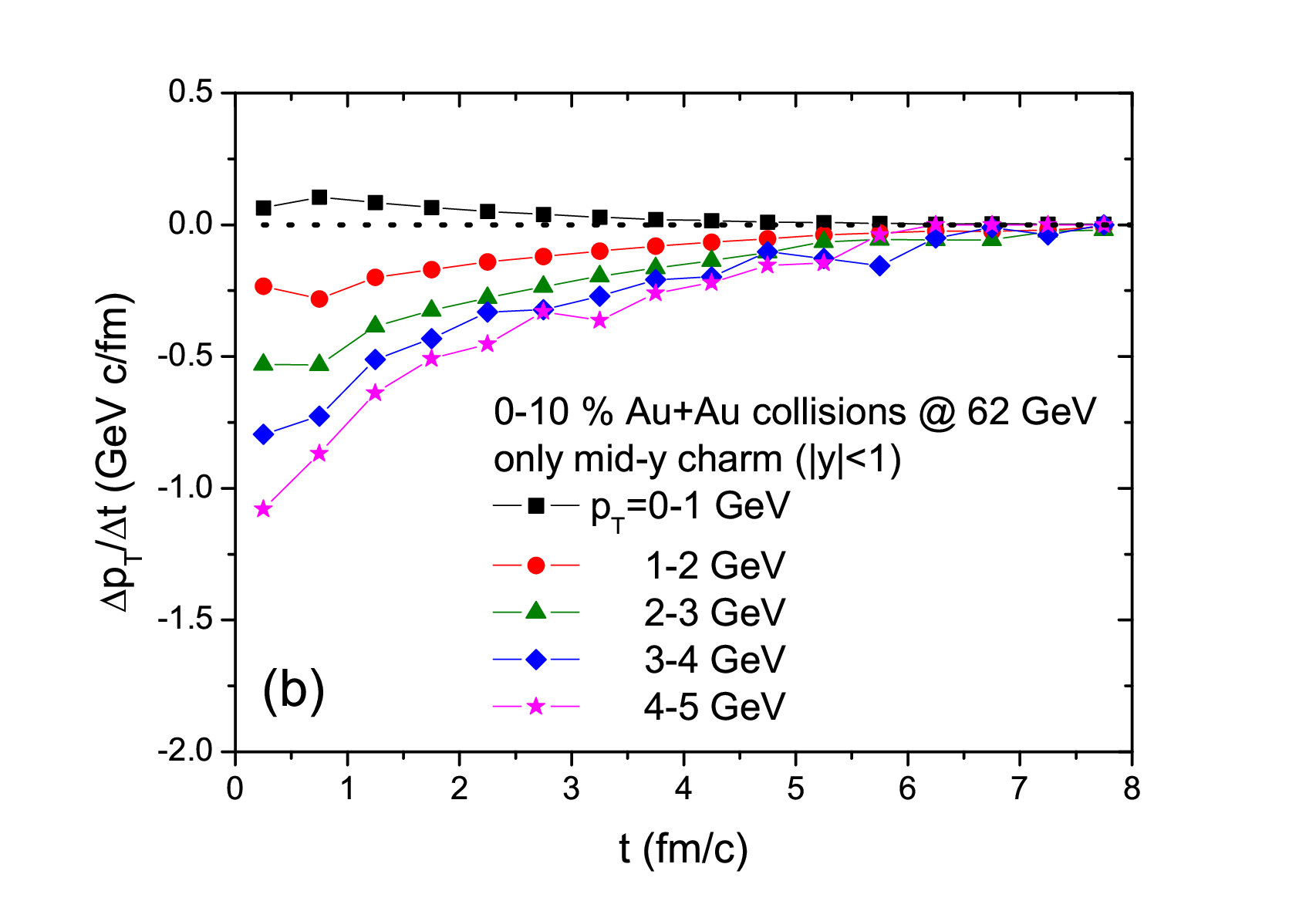}}
\centerline{
\includegraphics[width=9.5 cm]{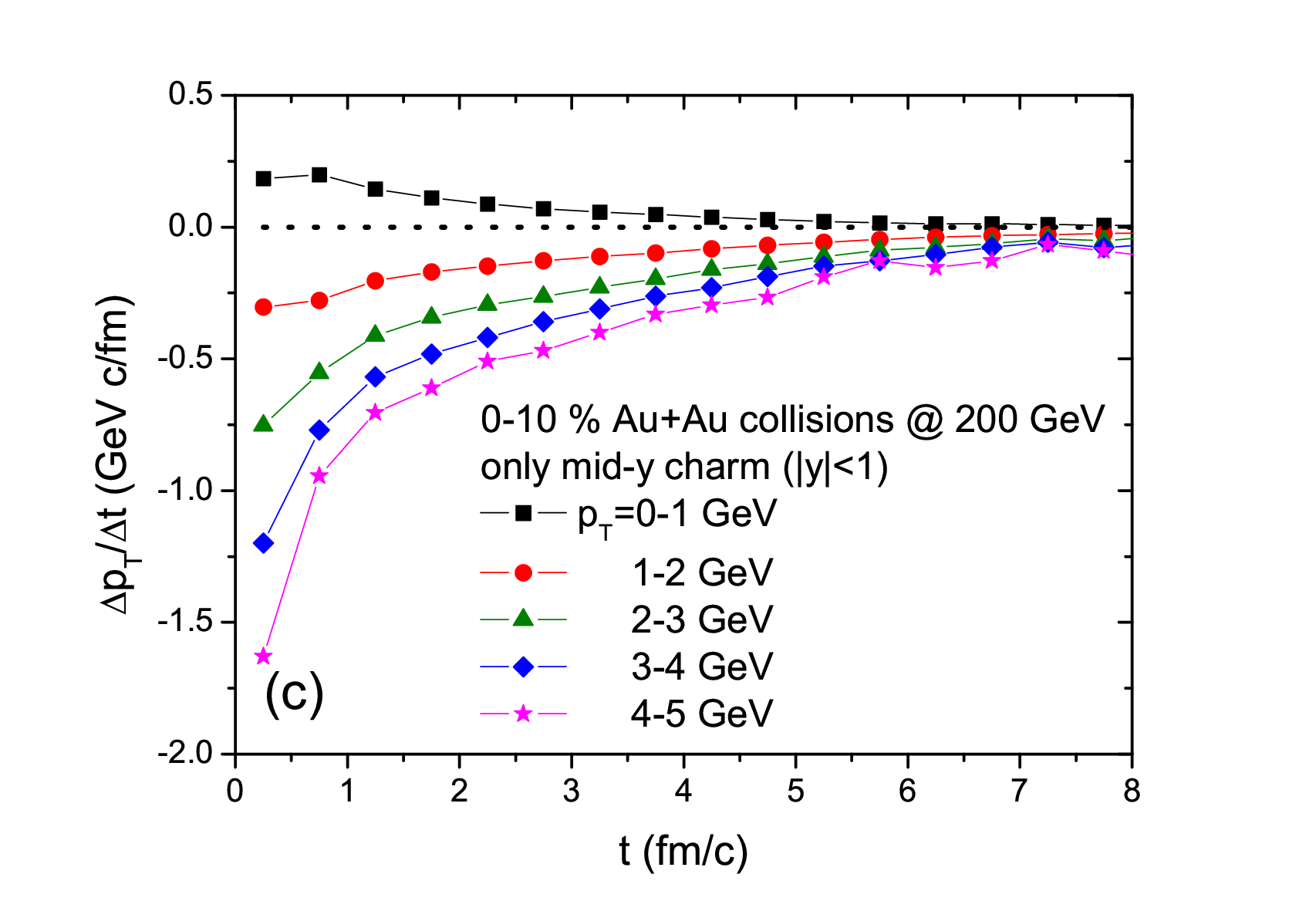}
\includegraphics[width=9.5 cm]{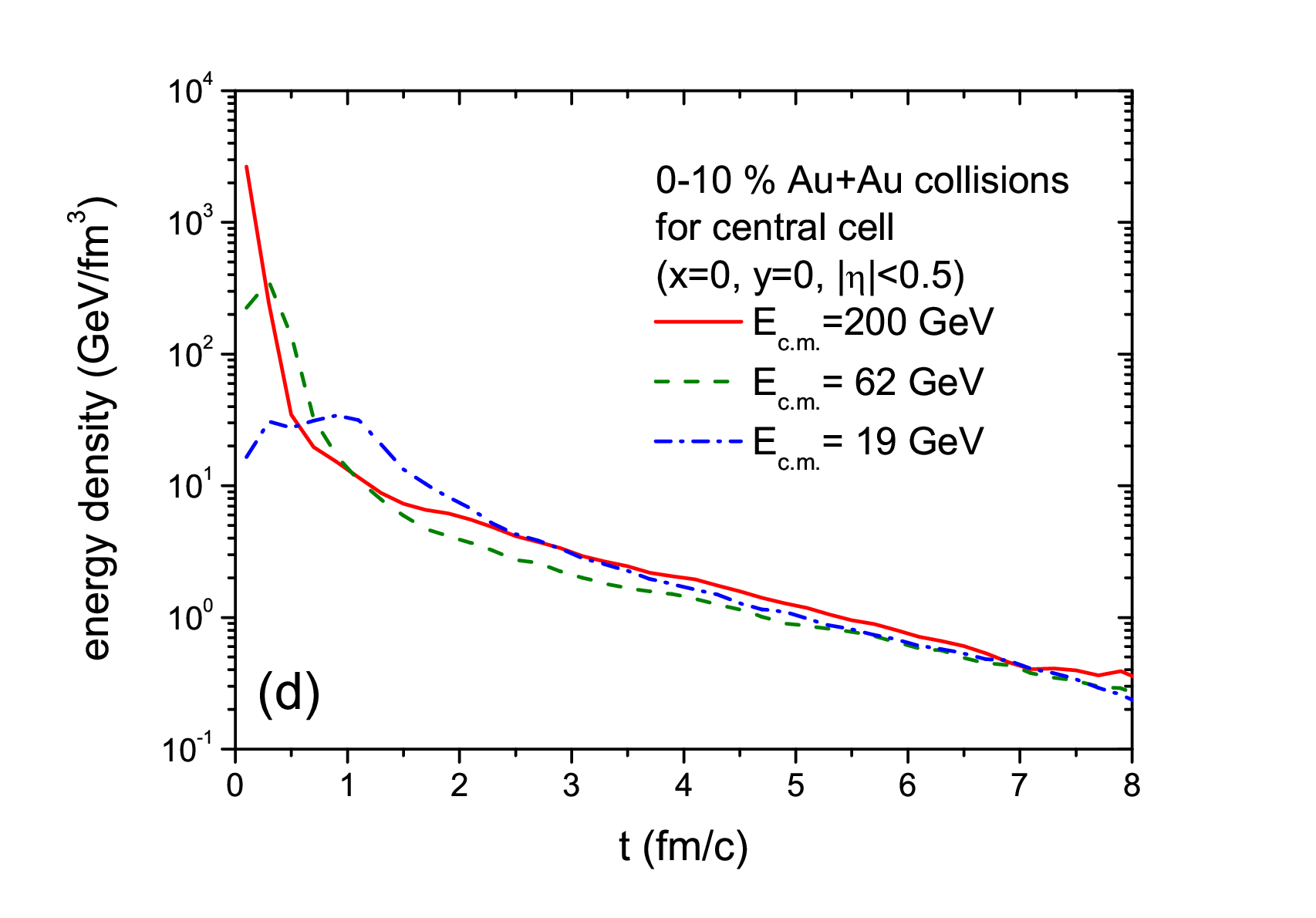}}
\caption{(Color online) Transverse momentum gain or loss of charm quarks per unit time at mid-rapidity ($|y|<1$) in 0-10 \% central Au+Au
collisions at $\sqrt{s_{\rm NN}}=$19.2 GeV (a), 62.4 GeV (b) and 200 GeV (c) and the energy density of the central cell as functions of time at each collision energy (d) from the PHSD approach.}
\label{dptf}
\end{figure*}

Figure~\ref{dptf} shows the transverse momentum gain or loss of charm quarks per unit time at mid-rapidity ($|y|<1$) and energy density of central regions as functions of time in 0-10 \% central Au+Au collisions at $\sqrt{s_{\rm NN}}=$19.2, 62.4, and 200 GeV.
We can see that a considerable energy and transverse momentum loss happens in the initial stage of heavy-ion collisions, because the energy density is extremely large as shown in panel (d).
For $\sqrt{s_{\rm NN}}=$19.2 GeV, however, the momentum loss is delayed, since it takes some time for two Au nuclei to pass through each other and produce charm quarks.
On the other hand, charm quarks, which have initially low transverse momentum, gain momentum due to the thermal motion of nuclear matter, which is larger at higher collision energies.

\subsection{Azimuthal angular correlations}

Finally we analyze the azimuthal angle between the transverse
momentum of a heavy-flavor meson and that of an antiheavy-flavor
meson for each heavy flavor pair before and after the interactions
with the medium in relativistic heavy-ion collisions. It is
suggested that the analysis of the azimuthal angular correlation
might provide information on the energy loss mechanism of heavy
quarks in the QGP~\cite{Cao:2015cba} because stronger interactions
should result in less pronounced angular correlations. Since in the
PHSD we can follow up the fate of an initial heavy  quark-antiquark
pair throughout the partonic scatterings, the hadronization and
final hadronic rescatterings, the microscopic calculations allow to
shed some light on the correlation between the in-medium
interactions and the final angular correlations.

\begin{figure*}[t]
\centerline{
\includegraphics[width=6.0 cm]{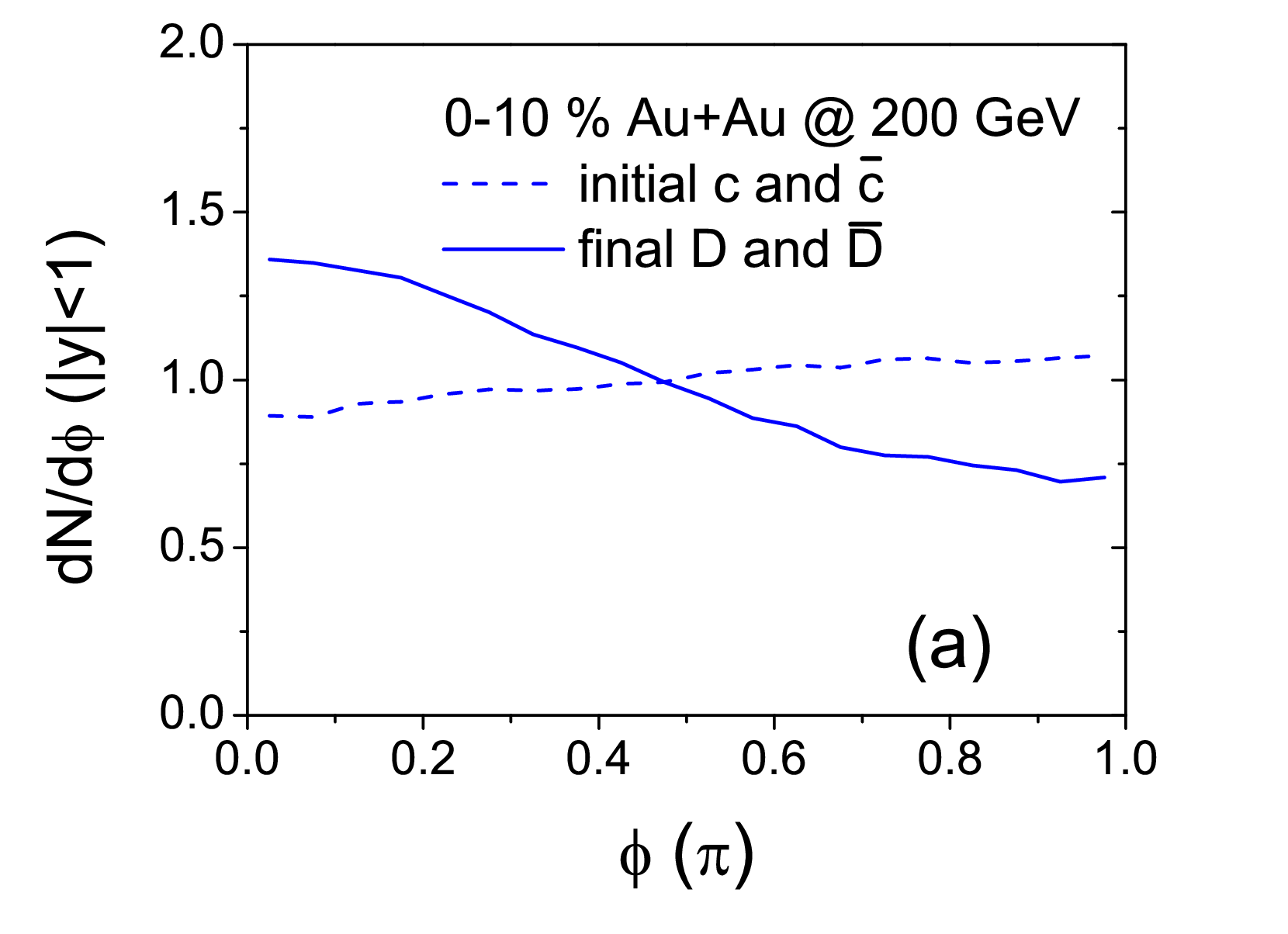}
\includegraphics[width=6.0 cm]{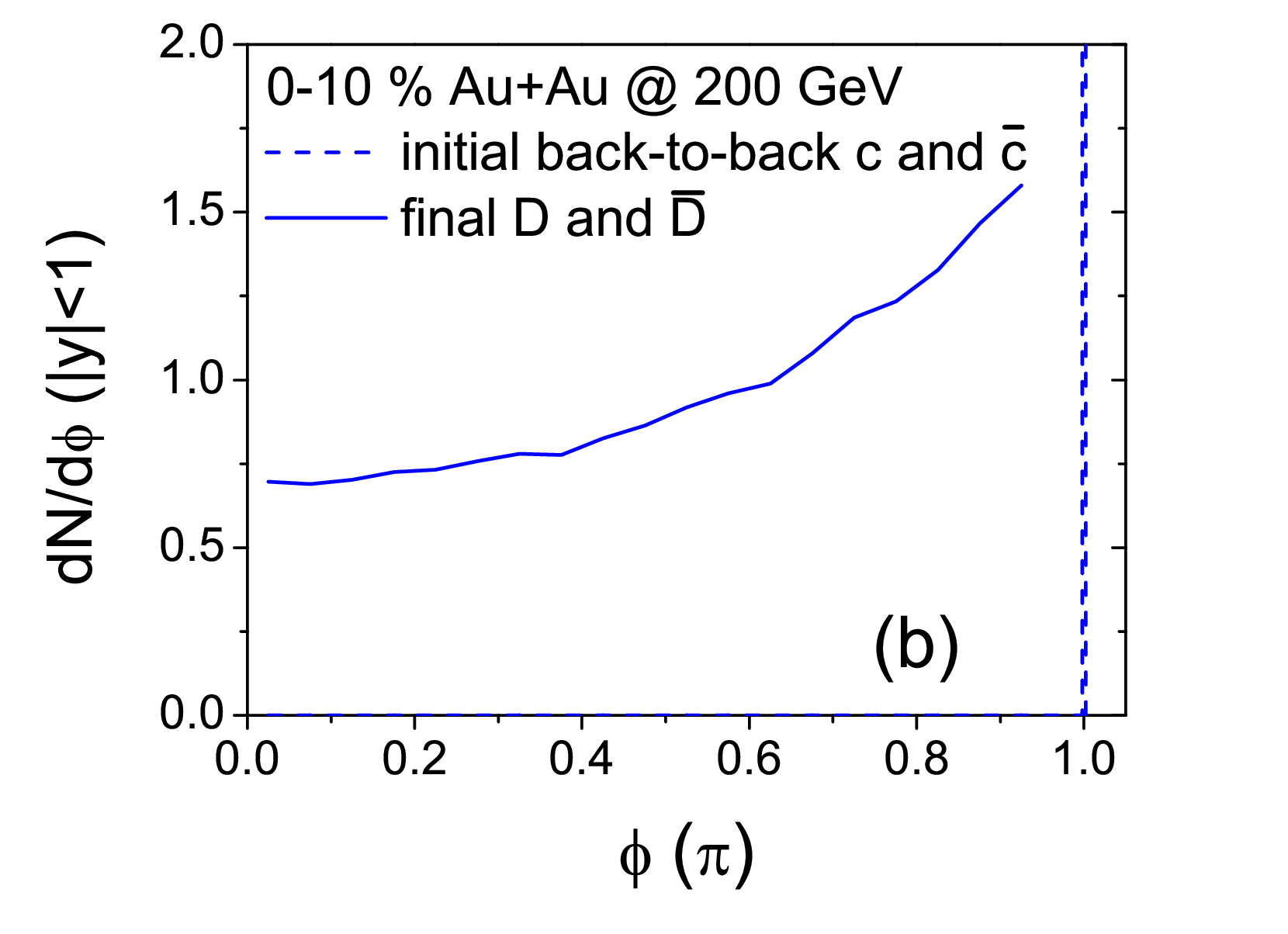}
\includegraphics[width=6.0 cm]{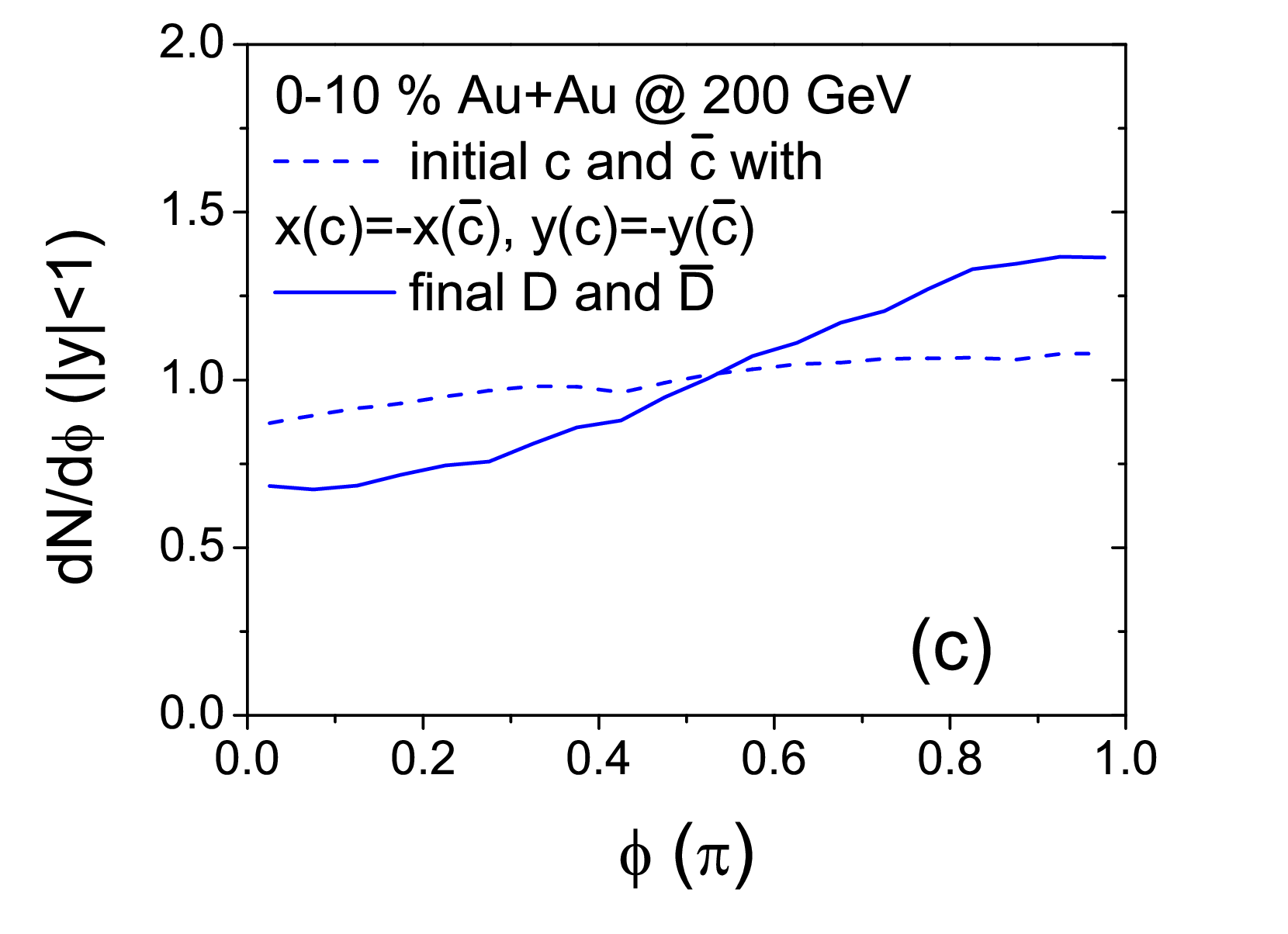}}
\centerline{
\includegraphics[width=6.0 cm]{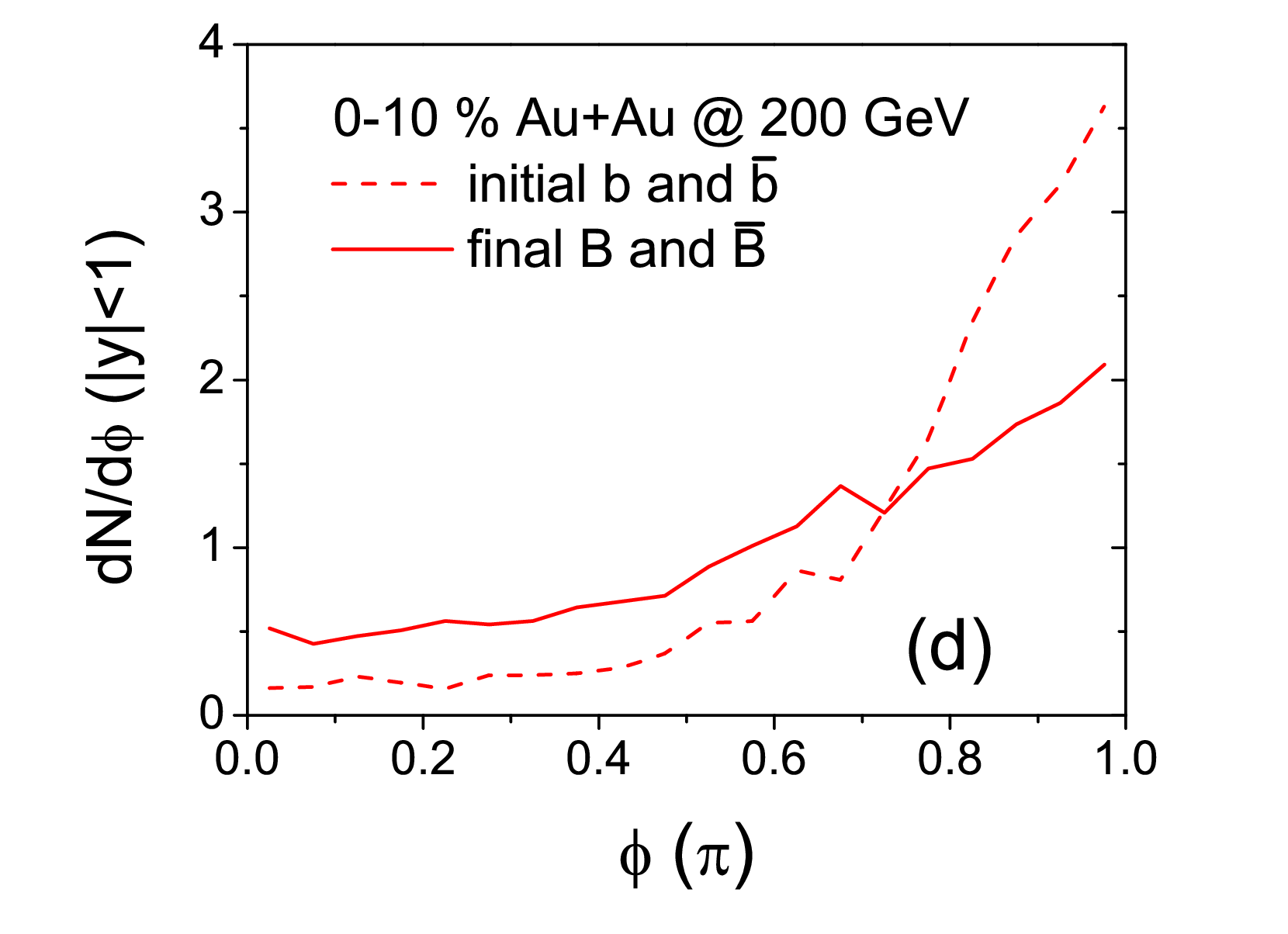}
\includegraphics[width=6.0 cm]{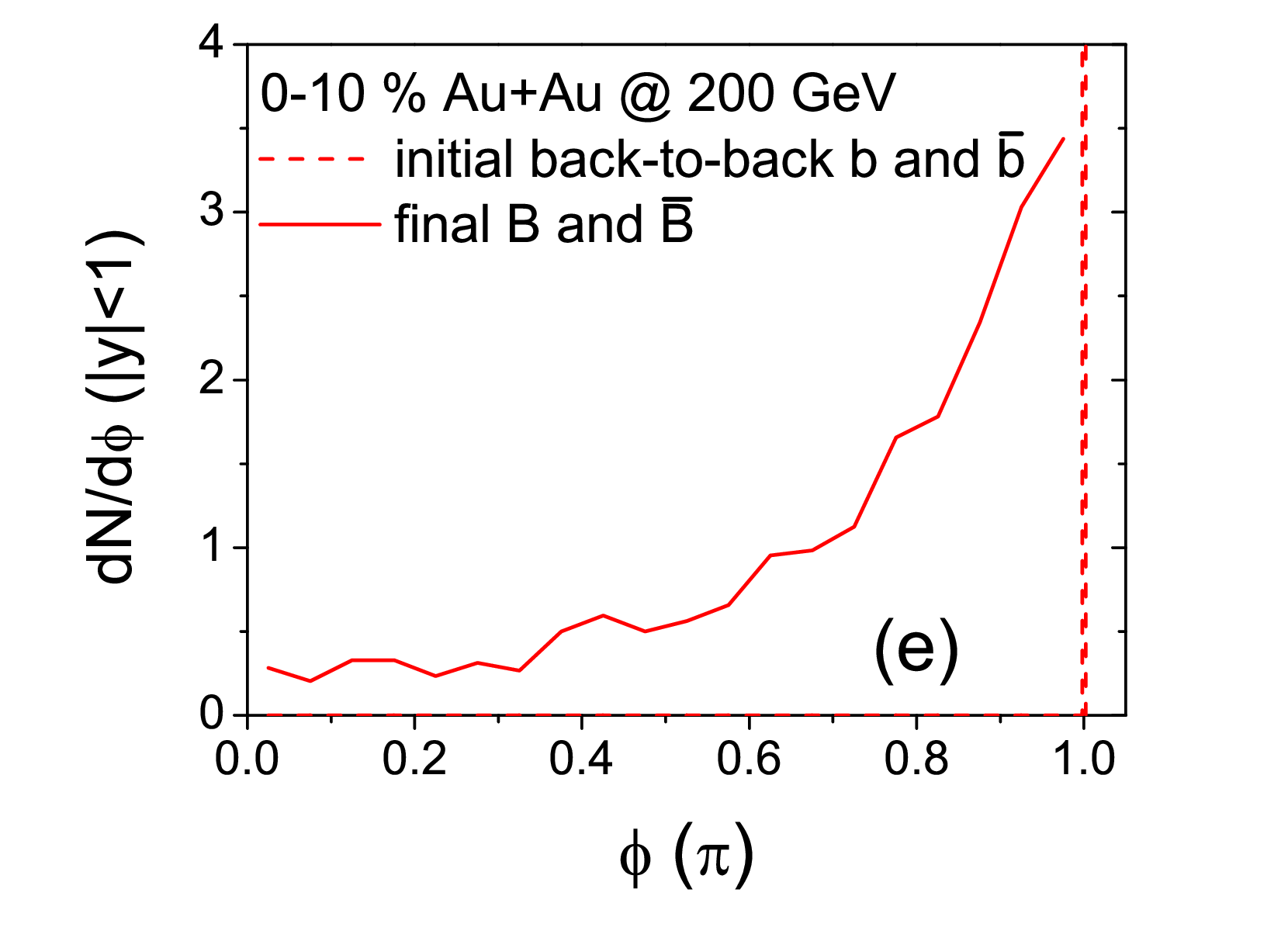}
\includegraphics[width=6.0 cm]{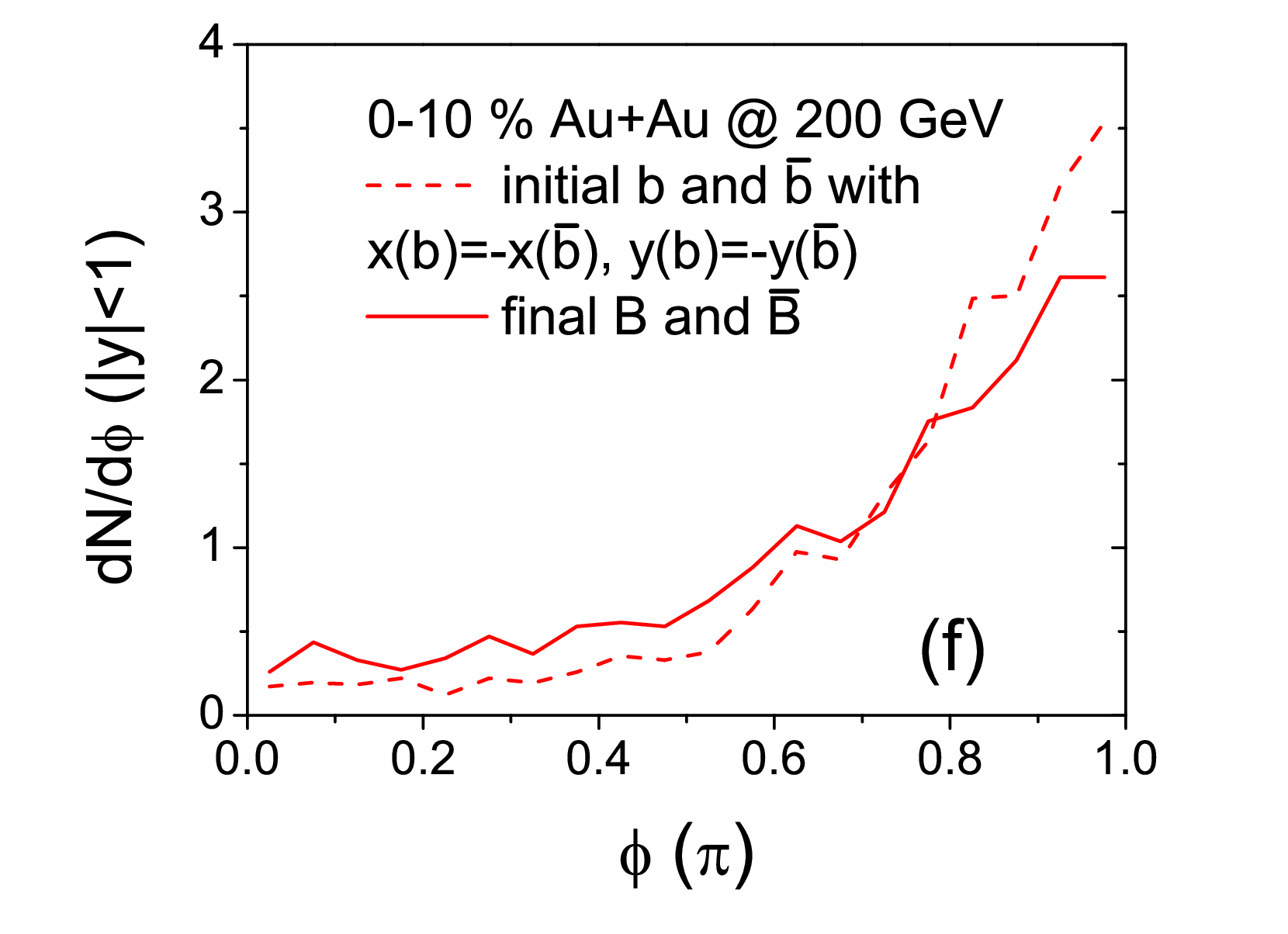}}
\caption{((Color online) The azimuthal angular correlation of initial
$c$ and $\bar{c}$ quarks (upper dashed) and final $D$ and $\bar{D}$
mesons (upper solid) and that of initial $b$ and $\bar{b}$ quarks
(lower dashed) and final $B$ and $\bar{B}$ mesons (lower solid) at
midrapidity ($|y|<1$) in 0-10 \% central Au+Au collisions at
$\sqrt{s_{\rm NN}}=$200 GeV.
(a) and (d) are from normal initial conditions, (b) and (e) from initial back-to-back heavy quark pairs, and (c) and (f) with the initial transverse position of
the heavy antiquark being opposite to that of the heavy quark in order
to investigate the flow effect on the angular correlation}
\label{corr}
\end{figure*}


Figure~\ref{corr} shows the azimuthal angular distributions of charm
and bottom pairs, respectively, in the upper left (a) and lower left (d)
panels at midrapidity ($|y|<1$) in 0-10 \% central Au+Au collisions
at $\sqrt{s_{\rm NN}}=$200 GeV. The dashed lines are the
correlations before the interactions with the medium in heavy-ion
collisions and the solid lines are those after freeze out of the
final heavy mesons. The initial azimuthal correlation of charm pairs
-- produced by the PYTHIA event generator -- is far from
back-to-back due to the associated production of further
quark-antiquark pairs. The distribution in the azimuthal angles
spreads widely from $0$ to $\pi$ although slightly more
populated close to $\phi=\pi$ (back-to-back). We recall that if a
heavy-quark pair is produced to the leading order in pQCD, the heavy
quark and heavy antiquark are back-to-back (or close to it) in the
transverse plane, assuming the transverse momentum of partons --
producing the pair -- to be small. However, the PYTHIA event
generator also takes into account the gluon splitting $(g\rightarrow
Q\bar{Q})$ which is populated near $\phi=\pi$ and the heavy quark
excitation $(gQ\rightarrow gQ)$ where a heavy quark or heavy
antiquark is produced from the parton distribution of the colliding
nucleon~\cite{Sjostrand:2006za}. As a result, the initial charm
pairs from PYTHIA have a mild dependence on $\phi$ at the  energy
$\sqrt{s_{\rm NN}}=$200 GeV. In the case of bottom quarks, however,
the initial pairs from PYTHIA are manifestly peaked near $\phi=\pi$,
as shown in the lower left panel (d) of figure~\ref{corr}, because the
contribution from gluon splitting and heavy quark excitation is
small compared to charm at $\sqrt{s_{\rm NN}}=$200 GeV.


After the initial production the heavy flavor partons strongly
interact with the medium in relativistic heavy-ion collisions.
Figure~\ref{interaction} shows that more than 97 \% of the heavy
quarks interact by scattering or coalescence with other partons at
$\sqrt{s_{\rm NN}}=$200 GeV. Accordingly, the azimuthal angular
correlation between the initial heavy quark and heavy antiquark is
washed out to a large extend.

In order to investigate the effect of the in-medium
interactions on the initial heavy quark-antiquark correlations we
have performed a model study where the initial heavy-quark pairs are
always produced back-to-back ($\phi=\pi$). Although the initial
correlations are all located at $\phi=\pi$, the correlations for
charm pairs disappear after the interactions with the medium in
heavy-ion collisions as seen from the upper middle panel (b) of
figure~\ref{corr}. However, the lower middle panels (e) of figure~\ref{corr} show that the initial angular correlation of bottom
pairs survives to some extend, because the bottom quark is too heavy
to change the direction of motion in  elastic scattering.
%

In the upper left panel of figure~\ref{corr} (a), we can see that the
azimuthal angular correlation is remarkably enhanced near $\phi=0$,
which implies that the $D$ and $\bar{D}$ mesons, which are produced
as one pair in the initial stage, move in a similar transverse
directions at freeze out. A possible reason for this behavior is the
transverse flow: A pair of charm and anticharm quarks are produced
at the same point through a nucleon-nucleon binary collision. In
case the scattering cross sections of charm and anticharm quarks are
large such that they are stuck in the medium and not separated far
from each other, they will be affected by similar transverse flows,
because the flow will depend on the position of the particles. Only
in case the charm and anticharm quarks are separated far enough to
be located at completely different transverse positions until the
transverse flow is generated in heavy-ion collisions, the collective
flows of charm and anticharm quarks, respectively,  will be
independent.


In order to investigate in particular the flow effect on the angular
correlation, we reflect the transverse position of the initial heavy
antiquark with respect to the origin in right panels of figure~\ref{corr}. Since it
is a reflection of the transverse position, the initial azimuthal
angular correlation in momentum space does not change. However, we
find that the distribution of azimuthal angles between the final $D$
and $\bar{D}$ mesons from the same pair is completely opposite to
that without the reflection, which are  shown, respectively, in the
upper panels (a) and (c) of figure~\ref{corr}. If the charm and
anticharm quarks from one pair can move a considerable distance and
be separated far enough from each other before the transverse flow
is developed, the results with and without the reflection should be
similar. But the results in figure~\ref{corr} indicate that the
interaction of charm and anticharm quarks with the medium is strong
and they get stuck in the nuclear matter and flow together.
Accordingly, the results in the panel (a) of figure~\ref{corr} --
indicating that the charm and anticharm quarks from one pair exhibit
similar flows depending on position -- are naturally explained due
to common collective flow.

\section{Summary}\label{summary}

We have studied single electron production through the semileptonic
decay of heavy mesons in relativistic heavy-ion collisions at
$\sqrt{s_{\rm NN}}=$200, 62.4, and 19.2 GeV within the PHSD
transport approach. The ratio of the initial scattering cross
section for bottom production to that for charm production at these
collision energies is less than 1 \%. However, since the $\rm p_T$
spectrum of bottom quarks is harder than that of charm quarks and
the single electrons from $B-$meson decay is much more energetic
than that from $D-$meson decay, it is essential to take $B-$meson
production into account in order to study the single electron
production, especially at high $\rm p_T$.

The Parton-Hadron-String Dynamics (PHSD) approach has been employed
since it successfully describes $D-$meson production in relativistic
heavy-ion collisions at RHIC and LHC
energies~\cite{Song:2015sfa,Song:2015ykw}. In this work, we have
extended the PHSD to $B-$meson production and compared single
electron production from heavy-meson decays  with the experimental
data from the PHENIX collaboration, because there are no
experimental data exclusively for $B-$mesons at the RHIC energies.

In analogy to the charm quark pairs, the bottom pairs are produced
by using the PYTHIA event generator which is tuned to reproduce the
$\rm p_T$ spectrum and rapidity distribution of bottom quark pairs
from the FONLL calculations. The (anti)shadowing effect, which is
the modification of the nucleon parton distributions in a nucleus,
is implemented by means of the EPS09 package. We have found that the
(anti)shadowing effect is not so strong at RHIC energies as compared
to  LHC energies \cite{Song:2015ykw}.

The  charm and bottom partons - produced by the initial hard
nucleon-nucleon scattering - interact with the massive quarks and
gluons in the QGP by using the scattering cross sections calculated
in the Dynamical Quasi-Particle Model (DQPM) which reproduces
heavy-quark diffusion coefficients from lattice QCD calculations at
temperatures above the deconfinement transition. When approaching
the critical energy density for the phase transition from above, the
charm and bottom (anti)quarks are hadronized into $D-$ and
$B-$mesons through the coalescence with light (anti)quarks. Those
heavy quarks, which fail in coalescence until the local energy
density is below 0.4 $\rm GeV/fm^3$, hadronize by fragmentation as
in p+p collisions. The hadronized $D-$ and $B-$mesons then interact
with light hadrons in the hadronic phase with cross sections that
have been calculated in an effective Lagrangian approach with
heavy-quark spin symmetry. Finally, after freeze-out of the $D-$ and
$B-$mesons they produce single electrons through semileptonic decays
with the branching ratios given by the Particle Data Group (PDG).

We have found that the coalescence probability for bottom quarks is
still large at high $\rm p_T$ compared to charm quarks, and the
$R_{\rm AA}$ of $B-$mesons is larger than that of $D-$mesons at the
same (high) $\rm p_T$. However, this can dominantly be attributed to
the much larger mass of the bottom quark. If the coalescence
probability and the $R_{\rm AA}$ are expressed as a function of the
transverse velocity of the heavy quark, both charm and bottom
coalescence become similar since both are comoving with the
neighbouring light antiquarks.

Furthermore, we found that the PHSD approach can roughly reproduce
the experimental data on single electron production in d+Au and Au+Au
collisions at $\sqrt{s_{\rm NN}}=$200 GeV and the elliptic flow of
electrons at $\sqrt{s_{\rm NN}}=$62.4 GeV from the PHENIX
collaboration. However, the $R_{\rm AA}$ at $\sqrt{s_{\rm NN}}=$62.4
GeV is clearly underestimated which presently remains as an open
puzzle. We have additionally made predictions for $D-$meson and
single electron production in Au+Au collisions at $\sqrt{s_{\rm
NN}}=$19.2 GeV which can be controlled by experiment in future.

Finally, we have studied the medium modifications of the azimuthal
angular correlation of heavy-flavor pairs in central Au+Au
collisions at $\sqrt{s_{\rm NN}}=$200 GeV. Here it has been found
that the initial azimuthal angular  correlation of charm pairs is
completely washed out during the evolution of the heavy-ion
collision, even in case they are assumed to be initially produced
back-to-back. This decoherence could be traced back to the
transverse flow which drives charm pairs (close in space) into the
same direction such that the azimuthal angular correlation is
enhanced around $\phi=0$. By considering that the direction of the
transverse flow essentially depends on position, the charm and
anticharm quarks from each pair apparently are not sufficiently
separated from each other before the transverse flow is developed.
This decorrelation thus can be attributed to the strong interactions
of charm with the medium produced in relativistic heavy-ion
collisions. On the other hand, the focussing of pairs around
$\phi=0$ is not observed for bottom pairs at RHIC energies due to
their significantly higher mass which prevents the bottom quarks
(and mesons) to change their momentum substantially in the
scattering processes.

\section*{Acknowledgements}
The authors acknowledge inspiring discussions with J.~Aichelin,
S. Brodsky, P. B. Gossiaux, and P. Moreau.
This work was supported by DFG under contract BR 4000/3-1 and
by the LOEWE center "HIC for FAIR". The computational resources have been
provided by the LOEWE-CSC.
LT acknowledges support from the Ram\'on y Cajal research programme and
FPA2013-43425-P Grant from Ministerio de Economia y Competitividad.
JMTR acknowledges the financial support from a Helmholtz Young Investigator Group VH-NG-822 from the Helmholtz Association and GSI.
DC acknowledges support from Grant Nr.
FIS2014-51948-C2-1-P from Ministerio de Economia y Competitividad, Spain.

\end{document}